\documentclass[a4paper,aps,preprintnumbers,showpacs,twocolumn,groupedaddress,nofootinbib]{revtex4-1}

\usepackage{amsmath}
\usepackage{amssymb}
\usepackage{graphicx}
\usepackage{epsfig}
\usepackage{color}
\usepackage{url}
\usepackage{times}
\usepackage{bm}
\usepackage{mathrsfs}
\usepackage[utf8]{inputenc}
\usepackage{hyperref}
\usepackage{enumerate}
\usepackage{amsthm}
\usepackage{slashed}
\usepackage{amsfonts}

\newcommand{\beq}{\begin{equation}}
\newcommand{\eeq}{\end{equation}}
\newcommand{\bea}{\begin{eqnarray}}
\newcommand{\eea}{\end{eqnarray}}
\newcommand{\bit}{\begin{itemize}}
\newcommand{\eit}{\end{itemize}}
\newcommand{\ben}{\begin{enumerate}}
\newcommand{\een}{\end{enumerate}}
\newcommand{\nn}{\nonumber}

\def\i{\textrm{i}}

\def\atanh{\textrm{atanh}}

\newcommand{\QMUL}{\affiliation{School of Mathematical Sciences, Queen Mary University of
  London, \\ Mile End Road, London E1 4NS, United Kingdom}}
          
\theoremstyle{plain}

\newcounter{mnotecount}

\newcommand{\mnotex}[1]
{\protect{\stepcounter{mnotecount}}$^{\mbox{\footnotesize $\bullet$\themnotecount}}$ 
\marginpar{
\raggedright\tiny\em
$\!\!\!\!\!\!\,\bullet$\themnotecount: #1} }

\begin{document}
\title{End point of nonaxisymmetric black hole instabilities in higher dimensions}
\author{Hans Bantilan}
\author{Pau Figueras}
\author{Markus Kunesch}\thanks{Until August 2018}
\author{Rodrigo Panosso Macedo} 

\QMUL

\begin{abstract}
We report on the end state of nonaxisymmetric instabilities of singly spinning asymptotically flat Myers-Perry black holes.
Starting from a singly spinning black hole in $D=5,6,7$ dimensions, we introduce perturbations with angular dependence described by $m=2$, $m=3$, or $m=4$ azimuthal mode numbers about the axis of rotation. 
In $D=5$, we find that all singly spinning Myers-Perry black holes are stable, in agreement with the results from perturbation theory. 
In $D=6$ and $7$, we find that these black holes are nonlinearly stable only for sufficiently low spins.
For intermediate spins, although the $m=2$ bar mode becomes unstable and leads to large deformations, the black hole settles back down to another member of the Myers-Perry family via gravitational wave emission; surprisingly, we find that \textit{all} such unstable black holes settle to what appear to be the \textit{same} member of the Myers-Perry family. 
The amount of energy radiated into gravitational waves can be very large, in some cases more than $30\%$ of the initial total mass of the system. 
For high enough spins, the $m=4$ mode becomes the dominant unstable mode, leading to deformed black holes that develop local Gregory-Laflamme instabilities, thus forming a naked singularity in finite time, which is further evidence for the violation of the weak cosmic censorship conjecture in asymptotically flat higher-dimensional spacetimes. 
\end{abstract}

\maketitle
\section{Introduction}
Black holes play a central role in general relativity (GR), the currently accepted classical theory of gravity.  
The recent direct detections of gravitational waves from black hole binary mergers \cite{Abbott:2016blz} together with the image of the shadow of the black hole at the center of the M87 galaxy by the Event Horizon Telescope \cite{Akiyama:2019cqa}, have changed the perception of these objects from the purely mathematical to the ``tangible". 
All these observations are compatible with an equilibrium (or quasiequilibrium) state that is described by the Kerr black hole \cite{Kerr:1963ud}. 
However, the astrophysical relevance of the Kerr black hole hinges on whether it is nonlinearly stable or not. 
All the evidence indicates that it is indeed stable, but a rigorous proof of the nonlinear stability of the Kerr black hole is not yet available.  

Following the discovery of the Gregory-Laflamme (GL) instability~\cite{Gregory:1993vy} of black strings and black branes in dimensions 5 and higher, black holes have become useful as laboratories to study dynamical instabilities, particularly due to their simplicity and their fundamental role in GR.
The study of the evolution of black hole instabilities in the fully nonlinear regime has been a fruitful area of research;
indeed, the pioneering work of Lehner and Pretorius~\cite{Lehner:2010pn}, with numerical simulations of the evolution of the GL instability of black strings in five dimensions, provided evidence that the weak cosmic censorship conjecture is false in asymptotically Kaluza-Klein spaces.

In higher-dimensional asymptotically flat spaces, the situation is qualitatively similar. 
Black rings \cite{Emparan:2001wn,Pomeransky:2006bd} are asymptotically flat rotating black holes with horizon topology $S^1\times S^n$. 
In the limit of large angular momentum along the $S^1$, black rings resemble thin (boosted) black strings, and hence they ought to be unstable under a GL type of instability. 
This was confirmed in Ref.~\cite{Santos:2015iua}. 
Reference~\cite{Figueras:2015hkb} used numerical general relativity to simulate the nonlinear evolution of black ring instabilities and showed that thin enough black rings evolve into naked singularities in finite time, thus potentially violating the weak cosmic censorship in higher-dimensional asymptotically flat spaces.\footnote{There are several issues that still need to be addressed, in particular, the structure of null infinity. In these dynamical spacetimes that develop naked singularities, it is not clear whether null infinity is incomplete. }  
This situation is not unique to black rings.
In fact, a general picture of the stability/instability properties of higher-dimensional black holes has emerged. 
Emparan and Myers~\cite{Emparan:2003sy} noticed that rapidly rotating black holes in higher dimensions, i.e., Myers-Perry (MP)  black holes (BHs), resemble black membranes and hence should also be GL unstable. 
Higher-dimensional black holes generically admit a regime of large angular momentum with highly deformed horizons described by largely separated length scales, captured by the so-called black folds \cite{Emparan:2009cs}, which are expected to be dynamically unstable.
To complete this general picture, it is thus important to determine the end point of these black hole instabilities in the rapidly spinning regime. 

The instability of MP BHs was confirmed in Ref.~\cite{Dias:2009iu} for linearized axisymmetric perturbations on top of a MP BH background.\footnote{MP BHs with equal spins on all rotation planes are also known to be unstable \cite{Dias:2010eu}. For these black holes, unlike the singly spinning ones, the total angular momentum is bounded and hence the endpoints of their instabilities can potentially be quite different. We will not consider MP BHs with equal spins in this article.} 
A fully nonlinear evolution of this axisymmetric instability of MP BHs revealed a sequence of concentric rings connected by black brane segments that became progressively thinner, eventually leading to a naked singularity in finite asymptotic time \cite{Figueras:2017zwa}. 
In the MP case, the dynamics of the horizon is not self-similar, unlike what was seen in the black string.
The study of nonaxisymmetric instabilities was initiated by Shibata and Yoshino in Refs.~\cite{Shibata:2009ad,Shibata:2010wz}. 
For a MP BH in $D$ dimensions with mass parameter $\mu$ and spin parameter $a$, these nonaxisymmetric instabilities set in at smaller dimensionless spin $a/\mu^{1/(D-3)}$ than the axisymmetric instabilities. 

Rotating black holes become dynamically unstable for values of $a/\mu^{1/(D-3)} \sim O(1)$.
Indeed, from thermodynamic considerations of the case in which a MP BH fragments and expels two nonspinning BHs, Emparan and Myers~\cite{Emparan:2003sy} estimated that nonaxisymmetric instabilities should set in at around $a/\mu^{1/(D-3)} \approx 1$. References~\cite{Shibata:2009ad,Shibata:2010wz} found that MP BHs with $a/\mu^{1/(D-3)} \lesssim 0.7$ are stable in $D=5,6,7$, but for higher spins, the MP BHs are  linearly unstable to a deformation of which the azimuthal angle dependence is $e^{i m\phi}$ for $m=2$, i.e., a bar-mode deformation.
In six and seven dimensions~\cite{Shibata:2010wz}, they found that, due to gravitational wave emission, this bar-mode instability saturates and eventually damps, and the black hole settles down to a MP BH with a lower spin.
A study of quasinormal modes of MP BHs~\cite{Dias:2014eua} corroborates these results, except in the five-dimensional (5D) case; while Ref.~\cite{Shibata:2009ad} found an instability, Ref.~\cite{Dias:2014eua} did not find an exponentially growing mode in the linear regime. 
These two results could be compatible with each other if in five dimensions the instability were nonlinear. 
In this article, we resolve the apparent tension between these linear and the nonlinear results.  
Analogous to the GL instability of black strings, for which higher harmonics become unstable for sufficiently thin strings, one would expect that for sufficiently large $a/\mu^{1/(D-3)}$ modes with $m>2$ become the dominant unstable modes.  
This is indeed the case for black rings \cite{Figueras:2015hkb}, and we show here that the same happens for MP BHs.

In this paper, we investigate the nonlinear evolution of MP BHs in $D=5,6,7$ dimensions with dimensionless spins of $0.7 \leq a/\mu^{1/(D-3)} \leq 1.5$, and perturbed by nonaxisymmetric deformations described by the $m=2$, $m=3$, and $m=4$ azimuthal mode numbers. 
We use the cartoon method \cite{Pretorius:2004jg, Shibata:2010wz,Cook:2016soy} to impose an SO($D-3$) symmetry that still captures the nonaxisymmetric instability while allowing us to restrict to $3+1$ dynamics in $D$ dimensions. 
We believe that this symmetry assumption should still be enough to capture the essential physics of the nonaxisymmetric instabilities and their end points. 

The rest of this article is organized as follows: in Sec.~\ref{sec:NumMeth}, we provide an overview of the numerical methods that we used in our simulations. 
Section~\ref{sec:results} constitutes the bulk of the article, and there we present our results for the nonlinear evolution of unstable MP BHs in various spacetime dimensions and for different values of the dimensionless spin parameter. 
We summarize our results and conclude in Sec~\ref{sec:discussion}. 
Technical results are collected in the Appendixes.   
In Appendix~\ref{sec:AFhigherD} we review the notion of asymptotic flatness in higher dimensions introduced in Ref.~\cite{Godazgar:2012zq} and how the matrix of Weyl scalars captures gravitational radiation. 
In Appendix~\ref{sec:Snharmonics}  we collect several properties of tensor spherical harmonics. 
Appendices~\ref{sec:AppS3} and~\ref{sec:AppS4} contain the tensor harmonics on the $S^3$ and the $S^4$, respectively, that we have used in our extractions. 
In Appendix~\ref{sec:WeylTransf} we review the transformation properties of the multipoles of the Weyl tensor under changes of basis. 
In Appendix~\ref{sec:AppHrznChi} we compare the contours of $\chi$ with the apparent horizon, in Appendix~\ref{sec:Pert_m3} we display contours of $\chi$ from the evolution of six dimensions with an $m=3$ perturbation, and Appendix~\ref{sec:AppConv} contains some convergence tests.

\section{Numerical Methods}
\label{sec:NumMeth}

In this section, we summarize the numerical methods that we have employed in our simulations. In Sec.~\ref{sec:evol} we describe our evolution scheme together with the our choice of initial data and gauge evolution equations. In Sec.~\ref{sec:gws} we provide some details about our approach to extracting gravitational waves in higher dimensions. 

\subsection{Evolution scheme}
\label{sec:evol}

The results presented here are obtained by solving the Einstein field equations in the conformal and covariant Z4 (CCZ4) formulation~\cite{Alic:2011gg,Weyhausen:2011cg}.
We use Cartesian coordinates to solve for the evolution of asymptotically flat black hole solutions in $D=5,6,7$ dimensions while imposing an SO($D-3$) symmetry using the modified cartoon method~\cite{Pretorius:2004jg,Shibata:2010wz,Cook:2016soy}. 
We redefine the constraint damping parameter $\kappa_1 \rightarrow \kappa_1/\alpha$ as in Ref.~\cite{Alic:2013xsa}, in which $\alpha$ is the lapse function, and we typically use constraint damping values $\kappa_1=0.4,\kappa_2=0$.  
As in Ref.~\cite{Figueras:2017zwa}, we use initial data for a singly spinning MP BH,
\begin{align}
\label{eq:MPdata}
ds^2 =& -dt^2 + \frac{\mu}{r^{D-5}\Sigma} (dt-a\,\sin^2\theta\,d\phi)^2  + \frac{\Sigma}{\Delta}\,dr^2 \\
&+ \Sigma\,d\theta^2 + (r^2+a^2)\sin^2\theta\,d\phi^2 + r^2\cos^2\theta\,d\Omega^2_{(D-4)}\,,\nn
\end{align}
where $\Sigma = r^2+a^2\cos^2\theta$, $\Delta = r^2 + a^2 - \mu\,r^{5-D}$, and $\mu$ and $a$ are the mass and spin parameters of the MP BH. 
In analogy with the transformation from Schwarzschild coordinates to isotropic coordinates in $D$ dimensions, we define a new radial coordinate $\rho$,
\beq
r = \rho \left[1 + \frac{1}{4}\left(\frac{r_h}{\rho}\right)^{D-3}\right]^\frac{2}{D-3}
\eeq
where $r_h$ is the largest real root of $\Delta(r_h)=0$. 
Then, the Cartesian coordinates are given by
\begin{equation}
\begin{aligned}
&x = \rho \sin\theta\cos\phi\,,\\
&y = \rho \sin\theta\sin\phi\,,\\
&z = \rho \cos\theta\cos\varphi_1\,,\\
&w_1= \rho \cos\theta\sin\varphi_1\cos\varphi_2\,,\\
&\ldots\\
&w_{D-4} = \rho \cos\theta\sin\varphi_1\dots \sin\varphi_{D-3}\,.
\end{aligned}
\end{equation}
Imposing the SO$(D-3)$ symmetry corresponds to working on the slice $w_1 =\ldots = w_{D-4}=0$; see Refs.~\cite{Pretorius:2004jg, Shibata:2010wz,Cook:2016soy}.

For values of $a/\mu^{1/3}\sim 1.5$, numerical noise is enough to trigger the instability in the $m=4$ sector.  
Note that although the initial (rapid) gauge adjustment induces a small burst of radiation that in practice contains modes with different $m$, this initial burst is induced by truncation error and hence is under control and small. 
For smaller values of the dimensionless spin, we trigger the instability by hand through an $m=2$ or $m=4$ deformation of the conformal factor $\chi$:
\begin{align}
  \chi = & \chi_0\Bigg\{1 + A\,f_{m}(x,y)  \nn  \\
 & \times \exp\left[-\left(\frac{\atanh(\chi_0)}{\atanh(\chi_H)} - \frac{\atanh(\chi_H)}{\atanh(\chi_0)}\right)^2\right] \Bigg\}\,,
\label{eq:chi_deformation1}
\end{align}
where $\chi_0$ is the conformal factor computed from the analytic initial data \eqref{eq:MPdata}, $\chi_H$ is some value of the conformal factor close to the apparent horizon, $A$ is the amplitude of the perturbation, and $f_m(x,y)$ is the function that induces the desired deformation:
\begin{equation}
\begin{aligned}
&m=2:\quad f_2(x,y) = x^2 - y^2\,,\\
&m=3:\quad f_2(x,y) = x^3 - 3 x y^2\,,\\
&m=4:\quad f_4(x,y) = x^4 - 6\,x^2\,y^2 + y^4\,.
\end{aligned}
\label{eq:chi_deformation2}
\end{equation}
Introducing an $m=2$ perturbation immediately at $t/\mu^\frac{1}{D-3}=0$ results in a dumbbell configuration that eventually settles down to a MP BH. 
Unless specifically stated, simulations described here feature perturbations introduced at $t/\mu^\frac{1}{D-3}=0$. 
For some simulations, we also introduce this deformation in $\chi$ after the gauge adjustment period at around $t/\mu^\frac{1}{D-3}\sim10$. 
For higher spins, this amounts to choosing a different branch of the dynamical evolution.
Perturbing after the gauge adjustment period at around $t/\mu^\frac{1}{D-3}\sim10$ results in an elongated configuration that eventually develops sharp edges at the ends. 
This last case is discussed in detail near the end of Sec.~\ref{sec:mp6d_large_spin}.

This method of perturbing the black hole violates the constraints, so even though the CCZ4 formulation quickly\footnote{In practice this timescale is exponential and much faster than the physical timescale of the instabilities.} takes us back to the constraint surface, we do not have control over where on the constraint surface we land. 
Therefore, to make sure that the mass and angular momentum of the perturbed data are not too far from those of the original MP BH we keep $A$ small (a typical value that we use is $A=0.02$). 

We evolve the lapse with the standard $1+\log$ slicing condition \cite{Alic:2011gg}, 
\begin{equation}
(\partial_t -\beta^i\partial_i)\alpha = c_\alpha\,\alpha \left(K - 2\,\Theta\right)\,,
\end{equation}
where $c_\alpha$ is a freely adjustable coefficient. 
In our runs we found that $c_\alpha=1.5$ worked well for $D=5,6,7$.\footnote{The preferred choice in the typical four-dimensional astrophysical setting is $c_\alpha=2$. } To evolve the shift, we used the modified Gamma-driver condition introduced in Ref.~\cite{Figueras:2015hkb}, 
\begin{equation}
(\partial_t -\beta^j\partial_j)\beta^i = c_\beta\,\left(\hat\Gamma^i - f(t,\rho)\, \bar\Gamma^i\right) - \eta\, \beta^i\,,
\label{eq:shiftcond}
\end{equation}
where $\hat\Gamma^i$ is the usual CCZ4 evolution variable, $\bar\Gamma^i$ is the contracted Christoffel symbol computed from the (conformally rescaled) initial data \eqref{eq:MPdata}, and
\begin{equation}
f(t,\rho) = \exp\left[-\left(\frac{t}{\mu^{1/(D-3)}}\right)^2\left( \delta_1\left(\frac{\rho_h}{\rho}\right)^2 + \delta_2\right)\right]\,,
\label{eq:gammadriver}
\end{equation}
with $\rho_h = r_h/(4^\frac{1}{D-3})$ and where $\delta_{1,2}$ are two adjustable parameters that control the decay of the initial gamma function near the horizon. 
Since the initial data have $\tilde\Gamma^i\neq0$, the role of the extra term in \eqref{eq:shiftcond} is to drive the gauge towards $\tilde\Gamma^i=0$ while making sure that the right-hand side of the equation remains relatively small throughout the evolution.  
Typical values of these parameters in our runs are $\delta_1 =0.2$ and $\delta_2=0.075$.
We can freely adjust the coefficient $c_\beta$ in \eqref{eq:gammadriver}; $c_\beta = 0.6$ works well for us (note that the typical value in four-dimensional simulations is $c_\beta = 0.75$). 
Following Ref.~\cite{Figueras:2015hkb}, we introduce diffusion terms (well inside the apparent horizon) on the right-hand side of the equations of motion for those variables that appear with second spatial derivatives. 
This improves the stability of our code, especially in the rapidly spinning regime. 
See Ref.~\cite{Figueras:2015hkb} for more details. 
We use Kreiss-Oliger dissipation~\cite{kreiss1973methods} to damp unphysical high-frequency modes that can arise at grid boundaries or during regridding, with typical dissipation values of $\sigma=0.4$. 
We numerically solve the CCZ4 equations using the {\sc GRChombo} libraries~\cite{Clough:2015sqa,Adams:2015kgr}, using up to nine levels of refinement, with a 2:1 refinement ratio, and typically $140^3$ points on the coarsest grid. We consider a cubic computational domain, with side length $L=200$.  The $x$ and $y$ are standard Cartesian coordinates, with infinite range, which we cut off at some finite values; we typically impose periodic boundary conditions (BCs) for simplicity, but we have also experimented with Sommerfeld BCs, and the results are essentially the same. Needless to say, our choice of  BCs limits the duration of our simulations. The $z$ coordinate has range $0\leq z \leq L$; at $z=0$, we impose regularity as derived from the cartoon method, and at $z=L$, we impose either periodic or Sommerfeld BCs. 
We discretize the equations using fourth-order finite differences and integrate in time using 4th-order Runge-Kutta method (RK4). 
We obtain approximately third-order convergence, see Appendix~\ref{sec:AppConv} for convergence tests.

We also track the apparent horizon (AH) during several stages of the evolution, (see Ref.~\cite{Figueras:2015hkb}); our apparent horizon finder assumes that the AH is a star-shaped surface. 
While this is sufficient to describe the AH for the low-spin runs, this assumption breaks down when the deformations are too large. 
In this case, one would need to use a more general parametrization of the AH along the lines of Ref.~\cite{Figueras:2017zwa} or more recently Ref.~\cite{Pook-Kolb:2018igu}. 
We leave this for future work. 
However, as in Ref.~\cite{Figueras:2015hkb}, we verify that in our working gauge certain contours of $\chi$ track the AH to within a few percent. 
Thus, we can use $\chi$ to get a good approximation of the location and shape of the AH. 
In Appendix~\ref{sec:AppHrznChi} we verify this claim with cases in which the AH satisfies our star-shaped assumption and can thus be calculated and compared to contours of $\chi$. 

\subsection{Gravitational wave extraction}
\label{sec:gws}

One of the aims of this article is to provide a general picture of higher-dimensional black hole instabilities and their end points.
To this end, the gravitational waves that are emitted during the evolution of these instabilities are a valuable source of insight.
We are able to do this in two ways.
The first is to extract the $h_+$ and $h_\times$ components of the metric perturbations at a certain radius along the $z$ axis \cite{Shibata:2009ad,Shibata:2010wz},
\begin{align}
&h_+ = \frac{\tilde \gamma_{xx} - \tilde\gamma_{yy}}{2}\,\left(\frac{z}{\mu^{1/(D-3)}}\right)^\frac{D-2}{2}\,,\\
&h_\times = \tilde\gamma_{xy}\,\left(\frac{z}{\mu^{1/(D-3)}}\right)^\frac{D-2}{2}\,.
\label{eq:metric_perturbation}
\end{align}
As noted in these references, $h_+$ and $h_\times$ basically contain the same information. 
This method has the advantage that it is very easy to implement and it accurately captures the $m=2$ modes, allowing us to compare our results with the existing results in the literature  \cite{Shibata:2010wz,Dias:2014eua}. 
However, this approach misses the higher $m$ modes and in particular the $m=4$ mode that becomes the dominant one at larger spins. 
Note that the perturbative calculations of Ref.~\cite{,Dias:2014eua} only consider modes with $m=2$. 
Therefore, for this work, we have implemented a completely general approach to gravitational wave extraction in higher dimensions based on the higher-dimensional analogues of the Newman-Penrose scalars introduced in Ref.~\cite{Godazgar:2012zq}. 
We follow the implementation of Ref.~\cite{Cook:2016qnt} to calculate the nonvanishing components of the Weyl tensor along the outgoing null rays, $\Omega'_{(A)(B)}$.  For the class of spacetimes that we consider, the nonvanishing components of $\Omega'_{(A)(B)}$ are \cite{Cook:2016qnt}
\begin{align}
&\Omega'_{(\hat \imath)(\hat \jmath)} =~ \frac{1}{2}\Big[R_{0k0l} m^k_{(\hat\imath)}m^l_{(\hat\jmath)} - R_{mk0l} m^m_{(\hat 1)}m^k_{(\hat\imath)}m^l_{(\hat\jmath)} \nn \\
&~ - R_{0kml} m^k_{(\hat\imath)}m^m_{(\hat 1)} m^l_{(\hat\jmath)} + R_{mknl} m^m_{(\hat 1)} m^k_{(\hat\imath)} m^n_{(\hat 1)} m^l_{(\hat\jmath)} \Big]\,,\\
&\Omega'_{(\hat a)(\hat b)} =~ \delta_{(\hat a)(\hat b)} \frac{1}{2\gamma_{ww}}\Big[ R_{w0w0} - 2\,R_{w0wk}m^k_{(\hat 1)} \nn \\
& - R_{wkwl}m^k_{(\hat 1)}m^l_{(\hat 1)}\Big]\,,
\end{align}
where the indices $\imath,\jmath =2,3$ run along the spatial angular dimensions of the computational domain and the indices $\hat a, \hat b,\ldots$ run along the transverse sphere. Here $R_{0klm}$, etc.; denote the components of the spacetime Riemann tensor, $m^i_{(\hat 1)}$ is the unit radial vector and $m^k_{(\hat\imath)}$  are the angular unit vectors. To construct an orthonormal basis of vectors, we start from 
\begin{equation}
\begin{aligned}
m_{(1)} &= x\,\partial_x + y\,\partial_y + z\,\partial_z\,,\\
m_{(2)} &= -(y^2 - z^2)\,\partial_x + x\,y\,\partial_y + x\,z\,\partial_z\,,\\
m_{(3)} &=  -z\,\partial_y + y\,\partial_z\,,\\
\end{aligned}
\end{equation}
and use the standard Gram-Schmidt orthornormalization procedure. Changing to spherical coordinates as in \eqref{eq:cartesian}, one can show that as $r\to\infty$, the basis orthonormalized vectors approach
\begin{equation}
\begin{aligned}
m_{(\hat 1)} &= \frac{1}{r}\partial_r\,,\\
m_{(\hat 2)} &= \frac{1}{r}\partial_{\theta_{n}}\,,\\
m_{(\hat 3)} &= \frac{1}{r\,\sin\theta_{n}}\partial_{\theta_{n-1}}\,.
\end{aligned}
\end{equation} 
The unit vectors along the transverse sphere directions are simply given by,
\begin{equation}
m_{(\hat a)} = \frac{1}{\sqrt{\gamma_{ww}}}\,\partial_{w_a}\,.
\end{equation}

We extract the individual modes in the gravitational wave signal  by projecting $\Omega'_{(A)(B)}$ onto a basis of tensor harmonics on the sphere at infinity:
\begin{equation}
\label{eq:Omega_spheHarm}
\Omega'_{\ell\ldots} = \lim_{r\to\infty} r^\frac{D-2}{2} \int d\Omega_{(n)} Y_{\ell\ldots}^{(A)(B)\ast} \,\Omega'_{(A)B)}
\end{equation}
where  $\ell\ldots$ denotes the set of quantum numbers that identify each of the harmonics, and $d\Omega_{(n)}$ is the volume element on the round unit sphere $S^n$.  In this article we focus on the $n=3,4$ cases but the procedure is completely general. 
In practice, we carry out the extraction at several radii to verify that we are in the wave zone. 
While a basis for tensor harmonics on $S^3$ is known, see, e.g., Ref.~\cite{Lindblom:2017maa}, to the best of our knowledge, our results for tensor harmonics on $S^4$ are new. 
See Appendix~\ref{sec:AppS4}. 
Needless to say, for the $m=2$ modes, both methods give the same frequencies and growth/decay rates within our numerical errors.\footnote{\label{fn:QNM_error}To calculate the QNMs, we first extract the exponential growth/decay rates by fitting a straight line over the envelope of local extrema in the log-linear plot of the given data. The errors associated with this step are mainly due to the particular time interval chosen. The early and late data are contaminated by junk radiation and accumulated numerical error, respectively, whereas the signal shows a transition from growing to decaying regimes for intermediary times.  Next, we use the information of the growth/decay rates to obtain a purely sinusoidal signal, from which we read the oscillatory  frequency via a Fourier transformation. Here, the accuracy is restricted to the wavelengths available in the signal.}

\subsubsection{Energy flux}
Following Ref.~\cite{Cook:2016qnt} (and references therein), the energy emitted in form of gravitational waves is given by
\beq
\label{eq:EnergyFlux}
\dot {M}(u) = - \lim_{r \rightarrow \infty} \dfrac{r^{D-2}}{8\pi} \int d\Omega_{(n)}  I_{(A)(B)} I^{(A)(B)}
\eeq
with
\beq
\label{eq:Integrand}
I_{(\alpha)(\beta)}  = \int_{-\infty}^{u} \Omega'_{(A)(B)} d\tilde{u}.
\eeq
In the expressions above, $u$ is the retarded time coordinate associated to the line element expressed in the Bondi-Sachs form (see Appendix~\ref{sec:AFhigherD}).

Taking into account the decomposition in the tensor spherical harmonics \eqref{eq:Omega_spheHarm}, Eq.~\eqref{eq:EnergyFlux} is expressed as
\beq
\dot{M}(u) = -\dfrac{1}{8\pi} \sum_{\ell\ldots} \left( \int_{-\infty}^{u} \Omega'_{\ell\ldots}(\tilde{u},r) d\tilde{u} \right)^2 \label{eq:EnergyFluxHarmDecomp}.
\eeq
In practice, we neglect the gravitational waves before the start of the simulation and we perform the time integral in terms of the ``computational" time $t$, starting at $t=0$. Total energy radiated, $E_\text{rad}$, can be computed by performing a final time integration.

\section{Results}
\label{sec:results}
We now present results for the nonlinear evolution of nonaxisymmetric perturbations of singly spinning black holes in five and six dimensions.  
The dynamics of MP black holes in seven dimensions is qualitatively similar to the six dimensional (6D) case.
We start by discussing the 5D case in Sec.~\ref{sec:mp5d}, in which we resolve the existing tension in the literature regarding the stability/instability of MP BHs with a sufficiently large spin. 
We move on in Sec.~\ref{sec:mp6d} to study the nonlinear stability properties of 6D MP BHs of various spins: in Sec.~\ref{sec:mp6d_gw}, we study the gravitational waveforms and extract the quasinormal modes (QNMs), in Sec.~\ref{sec:mp6d_ah} we study the physics of the AH, and in Sec.~\ref{sec:mp6d_large_spin} we discuss the end point of the instabilities for large angular momentum.  

In the following, we will use several geometric measures of the AH to estimate the mass and spin of the black hole throughout the evolution, even though the relations between the geometry of the horizon and the physical quantities of the black hole are only valid for equilibrium configurations. 
The following relations between the MP parameters and the horizon are valid in the stationary configuration
\beq
\label{eq:parameters_MP_Hrz}
r_{\rm h} =\left( \frac{A}{C}\frac{ 2\pi}{\Omega_{D-2}}\right)^{\frac{1}{D-3}}, \, \mu_{\rm h}=\frac{C}{2\pi} r_{\rm h}^{D-4}, \, a_{\rm h}=r_{\rm h}\sqrt{\frac{\mu_{\rm h}}{r_{\rm h}^{D-3}} -1}
\eeq 
with $A$ the horizon's area and $C$ its equatorial circumference. 
In the dynamical regime, Eq.~\eqref{eq:parameters_MP_Hrz} defines $r_{\rm h}$, $\mu_{\rm h}$, and $a_{\rm h}$. 
Note that our results scale with the chosen length scale $\mu$, related to the black hole mass of the initial configuration. 
As the evolution proceeds, we dynamically rescale the physical quantities according to the length $\mu_{\rm h}$. 
For instance, nonlinear dynamics that settles to a final MP BH in six dimensions results in a final, dimensionless spin of $(a/\mu^{1/3})_{\rm final} = a_{\rm h}/\mu^{1/3}_{\rm h}$.

Before we proceed, we remind the reader of some well-known properties of singly spinning MP BHs in $D\geq 5$ spacetime dimensions. 
See Ref.~\cite{Emparan:2008eg} and references therein for more details.  
For fixed total mass, in five dimensions, the angular momentum of a MP BH is bounded from above by an extremal bound; configurations saturating this bound have a naked ring singularity. 
This situation should be contrasted with the Kerr family of solutions, for which the solution saturating the Kerr bound corresponds to an extremal black hole with a degenerate horizon. 
On the other hand, for $D\geq 6$, the angular momentum is unbounded from above, and for any finite angular momentum, the black hole horizon is nondegenerate and has finite area. 

\subsection{Case $D=5$}
\label{sec:mp5d}
In Ref.~\cite{Shibata:2009ad}, the nonlinear dynamics of rapidly spinning MP BHs in five spacetime dimensions were studied for the first time, under perturbations that break the axial symmetry along the rotation plane while preserving the transverse U$(1)$. 
They found that for values of $a/\mu^{1/2}\sim 0.87$, there exists an exponentially growing mode that induces a bar-shaped deformation on the AH, i.e., an unstable $m=2$ mode. 
The authors found that the growth of the instability appeared to be stronger for larger values of $a/\mu^{1/2}$. 
However, the nonlinear evolution of the instability was not followed, so determining the end point of the instability remained an open problem.
On the other hand, Ref.~\cite{Dias:2014eua} studied perturbations of singly spinning MP BHs in $D\geq 5$ with $m\geq 2$.
In the $D=5$ case, the results of Ref.~\cite{Dias:2014eua} indicate that all linear modes decay exponentially, which would lead one to conclude that MP BHs are linearly stable in five dimensions, at least for the values of the spin parameters that were studied numerically. 
The results of Refs.~\cite{Shibata:2009ad} and  \cite{Dias:2014eua} would be compatible if the instability were nonlinear. 

Here, we revisit the results from Ref.~\cite{Shibata:2009ad} by carrying out analogous fully nonlinear simulations of perturbed MP BHs with various values of the spin parameter, and in particular for values of $a/\mu^{1/2}$ for which Ref.~\cite{Shibata:2009ad}  observed an instability. 
When comparing our results to those in Ref.~\cite{Shibata:2009ad}, one should note that, although our gauge choice is different, our method of perturbing the black hole is essentially the same as the one used in this reference. 

\begin{figure}[h!]
\begin{center}
\includegraphics[width=7.5cm]{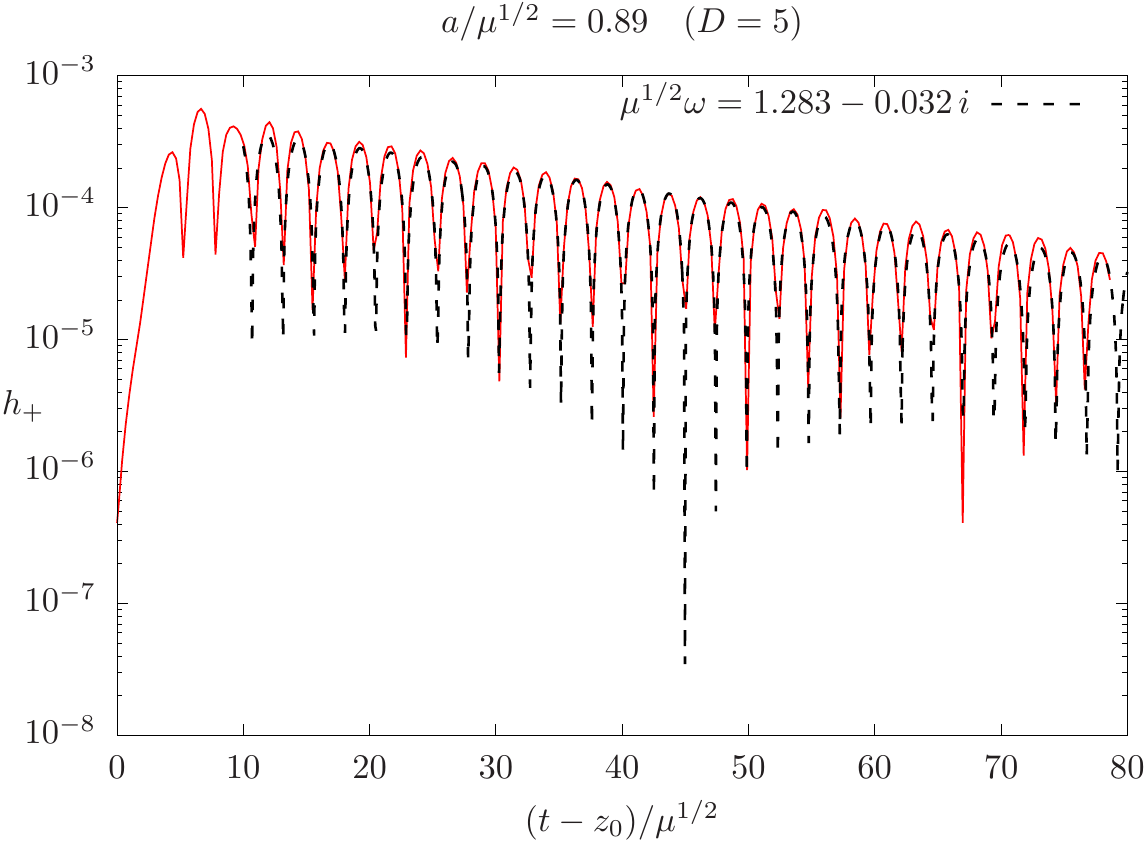}
\vspace{0.1cm}
\includegraphics[width=7.5cm]{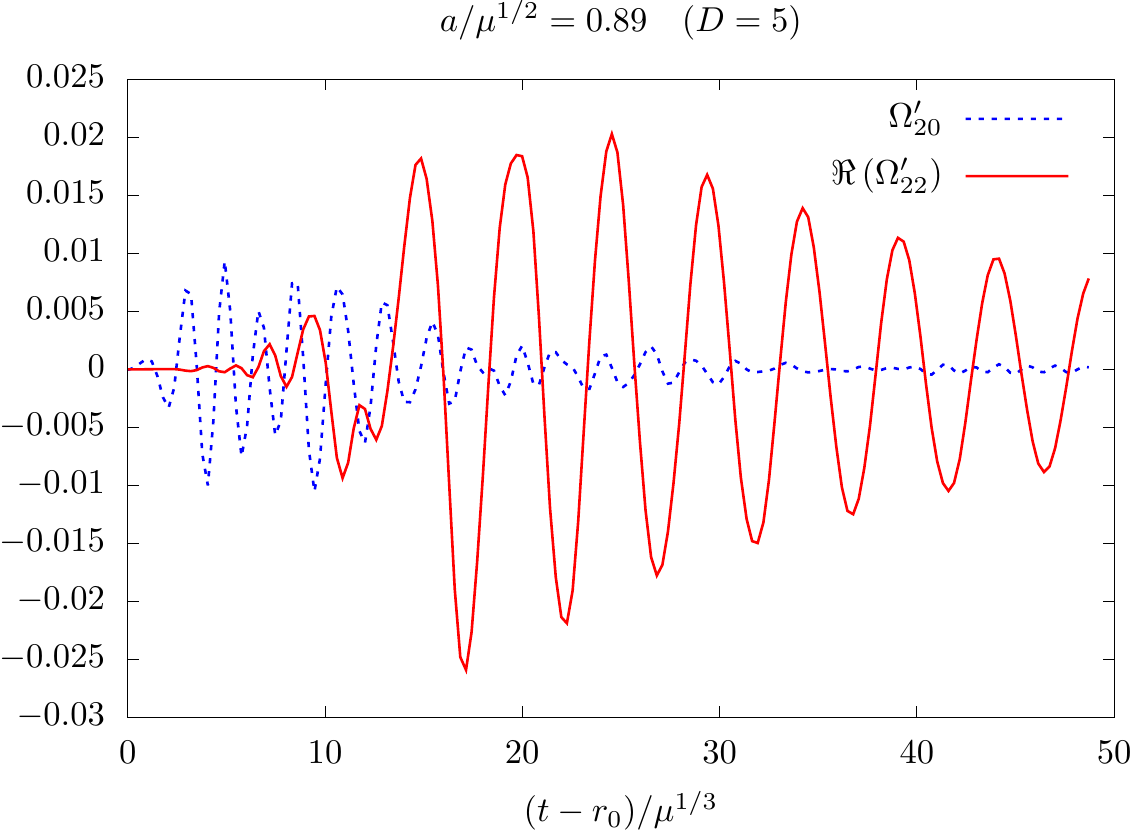}
\end{center}
\caption{\textit{Top}: gravitational waveform extracted from $h_+$ measured at a finite radius along the $z$ axis. The red line is the actual data while the dashed line is our best fit, corresponding to the values of the frequency and decay rate shown in the inset. \textit{Bottom}: $(\ell_3,m)=(2,0)$ and the real part of the $(\ell_3,m)=(2,2)$ scalar-derived multipoles of $\Omega'_{(A)(B)}$ extracted at a finite radius. Clearly, most of the signal is in the $m=2$ mode.
}
\label{fig:StableEvol_5D}
\end{figure}

Contrary to the results in Ref.~\cite{Shibata:2009ad}, we find that all rapidly spinning black holes in five dimensions that we are able to simulate are nonlinearly stable. 
For a perturbed black hole with $a/\mu^{1/2}=0.89$, for instance, Fig.~\ref{fig:StableEvol_5D} shows the gravitational waveform extracted on the $z$ axis~\cite{Shibata:2009ad} at $z_0/\mu^{1/2} = 21.5$. 
The plot below shows the $(\ell_3, m) = (2,2)$ scalar multipole of the gravitational waveform extracted from $\Omega'_{(A)(B)}$.  
From the damped oscillatory decay, we obtain a dominant QNM with frequency $\omega\mu^{1/2} = 1.283 -0.032\,i$, which agrees with the results from perturbation theory of ~\cite{Dias:2014eua}, i.e, $\omega r_{+} = 0.585 -0.015\,i$. 
For a further check, we computed the frequency and decay rate of the leading QNM from both $h_+$ and $\Omega'_{(A)(B)}$, and they agree within the truncation error.

In five dimensions, we find that gravitational waveform is essentially in the $(\ell_3,m)=(2,2)$ scalar-derived sector: see Fig. \ref{fig:StableEvol_5D}. 
The projection onto the other tensor harmonics, i.e., vector-derived and transverse traceless tensors, gives waveforms with amplitudes that are smaller than those in the $\ell_4=2$ scalar sector by 1 order of magnitude or more. 
Higher multipoles again give much smaller waveforms. 

Our results together with the results of Ref.~\cite{Dias:2014eua} suggest that 5D MP BHs are both perturbatively and nonperturbatively stable under perturbations that break the axial symmetry on the rotation plane. 
In this article, we do not study the stability properties of MP BHs with  values of $a/\mu^{1/2}$ that are arbitrarily close to extremality. 
As pointed out in Ref.~\cite{Yang:2014tla}, the dynamics of the Kerr black holes near extremality can exhibit turbulence, so it is conceivable that similar dynamics will arise for MP BHs in the limit $a/\mu^{1/2}\to 1$.

\subsection{Case $D=6$}
\label{sec:mp6d}
In this subsection we consider the nonlinear evolution of perturbed 6D MP BHs of different values of the spin parameter $a/\mu^{1/3}$. 
The results in seven dimensions are qualitatively similar, and we will not show them here. 

Recall that in $D\geq 6$ the dimensionless spin parameter $a/\mu^{1/3}$ of singly spinning MP BHs can be arbitrarily large. 
Reference~\cite{Emparan:2003sy} noted that the thermodynamics of spinning black holes exhibits a qualitative change in behavior as the angular momentum increases beyond a certain (dimension-dependent) critical value of $a_\textrm{crit}/\mu^{1/(D-3)}$; for values of $a/\mu^{1/(D-3)}$ smaller than this critical value, MP BHs behave similarly to the Kerr black hole, and for values of $a/\mu^{1/(D-3)}$ greater than this critical value, they behave like black membranes. 
In six dimensions, one finds $a_\textrm{crit}/\mu^{1/3}\simeq 1.29 $, and the ultraspinning instability kicks in at $a/\mu^{1/3}\simeq 1.57 $ \cite{Dias:2010maa}. 
The end points of certain ultraspinning instability were worked out in Ref.~\cite{Figueras:2017zwa}. 
On the other hand, the nonaxisymmetric instabilities of 6D MP BHs kick in at $a/\mu^{1/3}\simeq 0.74 $ \cite{Shibata:2010wz,Dias:2014eua}. 
In this article we study the endpoint of nonaxisymmetric instabilities for MP BHs with $0.7\leq a/\mu^{1/3} \leq 1.5$. 
The results presented here, taken in conjunction with those in Ref.~\cite{Figueras:2017zwa}, constitute a complete picture of the dynamics of singly spinning MP BHs. 

\subsubsection{Gravitational waves}
\label{sec:mp6d_gw}

\begin{table*}
\begin{center}
\begin{tabular}{|c|c||c|c||c|} 
      \hline
      \multicolumn{2}{|c||}{Initial state} & \multicolumn{2}{c||}{Final state}& \multicolumn{1}{c|}{ spin loss}\\
      \hline
      $a_{\rm h}/\mu_{\rm h}^{1/3}$  & $\omega_{22}\mu_{\rm h}^{1/3}$  & $a_{\rm h}/\mu_{\rm h}^{1/3}$  & $\omega_{22}\mu_{\rm h}^{1/3}$  & $\Delta\left(a_{\rm h}/\mu^{1/3}_{\rm h}\right)$ \\
      \hline
      $0.7$  & $1.07 - 0.03i$   & $0.7$  & $1.07 - 0.03i$   & ---\\
      \hline
      $0.8$  & $1.03 + 0.03 i$  & xxx & xxx   & xxx \\
      \hline
      $0.9$  & $1.03 + 0.07 i$ & $0.66$  & $1.04 - 0.03 i$    &  26.6\%\\
      \hline
      $1.0$  & $1.00 + 0.12i$  & $0.64$ & $1.08 - 0.05i$     & 36.0\%\\
      \hline
      $1.1$  & $0.95 + 0.17 i$  & $0.64$  & $1.06 - 0.05 i$     & 41.8\%\\
      \hline
      $1.2$  & $0.91 + 0.21 i$  & $0.63$  & $1.05 - 0.05 i$     & 47.5\%\\
      \hline
      $1.3$  & $0.88 + 0.24 i$  & $0.63$  & $1.05 - 0.05i$     & 51.5\%\\
      \hline
      $1.4$  & $0.82 + 0.26i$  & $0.63$  & $1.05 - 0.05i$     & 55.0\%\\
      \hline
      $1.5$  & $0.74 + 0.28i$  & $0.62$  & $1.03 - 0.05i$     & 58.6\%\\
      \hline
      \noalign{\vskip 0.5cm}   
\end{tabular}
\caption{Real and Imaginary parts of the leading quasinormal modes in the $(\ell_4,m)=(2,2)$ sector of gravitational perturbations. \textit{Left:} from perturbing the initial black hole. For $a/\mu^{1/3}=0.7$, black holes are stable, while for $a/\mu^{1/3}\geq 0.8$ black holes are unstable. \textit{Right}: leading QNM in the ring-down phase. For all initial values of $a/\mu^{1/3}$ that end up settling back to an MP BH, we find that the end state appears to be the same black hole. For $a/\mu^{1/3}=1.5$, the end point of the instability is a naked singularity. We estimate an error of $\pm 0.02$ in the real and imaginary parts of the frequencies (see footnote \ref{fn:QNM_error}). 
For the $a_{\rm h}/\mu_{\rm h}^{1/3} = 0.8$ case, which is the lowest-spin case we consider in which the spinning black hole is unstable, the timescale of the growing mode is the longest.
In practice, we find that the decay to the final state does not become evident until around $t/\mu^{1/3} \sim 200$, by which time the signal is already spoiled by accumulated numerical error at the typical resolutions with which we run. 
We found extracting the leading QNM for this case to be prohibitively expensive and have left its values as ``xxx''.
}
\label{tab:QNM}
\end{center}
\end{table*}

In Table \ref{tab:QNM}, we summarize the QNM frequencies in the $m=2$ sector that we have extracted from our fully nonlinear simulations. 
To obtain the results displayed on this table, we considered sufficiently small perturbations, with amplitudes $A=0.02$.
Although our simulations are fully nonlinear, thus making it difficult to obtain accurate results for linear waves, our results agree within the truncation error with the perturbative calculations of \cite{Dias:2014eua}. 
In the left column of Table \ref{tab:QNM} we display the normalized frequencies of the leading mode obtained from perturbing the initial black hole with an $m=2$ mode. 
This shows that black holes with  $a/\mu^{1/3}=0.7$ are stable, while for  $a/\mu^{1/3}\geq 0.8$, the sign of the imaginary part of the frequency indicates a linear instability. 
In the right column, we display the normalized frequencies of the leading QNM in the ring-down phase for those runs that settle back onto another MP BH. 
We measured the AH area and the length of the circumference of the AH on the equatorial plane of the final state to estimate the angular momentum parameter of the final MP black hole using \eqref{eq:parameters_MP_Hrz}. 
These results indicate that all runs settle down to what appear to be the \textit{same} black hole (within the numerical error), independent of the initial value of $a/\mu^{1/3}$ that we have considered. 

\begin{figure}
\begin{center}
\includegraphics[width=7.5cm]{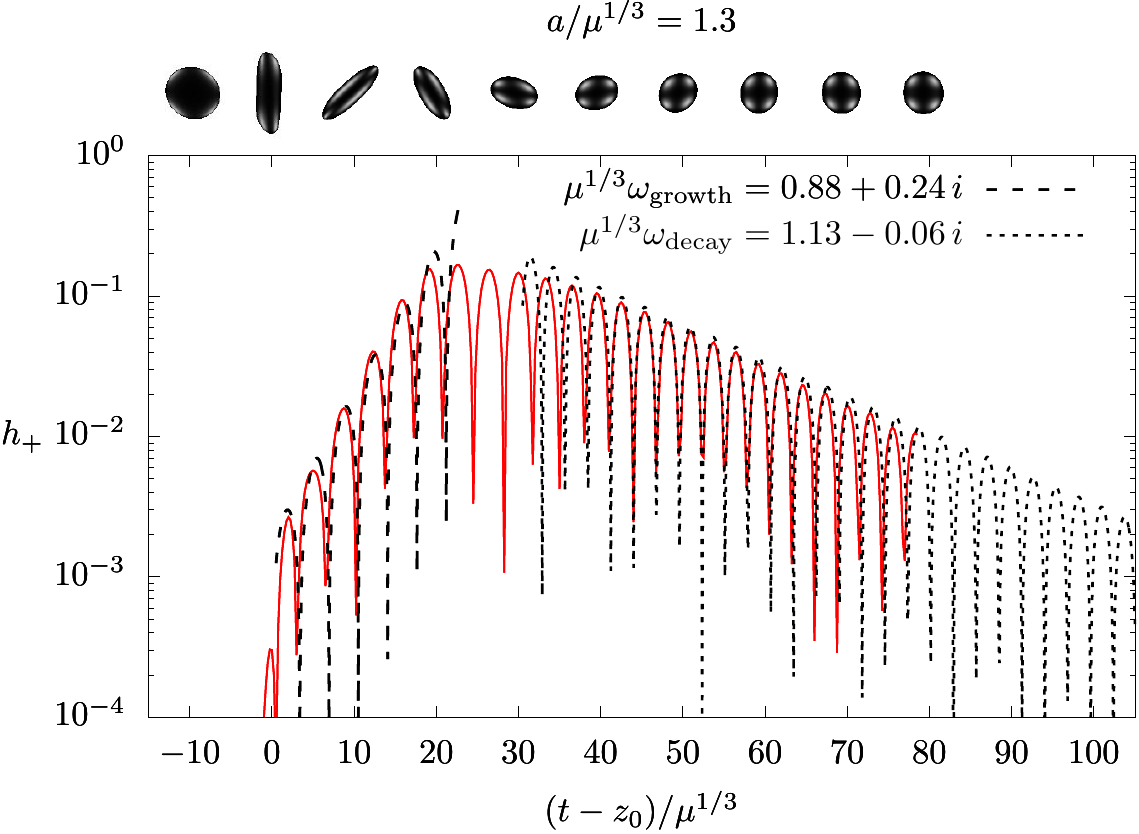}
\includegraphics[width=7.5cm]{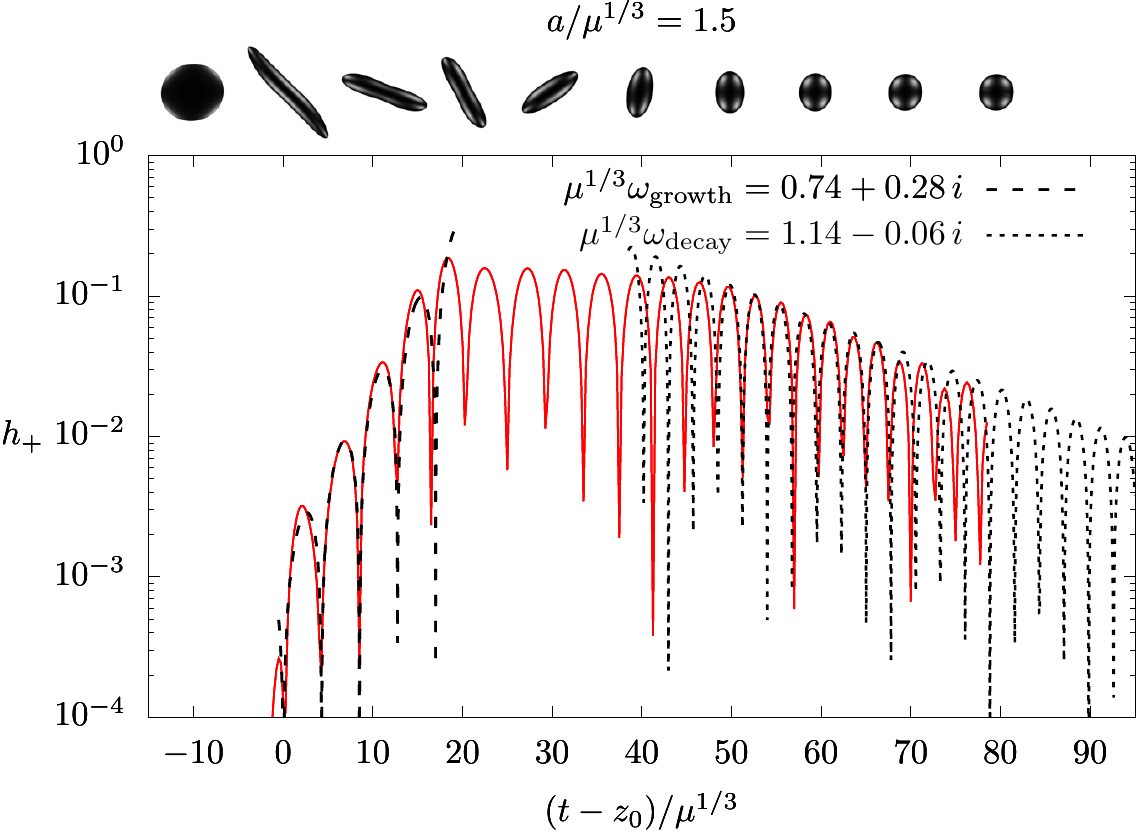}
\end{center}
\caption{Gravitational wave ($m = 2$ mode) at $z_0/\mu^{1/3} = 21.5$, together with snapshots of the apparent horizon for initial spin parameter $a/\mu^{1/3} = 1.3$ (\textit{top}) and $a/\mu^{1/3} = 1.5$ (\textit{bottom}).  The dashed and dotted lines provide the fit for the QNMs in the growing and decaying phases. Table \ref{tab:QNM} displays the frequencies normalized with respect to the initial/final horizon scales. For larger spins, the system stays longer in the transient regime between the initially growing phase and the final ring down.
}
\label{fig:StableEvol_GW}
\end{figure}

\begin{figure}
\begin{center}
\includegraphics[width=7.5cm]{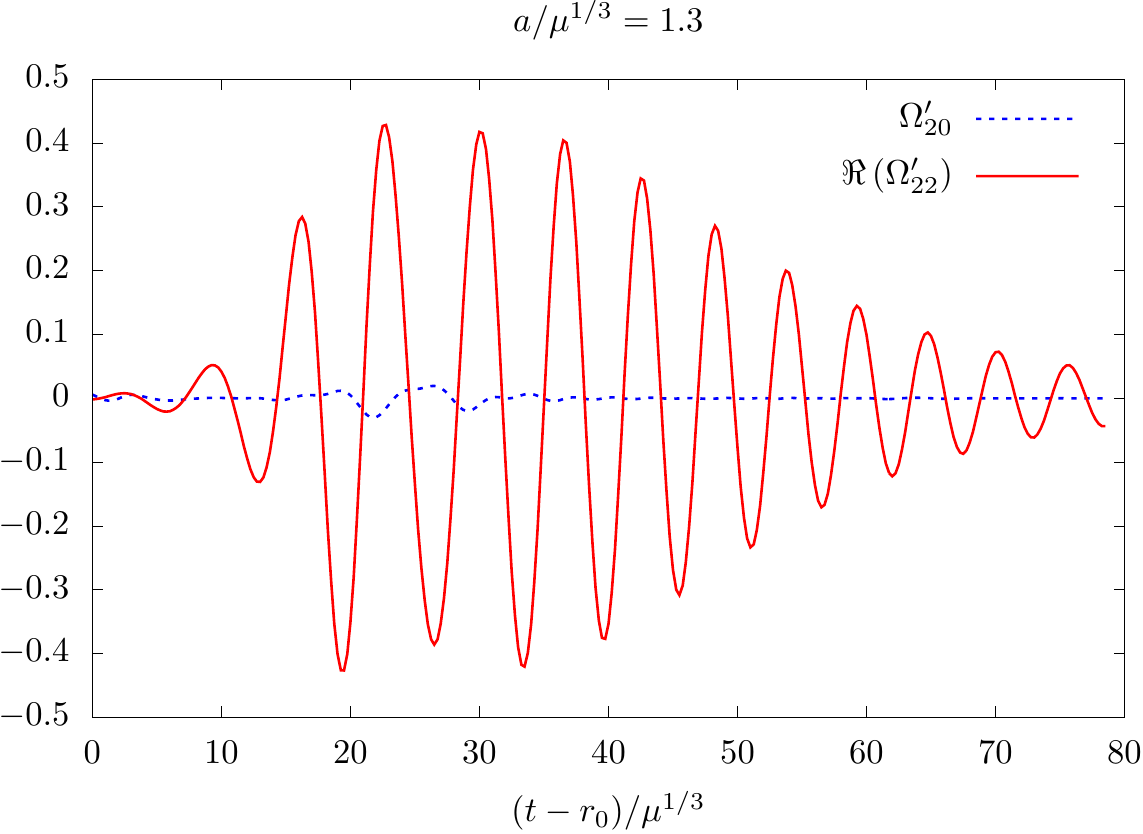}
\includegraphics[width=7.5cm]{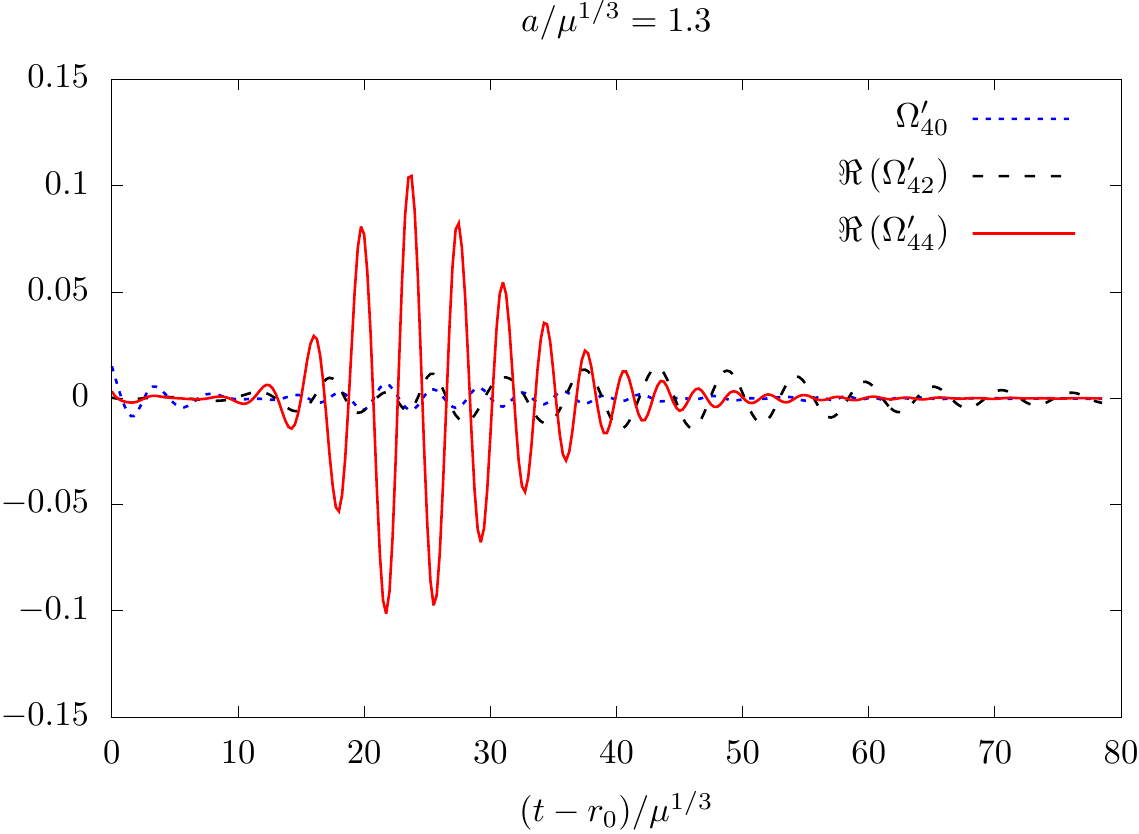}
\end{center}
\caption{
Gravitational waveforms for initial spin parameter $a/\mu^{1/3} = 1.3$ obtained from the higher-dimensional Weyl scalars projected on a sphere at finite radius: $(\ell_4,m) = (2,0)$ and the real part of the $(\ell_4,m) =(2,2)$ modes (\textit{top}) and $(\ell_4,m) = (4,0)$ and real parts of the $(\ell_4,m) = (4,2),(4,4)$ modes (\textit{bottom}).  Most of the signal lies on the $(\ell_4,m)=(2,2)$ mode, but higher harmonics are also excited due to the nonlinearities. 
}
\label{fig:StableEvol_Weyl}
\end{figure}

In Fig. \ref{fig:StableEvol_GW} we display the gravitational waveforms corresponding to black holes with initial spin $a/\mu^{1/3} = 1.3$ (top) and $a/\mu^{1/3} = 1.5$ (bottom), extracted using the metric perturbations. 
Together with the waveforms, we also display snapshots of the AH at different stages of the evolution as an inset above the figure. 
From these AH snapshots, it is clear that the dynamics is governed by an $m=2$ mode which deforms the AH into a bar shape. 
The black hole radiates mass and angular momentum until it spins down and settles down to an equilibrium MP BH, with lower mass and angular momentum. As we increase the spin parameter, the systems stays for a longer period in the transient regime between the initial dynamics described by perturbation theory and the final ring-down phase.
 
Figure~\ref{fig:StableEvol_Weyl} shows the gravitational waveforms extracted from the Weyl scalars for both $\ell_4=2$ (top) and $\ell_4=4$ (bottom) modes in the scalar-derived sector of tensor harmonics for $a/\mu^{1/3} = 1.5$. Most of the signal is in the $(\ell_4,m)=(2,2)$ mode, but the $\ell_4 = 4$ also displays a sizeable signal. 
However, for MP BHs with  $a/\mu^{1/3}=1.3$ the $(\ell_4,m)=(2,2)$ mode governs the nonlinear evolution, with qualitatively similar results for $a/\mu^{1/3}=1.5$. Table~\ref{tab:QNM_l4} displays the quasinormal frequencies in the $\ell_4 = 4$ sector for $a/\mu^{1/3}=1.3$ and $a/\mu^{1/3}=1.5$.  

\begin{figure}[t!]
\begin{center}
\includegraphics[width=7.5cm]{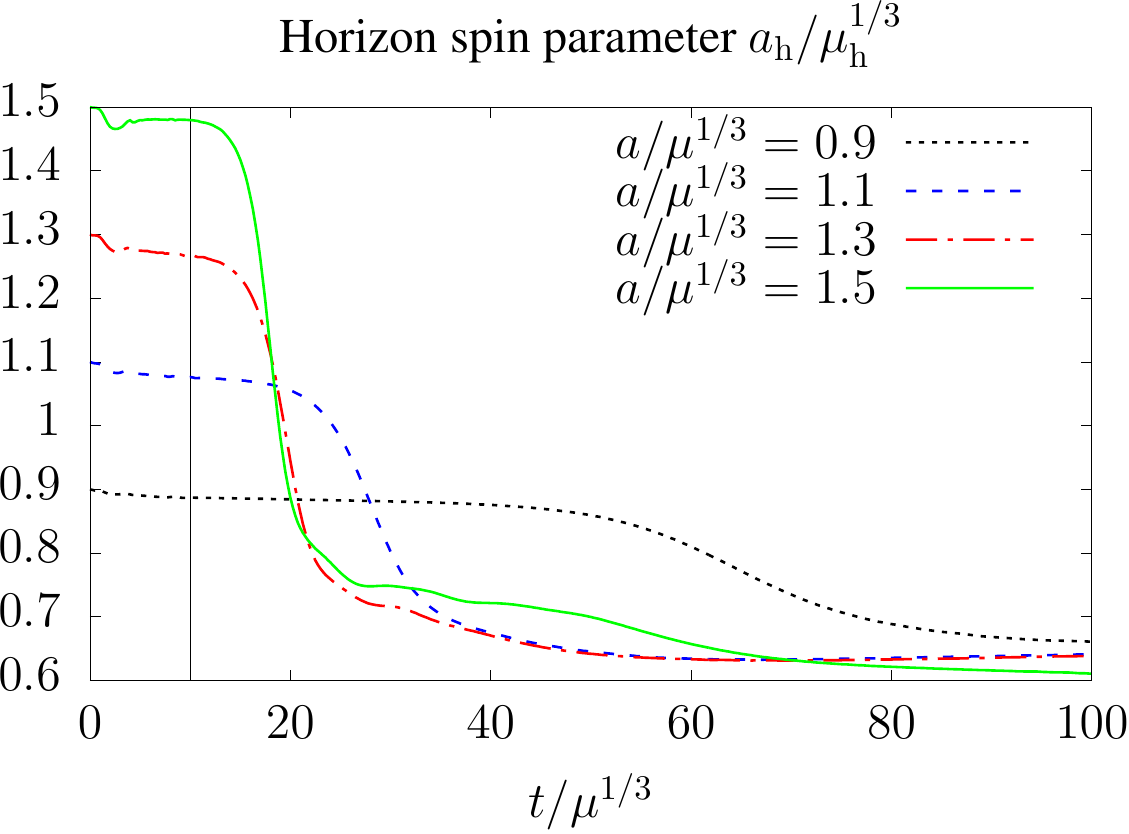}
\end{center}
\caption{Dynamical evolution of the normalized spin parameter $a_{\rm h}/{\mu_{\rm h}}^{1/3}$ for initial MP BHs with $a/\mu^{1/3} = 0.9$ (dotted black), $\mu_{\rm h}$ for $a/\mu^{1/3} = 1.1$ (dashed blue), $a/\mu^{1/3} = 1.3$ (dashed-dotted red), and $a/\mu^{1/3} = 1.5$ (continuous green). MP BHs radiate enormous amount angular momentum. The vertical line indicates the time around which the initial gauge dynamics settles down.}
\label{fig:StableEvol_Hor}.
\end{figure}

\begin{table*}
\begin{center}
\begin{tabular}{|c|c|c||c|c|c|} 
      \hline
      \multicolumn{3}{|c||}{Initial state} & \multicolumn{3}{c|}{Final state}\\
      \hline
      $a_{\rm h}/\mu_{\rm h}^{1/3}$  & $\omega_{42}\mu_{\rm h}^{1/3}$ & $\omega_{44}\mu_{\rm h}^{1/3}$  & $a_{\rm h}/\mu_{\rm h}^{1/3}$  & $\omega_{42}\mu_{\rm h}^{1/3}$ & $\omega_{44}\mu_{\rm h}^{1/3}$  \\
      \hline
       $1.3$  & $1.70 + 0.36i$  & $1.61 + 0.40i$  & $0.63$   & $1.68  - 0.12 i$  & $1.83 - 0.13i$    \\
      \hline
      $1.5$  & $1.45 + 0.40i$ & $1.57 + 0.45i$  & $0.62$  & $1.66 - 0.13i$ & $1.87 - 0.17i$     \\
      \hline
      \noalign{\vskip 0.5cm}   
\end{tabular}
\caption{
Real and imaginary parts of the subleading $(\ell_4,m)=(4,2)$ and $(\ell_4,m)=(4,4)$ quasinormal mode for $m=2$ perturbation of a MP BH with the initial configuration $a/\mu^{1/3}=1.3$ and $a/\mu^{1/3}=1.5$. We estimate an error of $\pm 0.04$ in the real and imaginary parts of the frequencies (see footnote\ref{fn:QNM_error}).
}
\label{tab:QNM_l4}
\end{center}
\end{table*}

The end point of the evolution of these instabilities depends, in a nontrivial way, on the amplitude of the initial perturbation and the value of $a/\mu^{1/3}$ for the initial black hole. 
For small enough perturbations and a small enough initial value of $a/\mu^{1/3}$, the end state of the nonaxisymmetric instabilities is another MP BH. 
Surprisingly, the perturbed black holes of this type that we have been able to simulate all appear to settle down to the same MP BH, the one with $(a/\mu^{1/3})_\textrm{final}=0.63$.  
We are not aware of the physical reason that singles out this particular member of the MP family, but our simulations suggest that it behaves as an attractor. 
It would be worth investigating this point further.  

\subsubsection{Apparent horizon}
\label{sec:mp6d_ah}

\begin{figure}[t!]
\begin{center}
\includegraphics[scale=0.3]{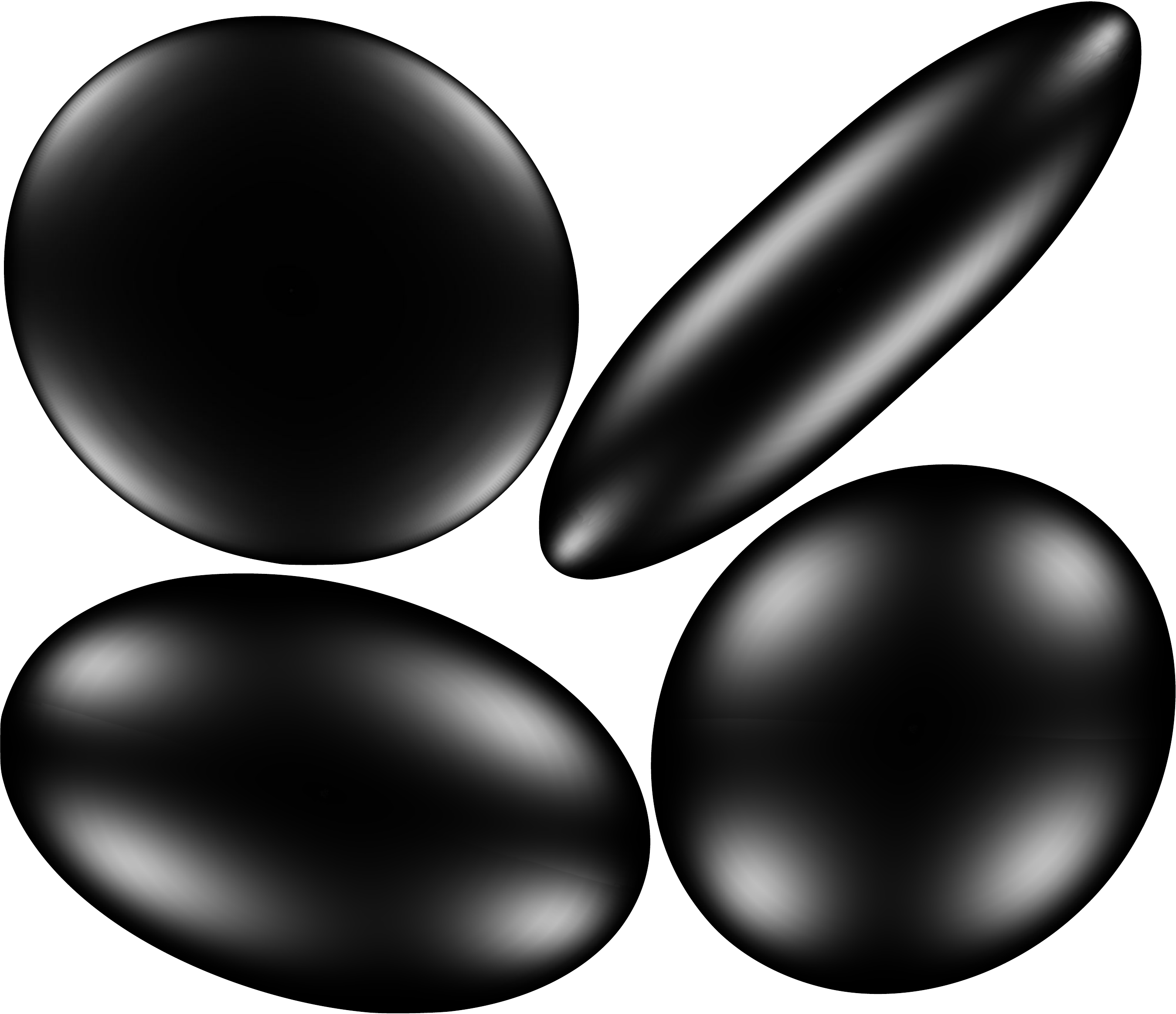}
\end{center}
\caption{ Snapshots of AH during the evolution of an unstable MP BH with initial $a/\mu^{1/3} = 1.3$ and perturbation amplitude $A=0.02$. The sequence of images should be read horizontally, from left to right, and from top to bottom.  There is a strong bar-shaped deformation during the highly nonlinear stages of the evolution. Gravitational waves eventually radiate away the nonaxisymmetric deformations, and the black hole settles to another member of the MP family, with lower angular momentum.}
\label{fig:Evol_AH_a1p3}
\end{figure}

In Fig. \ref{fig:StableEvol_Hor}, we display the evolution of the estimated black hole spin parameter from the geometry of the AH \eqref{eq:parameters_MP_Hrz}. This estimate only corresponds to the spin parameter of a black hole in equilibrium, and therefore it is exact only at the initial and final stages of the evolution. However, it indicates that highly spinning unstable black holes can radiate an enormous amount of angular momentum due to the large deformation of the horizon during the highly nonlinear stages of the evolution. For instance, black holes with an initial  $a/\mu^{1/3}=1.5$ can radiate up to $58\%$ of the initial spin. In Fig. \ref{fig:Evol_AH_a1p3} we display four snapshots of the AH during the evolution of a perturbed MP with initial  $a/\mu^{1/3}=1.3$. As can be seen from these snapshots, the AH develops a strong bar-shaped type of deformation, which leads to a very strong emission of gravitational radiation. As the plots in Fig. \ref{fig:StableEvol_Hor} show, the amount of angular momentum radiated increases with the initial value of $a/\mu^{1/3}$, so it is conceivable that one can achieve even higher emission percentages for larger initial values $a/\mu^{1/3}$. However, as we will discuss next, there may be a (dimension-dependent) dynamical upper bound due to the fact that the generic end point of the evolution of the instability of MP BH with a sufficiently large initial value of $a/\mu^{1/3}$ or a sufficiently large initial deformation is no longer another MP BH but is a naked singularity. 

\subsubsection{Large initial spin}
\label{sec:mp6d_large_spin}

Given the large deformations observed at moderate spins and small perturbations, e.g., see Fig. \ref{fig:Evol_AH_a1p3}, one may wonder what happens when the spin of the unperturbed black hole or the size of the deformation is increased. 
In this article, we are interested in the end points of perturbative instabilities of black holes, so we will always keep the initial deformation small enough to remain in the perturbative regime initially and vary the spin of the MP BH.
As we now show, the inevitable end point of the instabilities for sufficiently rapidly spinning black holes is a naked singularity. 
Before we proceed, note that as the deformations of the AH beyond spherical symmetry become increasingly large during the highly nonlinear stages of the evolution the AH ceases to be a star-shaped surface, which is an assumed property underlying the AH finder we use. 
The same situation was shown to occur in unstable black rings \cite{Figueras:2015hkb} or in the ultraspinning instability of MP black holes \cite{Figueras:2017zwa}. 
The latter reference managed to find non-star shaped AH using an intrinsic parametrization of the surface, while Ref.~\cite{Pook-Kolb:2018igu} proposed the use of a reference surface. 
We will leave the interesting problem of implementing one of these methods to the present situation for future work. 
However, as in Ref.~\cite{Figueras:2015hkb,Figueras:2017zwa}, we have verified that certain contours of the conformal factor $\chi$ provide a very accurate description of the AH, within less than a few percent. 
See Appendix~\ref{sec:AppHrznChi} for quantitative comparison between the contours of $\chi$ and the actual AH for various runs in which the latter can be found. 
Therefore, we will use contours of $\chi$ as approximations of AH for situations where the latter is not a star-shaped surface, and hence get intuition of the physics of the instabilities and their end points.\footnote{Note that since contours of $\chi$ provide accurate approximations of the AH, in principle one could use them as suitable reference surfaces to find the AH using the method of \cite{Pook-Kolb:2018igu}.}

\begin{figure}[t!]
\begin{center}
\includegraphics[width=7.5cm]{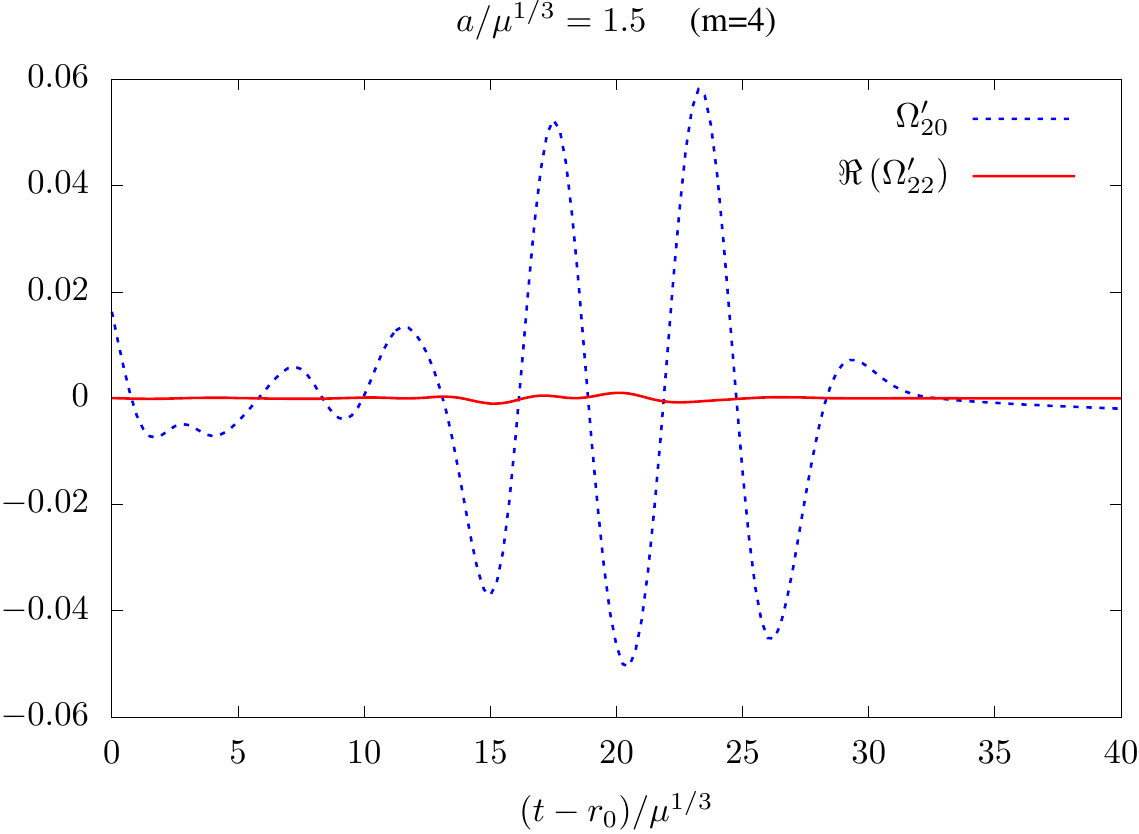}
\includegraphics[width=7.5cm]{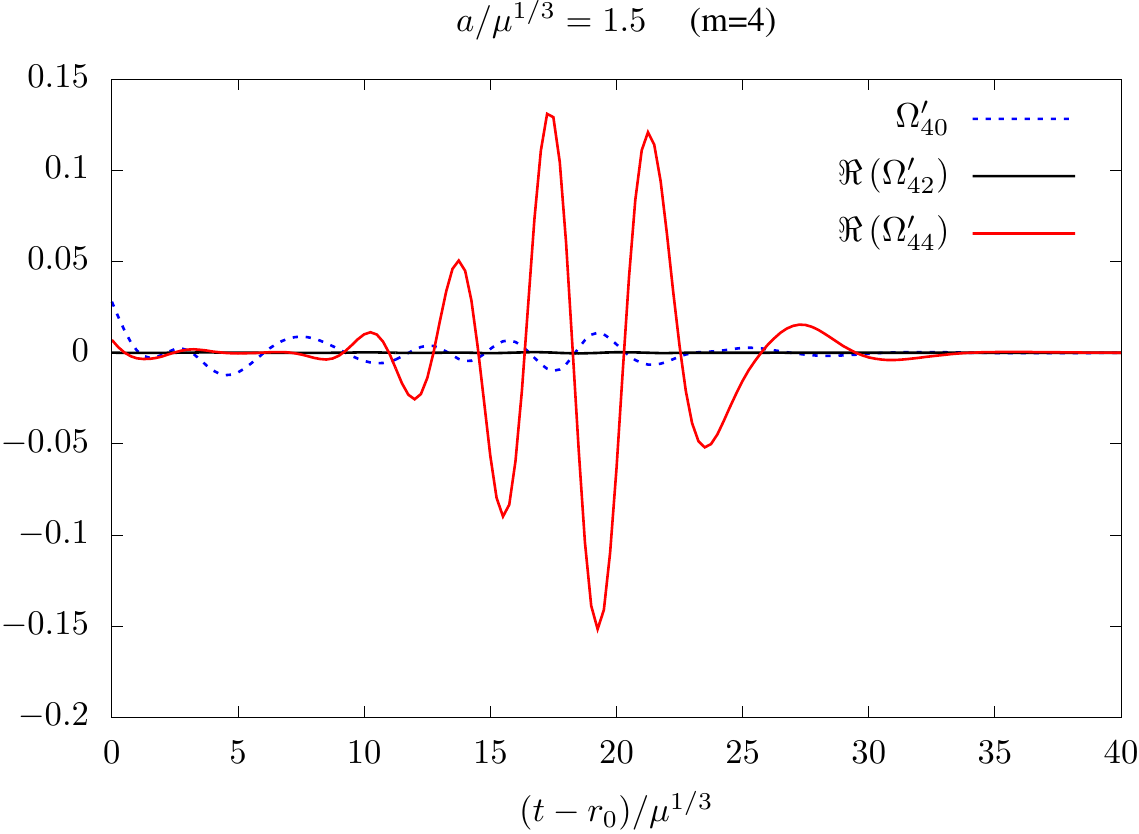}
\end{center}
\caption{ Gravitational waveforms obtained during the evolution of an unstable MP BH with initial $a/\mu^{1/3}=1.5$, with the instability triggered by numerical noise. \textit{Top:} $(\ell_4,m) = (2,0)$ and the real part of the $(\ell_4,m) = (2,2)$ modes. \textit{Bottom}:  $(\ell_4,m) = (4,0)$ and the real part of the $(\ell_4,m) = (4,2),\,(4,4)$ modes. Most of the signal is in the $m=4$ mode, but the $m=2$ modes are also present. Note the difference in amplitude between this run and the  $a/\mu^{1/3}=1.3$ one, Fig. \ref{fig:StableEvol_GW}. This suggests that at higher spins gravitational wave emission becomes less efficient and instabilities have time to develop and eventually form a naked singularity. This run crashes close to the singularity but we artificially froze the evolution around the black hole to extract the gravitational waves. Therefore, the part of the waveform after $(t-r_0)/\mu^{1/2}\sim 23$ is unphysical. }
\label{fig:Evol_GW_a1p5}
\end{figure}

For MP BH's with an initial $a/\mu^{1/3}=1.5$, we find that numerical noise is enough to excite the $m=4$ mode, which is the one that dominates the subsequent nonlinear evolution.\footnote{The same happens for black rings: moderately thin rings are unstable under a GL mode with $m=2$, while for sufficiently thin rings the fastest growing instability is in $m=4$ sector. Presumably, for even thinner rings, higher $m$ modes are more unstable, just as in the GL instability of black strings.} 

Modes with $m=2$ that are present in the noise also get excited but have much smaller amplitude. 
See Fig. \ref{fig:Evol_GW_a1p5}. 
Comparing the waveforms of the $m=4$ instability for $a/\mu^{1/3}=1.5$ in Fig. \ref{fig:Evol_GW_a1p5} with their counterpart waveforms for the $m=2$ instability for $a/\mu^{1/3}=1.3$ in Fig. \ref{fig:StableEvol_GW}, notice that the amplitude for the $m=2$ case is much larger, even though the AH deformation is quite extreme in both cases. 
This suggests that the efficiency of the gravitational wave emission during the evolution of the instabilities decreases as $a/\mu^{1/3}$ increases, hence giving more time for these instabilities to grow and eventually form a naked singularity. 
The simple physical reason for this is that the structures that form at large values of $a/\mu^{1/3}$, as deformed as they may be, contain only a small amount of the total mass.

\begin{figure}[t!]
\begin{center}
\includegraphics[scale=0.3]{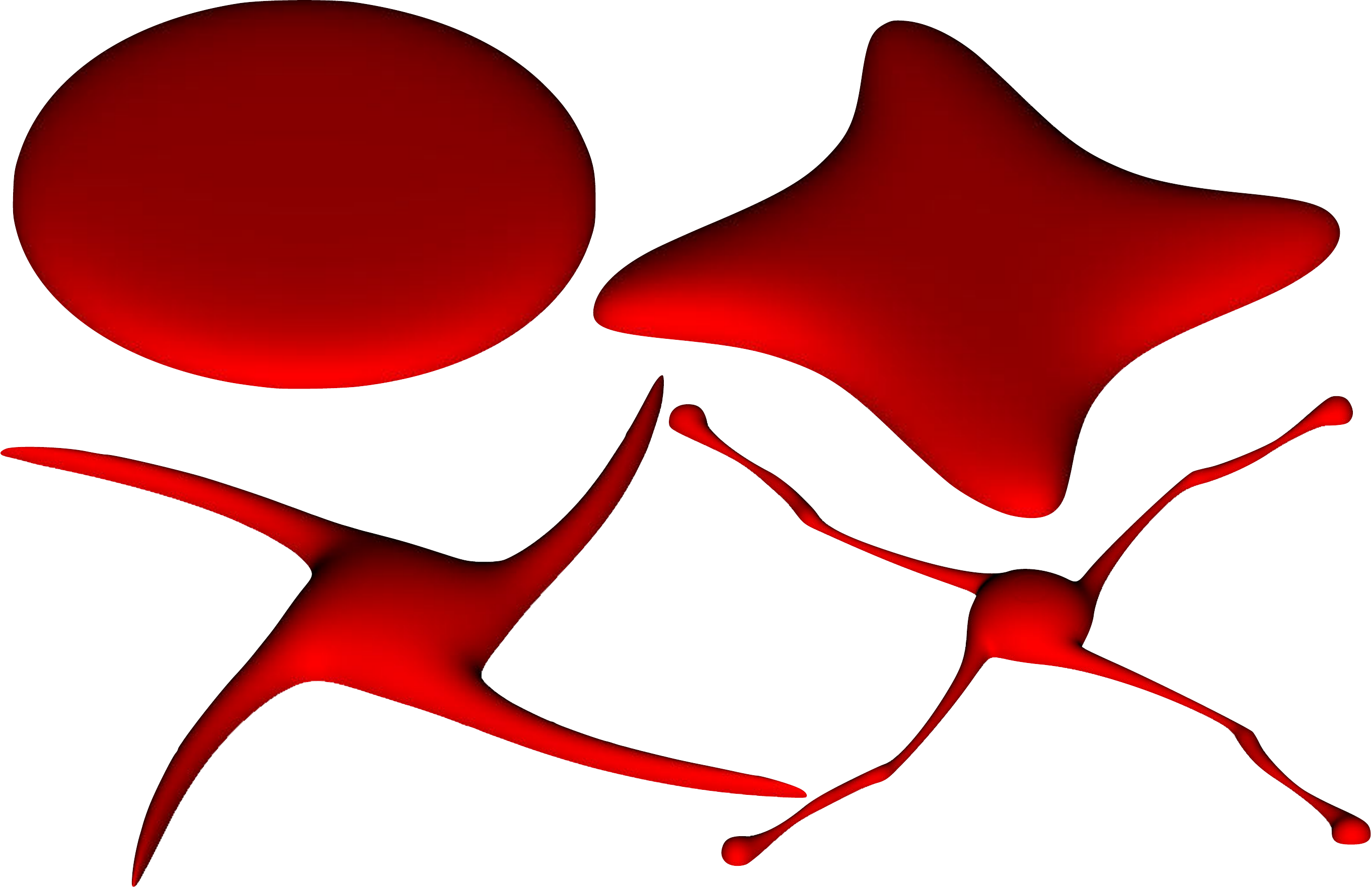}
\end{center}
\caption{ Snapshots of the $\chi=0.5$ contour during the evolution of an unstable MP BH with initial $a/\mu^{1/3} = 1.5$ and no initial perturbation. This contour tracks the AH very closely; see Appendix~\ref{sec:AppHrznChi}.  The evolution is dominated by an $m=4$ present in the numerical noise. During the evolution, the black hole develops a square shape, and the tips of the square eventually grow into long arms. The latter eventually become GL unstable. The last snapshot shows the second generation bulge propagating along the arm, a characteristic feature of GL dynamics.}
\label{fig:Evol_chi_a1p5_m4}
\end{figure}

In Fig. \ref{fig:Evol_chi_a1p5_m4}, we show representative snapshots of $\chi=0.5$ contours from the evolution of the MP BH with $a/\mu^{1/3}=1.5$.
The initial deformation of the black hole horizon into a square shape is characteristic of the $m=4$ mode.
Subsequently, as the instability develops, the corners of the square grow into four arms.
These arms become longer and thinner as time goes by, until at some point a local GL instability kicks in along the arms, signaled by the appearance of a bulge in the central part of each of the arms. 
These bulges travel outward towards the tip of the arms due to centrifugal force, which leads to a further thinning of arms, and mass accumulates at the tips, forming nearly spherical bulges. 
It becomes very difficult to continue the simulation beyond this point since resolving the new generations of bulges that form due to the local GL instability becomes computationally very expensive.\footnote{Recall that the partial differential equations (PDE) that we are solving are effectively $(3+1)$ dimensional, while in Ref.~\cite{Lehner:2010pn} or \cite{Figueras:2017zwa}, the system is $(2+1)$ dimensional, thus allowing one to get much closer to the singularity.} 
However, the local evolution of the instability along each of the arms appears to be analogous to the ultraspinning case of Ref.~\cite{Figueras:2017zwa}. 
In this case, it was observed that the interaction between the centrifugal force and mass accretion by the big bulges accelerates the formation of a naked singularity compared to the GL instability of black strings \cite{Lehner:2010pn}, and the process is no longer self-similar. 
Note that the central bulge becomes more spherical as the evolution proceeds and in fact contains most of the mass of the system. 
This explains why the gravitational wave emission is less efficient at this stage of the evolution, at sufficiently large values of $a/\mu^{1/3}$. 
We believe that this is a generic aspect of the dynamics of unstable black holes at large angular momentum. 
The most likely end point of this instability is that quantum gravity effects will govern the breakup of the arms, and the resulting four black holes will either fly off to infinity or be recaptured by the central black hole. We also performed simulation with $m=3$ perturbations, in which we observe essentially the same qualitative dynamics, but with the black hole developing three arms instead of four --- see Appendix~\ref{sec:Pert_m3}.

\begin{figure}[t!]
\begin{center}
\includegraphics[scale=0.2]{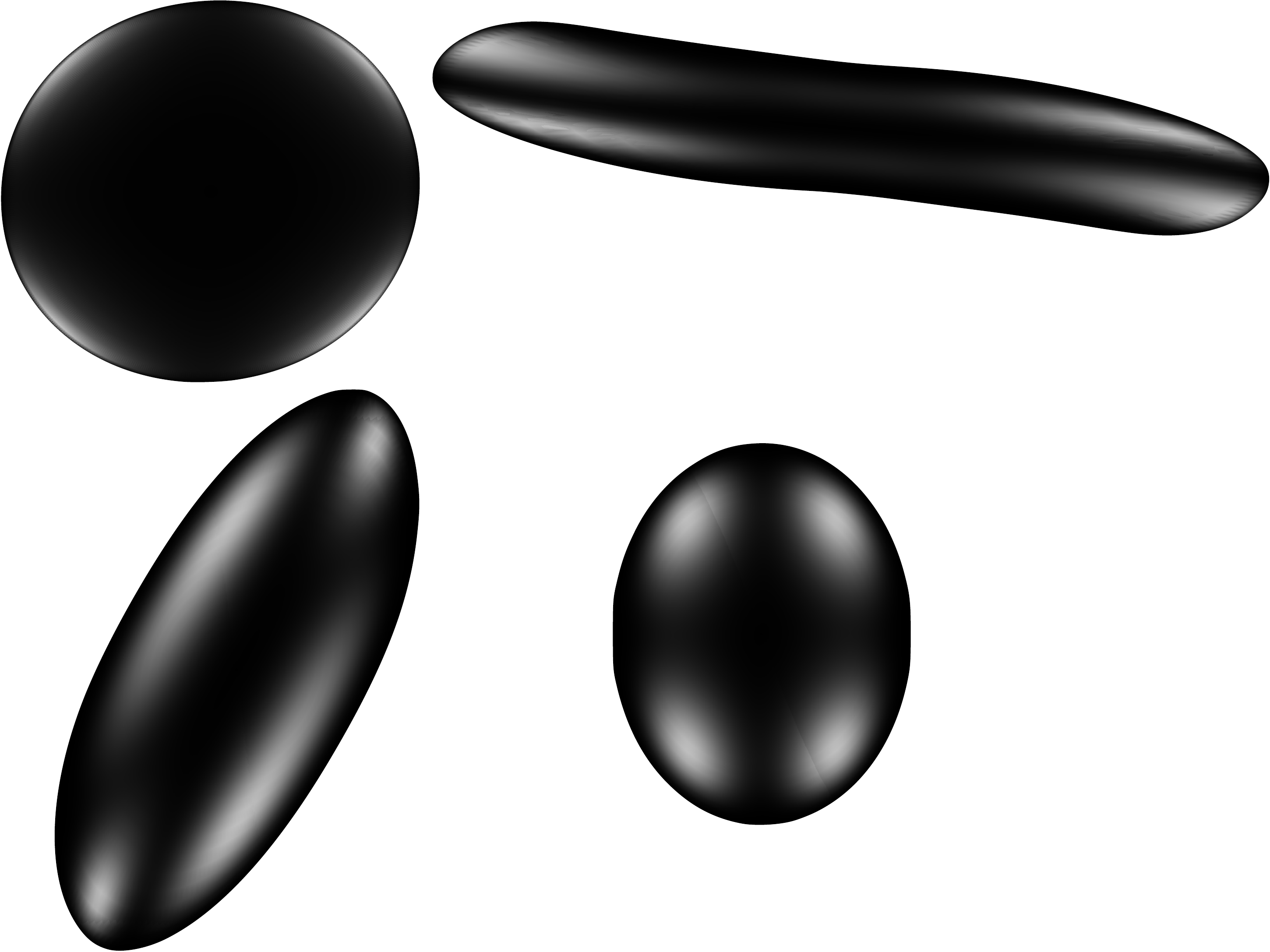}
\includegraphics[scale=0.3]{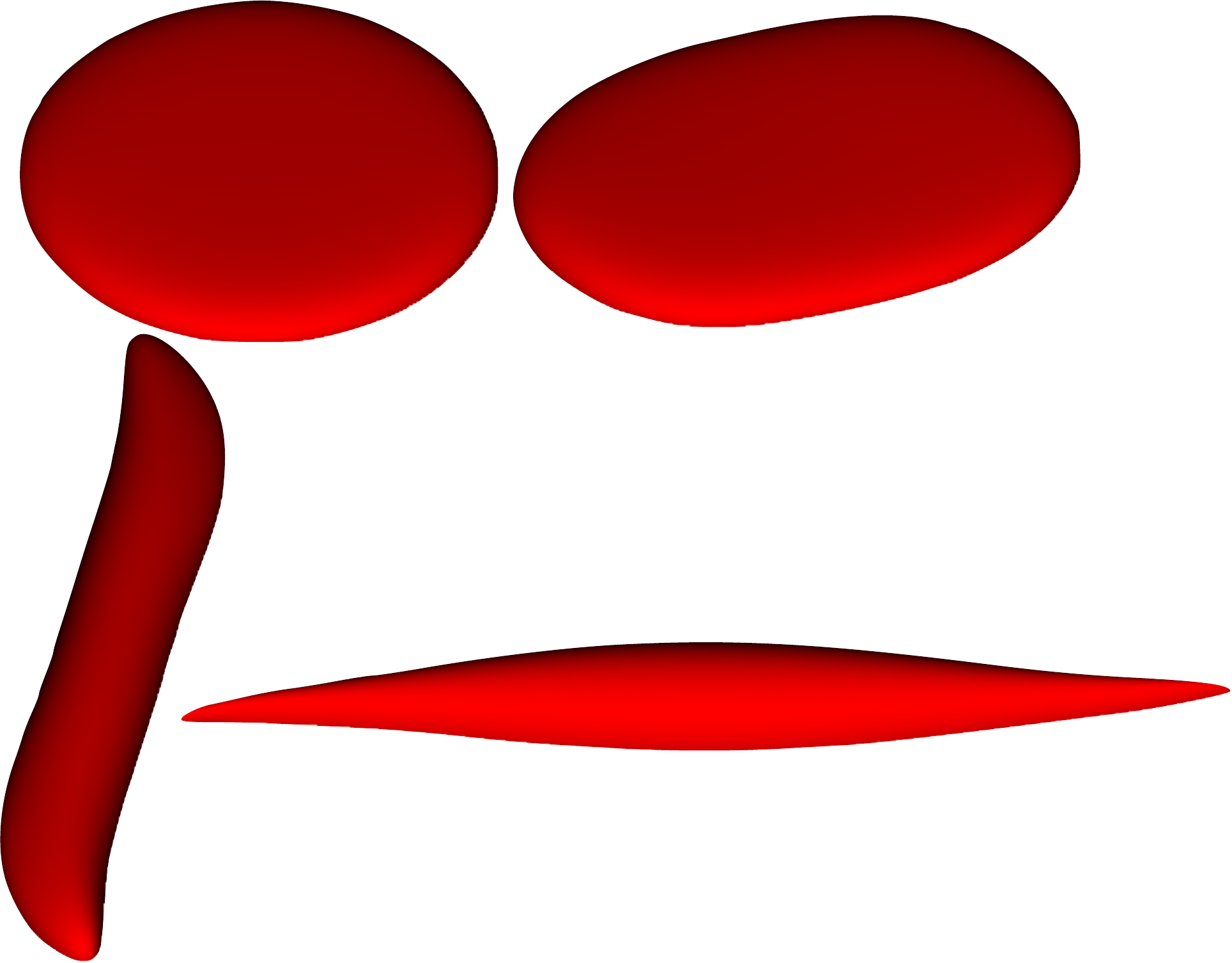}
\end{center}
\caption{ Unstable MP BH with initial $a/\mu^{1/3} = 1.5$ and an $m=2$ perturbation with amplitude $A=0.02$. \textit{Top}: AH snapshots for the evolution with perturbation introduced at $t/\mu^\frac{1}{D-3}=0$. Gravitational waves radiate away the nonaxisymmetric deformations and the black hole settles to another member of the MP family. \textit{Bottom}: Snapshots of the $\chi=0.5$ contour for evolution resulting from perturbation introduced after the gauge adjustment period at around $t/\mu^\frac{1}{D-3}\sim10$. The bar shape gets longer over time and, eventually, pointy at the tip. In analogy with the $m=4$ case, the appearance of pointy tips precedes the formation of long and thin arms, which eventually become GL unstable. We suspect that this is the most likely fate of this evolution. 
}
\label{fig:Evol_chi_a1p5_m2}
\end{figure}

If we start with a black hole with initial $a/\mu^{1/3}=1.5$ and perturb it with an $m = 2$ mode, we can in principle explore the nonlinear evolution of this sector of perturbations at large angular momentum. 
Doing so by perturbing at $t/\mu^{1/3} = 0$ only results in a dumbbell \footnote{We thank Roberto Emparan for telling us about the existence of two sectors of instabilities for the $m=2$ perturbations \cite{RobertoPrivate}.}configuration that eventually settles back down to a MP BH, as shown in the top panel of Fig.\ref{fig:Evol_chi_a1p5_m2} for an initial $m=2$ perturbation with amplitude of $A=0.02$. 
However, by perturbing {\em after} the gauge adjustment period at around $t/\mu^{1/3} = 10$, the perturbation turns out to be in a sector where unstable modes dominate the nonlinear regime. 
In this case, we find that during the highly nonlinear regime of the subsequent evolution, the AH becomes more elongated and develops very sharp features at the edges; see the bottom panel in Fig. \ref{fig:Evol_chi_a1p5_m2}. 
The appearance of these sharp features is not a numerical artifact and it makes the numerical simulation of the system very difficult. 
While we were unable to continue this particular simulation beyond this point due to the cost of the computation, one is tempted to conjecture that the end point of the evolution may well be a naked singularity. 
Indeed, the spacetime curvature at these sharp tips grows large, as spatial gradients diverge and the timescales become very fast. 
In fact, in the $m=4$ case, the appearance of the sharp tips precedes the formation of long and thin arms, which eventually become GL unstable. 
We suspect that the same will happen in the present $m=2$ case: it is quite possible that the evolution will continue by forming two long arms, joined at a central region that is nearly spherical and contains most of the mass. 
These long arms will become thinner over time due to the centrifugal force and eventually will develop local GL instabilities. 
Therefore, the end point of the evolution is likely the formation of a naked singularity in finite asymptotic time.

\subsubsection{Energy flux}
\begin{figure}[h!]
\begin{center}
\includegraphics[width=7.7cm]{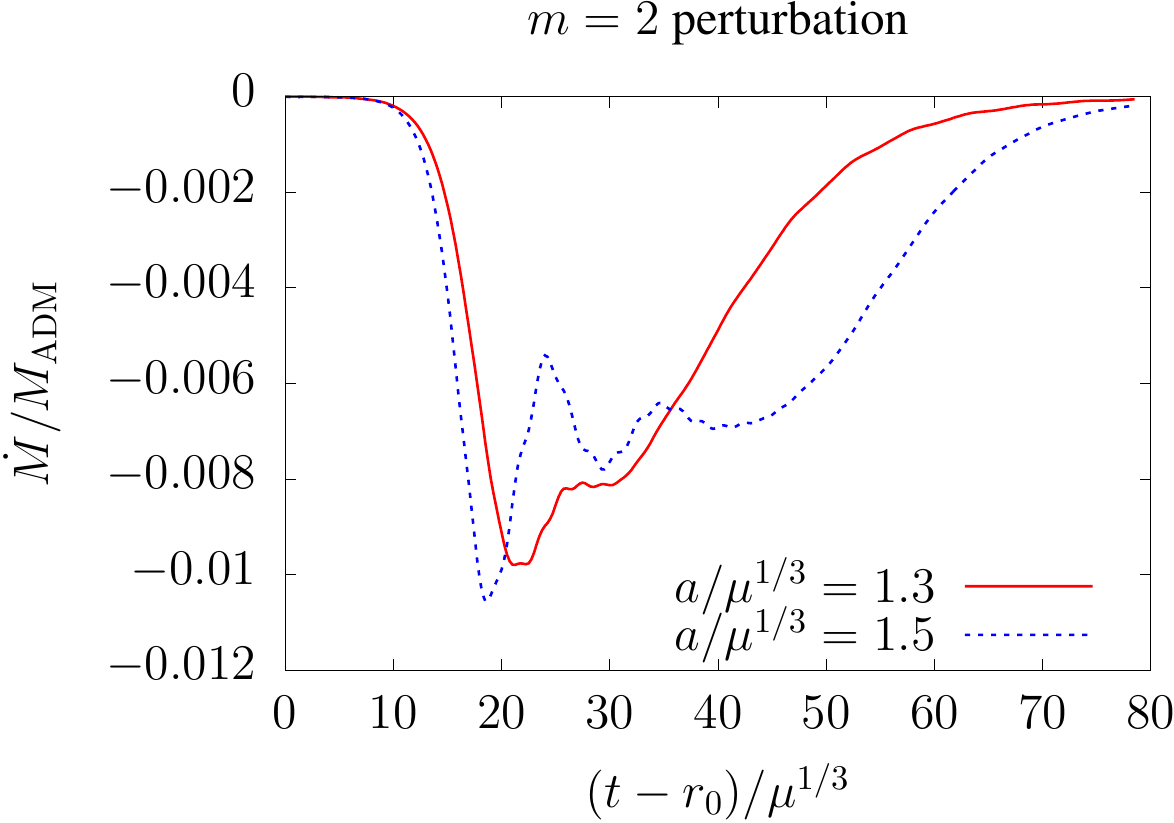}
\includegraphics[width=7.5cm]{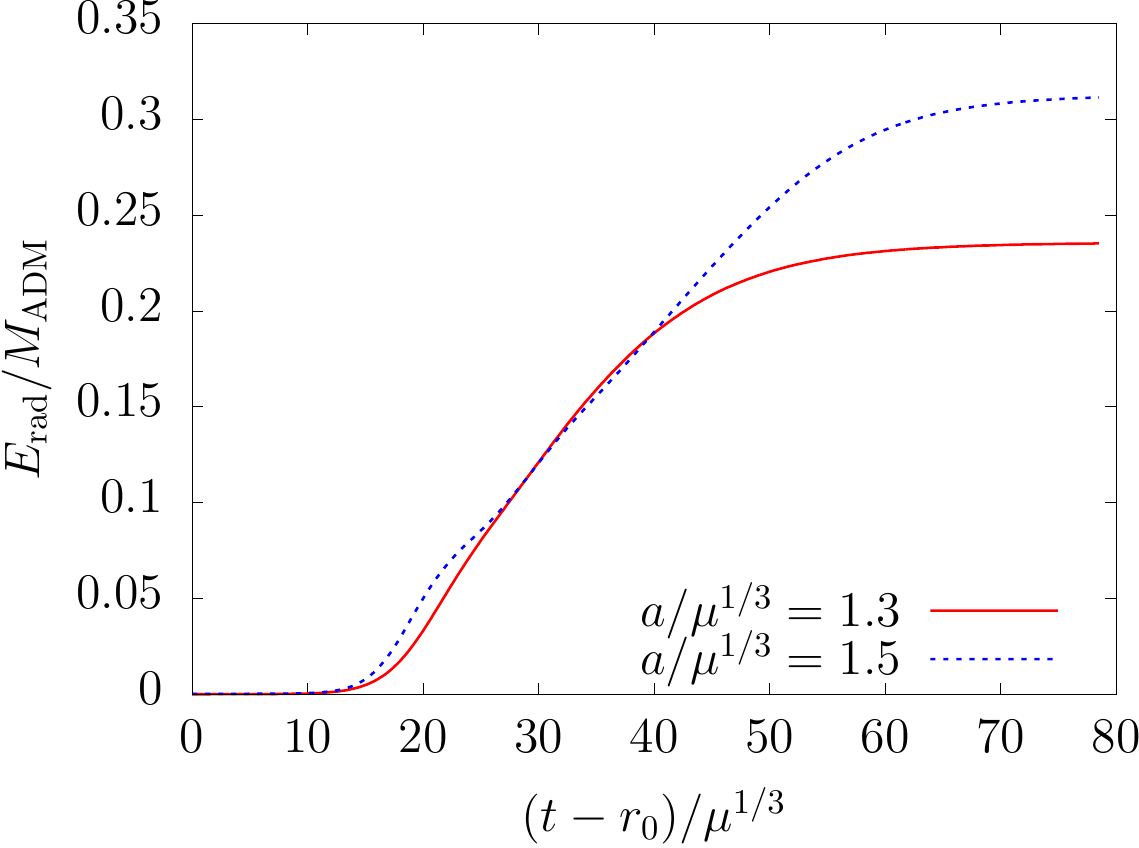}
\end{center}
\caption{\textit{Top}: energy flux for initial spins $a/\mu^{1/3} = 1.3$ (continuous red) and $a/\mu^{1/3} = 1.5$ (dotted blue). The emission reaches its peak during the initial growing phase, followed by a rather constant emission during the transient period and finally dropping to zero during the ring down --- cf.~Fig.\ref{fig:StableEvol_GW}. \textit{Bottom}: total radiated energy, corresponding to approximately $23.54\%$ ($a/\mu^{1/3} = 1.3$) and $31.15\%$ ($a/\mu^{1/3} = 1.5$) of the BH's mass.}
\label{fig:EnergyFlux_func_a}
\end{figure}

We end this section by calculating the energy emitted in the form of gravitational waves according to Eq.~\eqref{eq:EnergyFluxHarmDecomp}. As discussed in the previous sections (see also Appendix~\S\ref{sec:AppS4}), the main contribution to wave signal comes from the $\ell_4=2$ and $\ell_4=4$ projections of the Weyl tensor into the scalar-derived tensor harmonics. The top panel of Fig.~\ref{fig:EnergyFlux_func_a} compares the normalized energy flux for the $m=2$ perturbation of MP BH's with initial spin $a/\mu^{1/3} = 1.3$ and $a/\mu^{1/3} = 1.5$. The peak of emission occurs around $(t-r_0)/\mu^{1/3} = 15-20$, which corresponds exactly to the period dominated by the growing QNM phase. Then, the emission rate stays relatively constant during the transient period, before dropping to zero in the final ring down phase. These three phases of the emission are evident in the gravitational wave signal depicted in~Fig.\ref{fig:StableEvol_GW}. 

The orders of magnitude in the energy flux are essentially the same for both systems. However, as noticed in Sec.~\ref{sec:mp6d_gw}, the larger the initial spin, the longer the transient period. Hence, the system emits energy for a longer time, before settling down to the final BH. Indeed, the bottom panel of Fig.~\ref{fig:EnergyFlux_func_a} shows the normalized radiated energy, and we observe and emission of approximately $23.54\%$ and $31.15\%$ of the BH's mass with initial spin $a/\mu^{1/3} = 1.3$ and $a/\mu^{1/3} = 1.5$, respectively. This should be contrasted with the head-on collisions or the merger of binary black holes in higher dimensions, for which less than $1\%$ of the total mass is radiated \cite{Witek:2010xi,Cook:2017fec,Cook:2018fxg} (in $D=6$, the fraction of mass radiated is $(8.19\pm0.05)\times 10^{-4}$ in the head-on collisions and $(1.99 \pm 0.05) \times 10^{-1}$ in the binary merger), or in the first black hole binary merger recorded by LIGO \cite{Abbott:2016blz}, which radiated roughly $4.6\%$ of the initial mass into gravitational waves. 

It is instructive to compare the energy emitted by the gravitational waves against the total available energy given by
\beq
E_{\rm avail} = 1 - \dfrac{M_{\rm irr}}{M},
\eeq
with $M_{\rm irr}$ the irreducible mass
\beq
\frac{M_{\rm irr}}{M} = \frac{ r_+^{d-3}}{\mu} \Bigg(1 + \left(\dfrac{a}{r_+}\right)^2 \Bigg)^{\frac{d-3}{d-2}}  \,,
\eeq
where $r_+$ is the horizon location of the MP BH in the Boyer-Lindquist type of coordinates \eqref{eq:MPdata}. The energy carried out by gravitational waves corresponds to $E_{\rm rad}/E_{\rm avail} \sim 0.60$ and $E_{\rm rad}/E_{\rm avail} \sim 0.64$ for $a/\mu^{1/3} = 1.3$ and $a/\mu^{1/3} = 1.5$, respectively.

\begin{figure}[t!]
\begin{center}
\includegraphics[width=7.7cm]{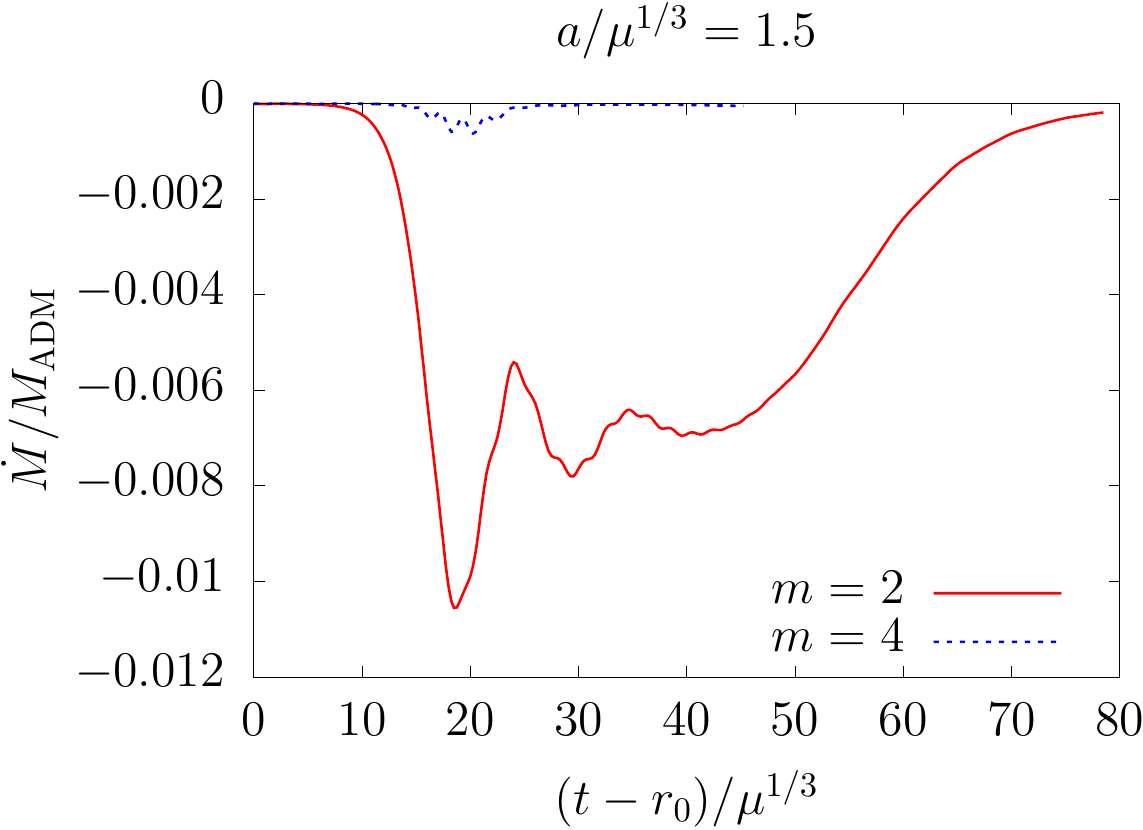}
\includegraphics[width=7.5cm]{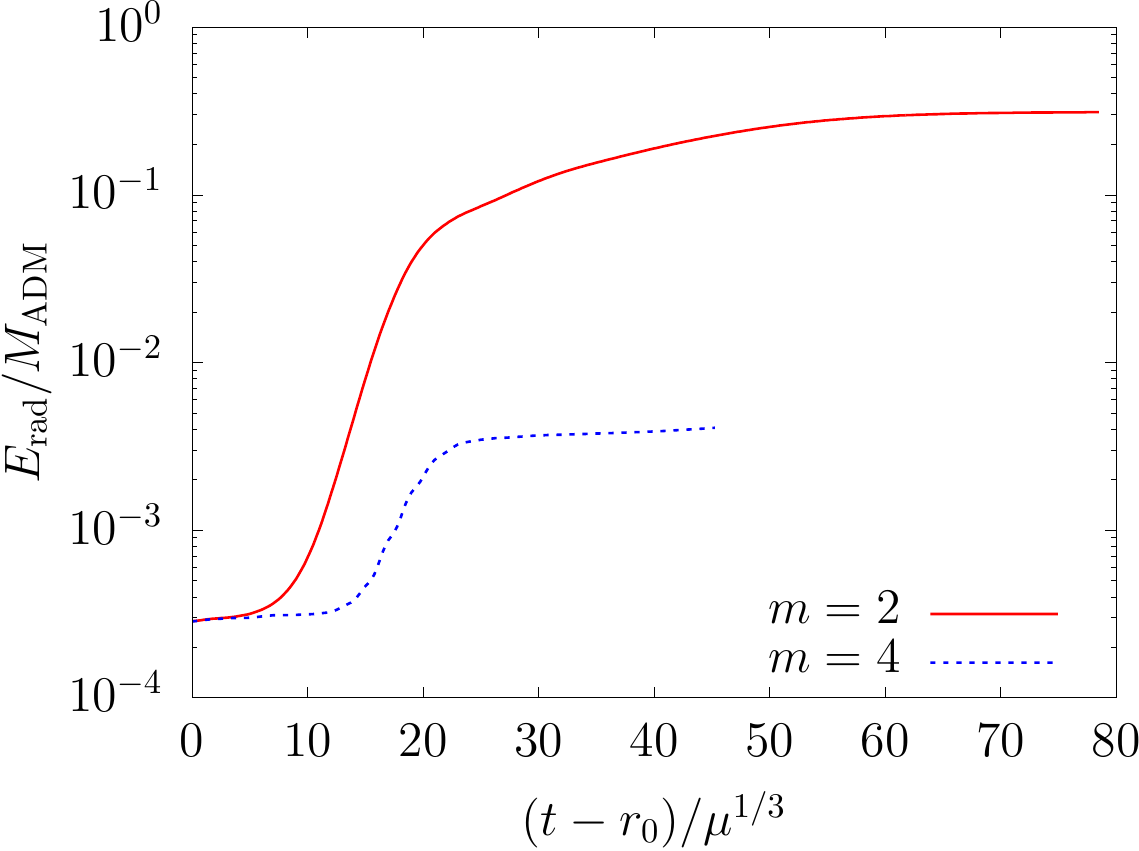}
\end{center}
\caption{Initial spin $a/\mu^{1/3} = 1.5$ perturbed with $m=2$ (continuous red) and $m=4$ (dotted blue) deformation. The emission rate \textit{top} and total radiated energy \textit{bottom} for the $m=4$ dynamics is less efficient than the $m=2$ evolution by at least 2 orders of magnitude. }
\label{fig:EnergyFlux_a150}
\end{figure}

Finally, we compare the energy emission for the MP BH with spin $a/\mu^{1/3} = 1.5$ perturbed with $m=2$ and $m=4$ deformations. 
Figure~\ref{fig:EnergyFlux_a150} confirms that the efficiency of the emission is lower by 2 orders of magnitudes in the $m=4$ dynamics, which gives more time for the instabilities to grow, as discussed in the previous section.
In particular, the $m=4$ perturbation depicted in Fig. \ref{fig:Evol_chi_a1p5_m4} evolves towards a configuration with a large central bulge that contains most of the black hole's mass and becomes progressively more spherical as the evolution proceeds.
This mostly spherical configuration is thus expected to be less efficient at emitting gravitational waves than the configuration that results from the $m=2$ perturbation depicted in the top panel of Fig. \ref{fig:Evol_chi_a1p5_m2}.

\section{Conclusion}
\label{sec:discussion}

In this article we describe the evolution and end points of nonaxisymmetric perturbations of singly spinning MP BHs in five and six dimensions. 
In the 5D case, we find that all black holes that we have considered are nonlinearly stable, thus resolving the tension between the numerical simulations of Ref.~\cite{Shibata:2009ad} and the perturbation theory results of \cite{Dias:2014eua}. 
In the 6D case we found that slowly spinning black holes are stable. 
As the initial spin increases, at around $a/\mu^{1/3}\approx 0.8$ MP BHs become unstable under perturbations with an $m=2$ azimuthal mode number (i.e., bar mode), as Ref.~\cite{Shibata:2010wz} had previously observed and Ref.~\cite{Dias:2014eua} confirmed. 
We have been able to evolve MP BHs with significantly larger initial spins than previous works, and the picture that has emerged is the following: in the nonlinear regime, unstable black holes develop a bar shape which leads to a strong emission of gravitational waves, through which the black hole eventually settles back down to another member of the MP family with a lower angular momentum.  
See Fig. \ref{fig:Evol_AH_a1p3}. 
As the initial spin of the black hole grows, the amount of mass and angular momentum emitted by the gravitational waves during the evolution also grows, reaching a staggering $31.15\%$ of radiated mass for a black hole with initial spin $a/\mu^{1/3}=1.5$. 
This is much larger than previously observed in head-on collisions and merger of black holes in higher dimensions. 
It would be interesting to explore whether these high emission rates have implications for scenarios with large extra dimensions. 
Another surprising fact uncovered in this work is that \textit{all} unstable MP BHs that settle back down to another member of the MP family in fact seem to settle onto the same black hole, the MP BH with $a/\mu^{1/3}=0.63$. 
Therefore, this particular member of the MP family seems to be an attractor, at least for the evolution of unstable BHs under a bar-mode type of deformation. 
We do not have a physical explanation for this behavior since the  $a/\mu^{1/3}=0.63$ black hole does not to have any particular physical properties that single it out among nearby solutions. 
It would be very interesting to understand why this particular solution seems to be special. 

For initial spins of $a/\mu^{1/3}\approx1.5$, it is the $m=4$ mode that has the fastest growth rate and will generically dominate the subsequent evolution, even though lower $m$ modes are still present. 
The same situation was observed in the case of the black ring in 5D \cite{Figueras:2015hkb}. 
The evolution of these instabilities is different than those at lower spins; the $m=4$ mode results in a square-type deformation, which then evolves into four arms (see Fig. \ref{fig:Evol_chi_a1p5_m4}). 
These arms grow longer and thinner over time, until they eventually develop a local GL instability. 
In this article we have been able to follow the evolution of this local instability beyond the formation of the first generation of bulges. 
After this point, the simulation becomes too computationally expensive. 
However, at this point the fate of these black holes is sealed: the development of the GL instability is at least exponential in time, while the emission of gravitational waves only follows a power law. 
Therefore, once the local GL instability kicks in, a naked singularity will inevitably form in finite asymptotic time before the unlucky black hole has had time to radiate away enough mass and angular momentum. 
Since this mechanism is generic, one may interpret this as further evidence that the weak cosmic censorship conjecture is false in higher-dimensional asymptotically flat spacetimes. 
Note that it is not clear whether the type of singularity that forms is visible to external faraway observers and hence whether it constitutes an honest violation of the conjecture. 
To settle this issue, one would need to analyze the (in)completeness of null infinity, which remains an open question for this class of spacetimes. 
However, arbitrarily large curvatures near the horizon do become visible. 
Since the mass contained in these regions of increasingly large curvature becomes negligible over time, the potential violation of the weak cosmic censorship conjecture would be of the mildest possible type in the language of Ref.~\cite{Andrade:2018yqu}. 
For these rapidly spinning black holes, one can also trigger an instability in the $m=2$ sector of perturbations.
Even though we were not able to follow the evolution to the point of seeing the appearance of local GL instabilities and hence naked singularities, it seems that this is the most likely end point. 

The instabilities discussed in this article are of the ``elastic" type \cite{Figueras:2015hkb}, as opposed to the GL type of instabilities. 
Intuitively, elastic instabilities deform the black hole horizon without introducing substantial thickness variations. 
GL instabilities in singly spinning MP BHs kick in at slightly larger spins, $a/\mu^{1/3}=1.572$ in $D=6$ \cite{Dias:2010maa}; the evolution and end points of the latter were studied in Ref.~\cite{Figueras:2017zwa}. 
Whether the elastic type of instabilities or the GL ones dominate the evolution of a given black hole is a question of suitably preparing the initial conditions. 
We do not see any obstruction in being able to construct open sets of initial conditions for which  either the elastic or GL instability dominates the subsequent evolution by exciting the desired mode with a sufficiently large initial amplitude. 
Therefore, this article, together with Ref.~\cite{Figueras:2017zwa}, provides a complete picture of the dynamics of some of the most notable instabilities that afflict MP black holes and their respective end points.\footnote{The superradiant instability is another important instability that affects these black holes, but we have not considered it here.}  

One important result in higher-dimensional black hole physics that emerges from our work is that sufficiently rapidly rotating black holes in higher-dimensions are unstable and evolve into a naked singularity in finite time as a result of local GL dynamics, regardless of the nature of the original instability. 
Both the elastic and GL instabilities are quite generic in the sense that they should afflict other types of higher-dimensional black holes apart from MP BHs and black rings. 
This leads us to the following conjecture. 
The GL instability is the only mechanism that higher-dimensional vacuum GR has to change the topology of a black hole horizon in dynamical spacetimes. 
If true, this conjecture would make the detailed understanding of the GL instability an even more pressing issue. 
The fact that the GL instability in black strings appears to be self-similar and perhaps controlled by an attractor offers hope that some analytic progress is possible in this case.

The dynamics of singly spinning MP BHs in seven dimensions is qualitatively and quantitatively very similar to the 6D ones. 
Namely, BHs are nonlinear stable for sufficiently small angular momentum. 
At larger spins, they first become unstable under a bar-mode type of deformation. 
The evolution of this instability again ends on another member of the MP family, with lower mass and spin. 
For larger angular momentum, the $m=4$ mode becomes the fastest-growing unstable mode, and the end point is a naked singularity, which forms in finite asymptotic time. 
The only difference between six and seven dimensions is that naked singularities appear to form at lower values of the angular momentum. 
For instance, in seven dimensions, the $a/\mu^{1/4}=1.3$ MP BH is unstable under an $m=4$ mode, and the end point of the instability is a naked singularity. 
As a rule of thumb, it seems that forming naked singularities is ``easier" as $D$ increases. 

One may object that unstable BHs should not be regarded as ``generic'' and consequently the potential violations of the weak cosmic censorship conjecture in this and previous works should not be considered generic. 
First, the question of whether it is possible or not to construct open sets of initial conditions sufficiently close to the unstable black holes considered in these works remains open. 
Second, in this article we have shown that black holes with sufficiently large angular momentum develop local GL instabilities which inevitably end up forming naked singularities. 
Last but not least, using the large $D$ limit of GR, Ref.~\cite{Andrade:2018yqu} provided compelling evidence that it should be possible to generically form single black holes with large angular momentum by colliding black holes with a nonzero impact parameter. 
Moreover, this reference also suggested that these black holes should evolve into naked singularities via local GL instabilities. 
In this article, we have shown that these local GL instabilities do indeed kick in when one expects them to do so. 
It now remains to show that indeed the scenario put forward in Ref.~\cite{Andrade:2018yqu} is realized in finite dimensions. 
$D=6$ should suffice. 

\begin{acknowledgements}
We would like to thank Tomas Andrade, Jay Armas, Roberto Emparan, Mahdi Godazgar, Harvey Reall, and Ulrich Sperhake for discussions and the referee for a careful reading of the manuscript.
We have benefited from being part of the {\sc GRChombo} Collaboration (\texttt{www.grchombo.org}), our special thanks go to the whole collaboration.
HB, PF, MK and RPM were supported by the European Research Council Grant No. ERC-2014- StG 639022-NewNGR ``New frontiers in numerical general relativity". 
PF is also supported by a Royal Society University Research Fellowship (Grant No. UF140319).  
We acknowledge that the results of this research have been achieved using the DECI resource Cartesius based in the Netherlands with support from the PRACE DECI-14-Tier-1 call. 
Some of the computations reported in this article were also performed at the MareNostrum4 supercomputer at the Barcelona Supercomputing Centre, Grants No. FI-2019-1-0010, FI-2018-3-0004, FI-2017-3-0012, FI-2017-2-0018, FI-2017-1-0042. 
We also acknowledge the use of Athena at HPC Midlands+, which was funded by the EPSRC on Grant No. EP/P020232/1, in this research, as part of the HPC Midlands+ consortium. 
Part of this work was performed using the Cambridge Service for Data Driven Discovery (CSD3), part of which is operated by the University of Cambridge Research Computing on behalf of the STFC DiRAC HPC Facility (www.dirac.ac.uk). The DiRAC component of CSD3 was funded by BEIS capital funding via STFC capital Grants No. ST/P002307/1 and ST/R002452/1 and STFC operations Grant No. ST/R00689X/1. 
DiRAC is part of the National e-Infrastructure.
\end{acknowledgements}

\bibliography{bibitems}

\begin{thebibliography}{46}%
\makeatletter
\providecommand \@ifxundefined [1]{%
 \@ifx{#1\undefined}
}%
\providecommand \@ifnum [1]{%
 \ifnum #1\expandafter \@firstoftwo
 \else \expandafter \@secondoftwo
 \fi
}%
\providecommand \@ifx [1]{%
 \ifx #1\expandafter \@firstoftwo
 \else \expandafter \@secondoftwo
 \fi
}%
\providecommand \natexlab [1]{#1}%
\providecommand \enquote  [1]{``#1''}%
\providecommand \bibnamefont  [1]{#1}%
\providecommand \bibfnamefont [1]{#1}%
\providecommand \citenamefont [1]{#1}%
\providecommand \href@noop [0]{\@secondoftwo}%
\providecommand \href [0]{\begingroup \@sanitize@url \@href}%
\providecommand \@href[1]{\@@startlink{#1}\@@href}%
\providecommand \@@href[1]{\endgroup#1\@@endlink}%
\providecommand \@sanitize@url [0]{\catcode `\\12\catcode `\$12\catcode
  `\&12\catcode `\#12\catcode `\^12\catcode `\_12\catcode `\%12\relax}%
\providecommand \@@startlink[1]{}%
\providecommand \@@endlink[0]{}%
\providecommand \url  [0]{\begingroup\@sanitize@url \@url }%
\providecommand \@url [1]{\endgroup\@href {#1}{\urlprefix }}%
\providecommand \urlprefix  [0]{URL }%
\providecommand \Eprint [0]{\href }%
\providecommand \doibase [0]{http://dx.doi.org/}%
\providecommand \selectlanguage [0]{\@gobble}%
\providecommand \bibinfo  [0]{\@secondoftwo}%
\providecommand \bibfield  [0]{\@secondoftwo}%
\providecommand \translation [1]{[#1]}%
\providecommand \BibitemOpen [0]{}%
\providecommand \bibitemStop [0]{}%
\providecommand \bibitemNoStop [0]{.\EOS\space}%
\providecommand \EOS [0]{\spacefactor3000\relax}%
\providecommand \BibitemShut  [1]{\csname bibitem#1\endcsname}%
\let\auto@bib@innerbib\@empty
\bibitem [{\citenamefont {Abbott}\ \emph {et~al.}(2016)\citenamefont {Abbott}
  \emph {et~al.}}]{Abbott:2016blz}%
  \BibitemOpen
  \bibfield  {author} {\bibinfo {author} {\bibfnamefont {B.~P.}\ \bibnamefont
  {Abbott}} \emph {et~al.} (\bibinfo {collaboration} {LIGO Scientific,
  Virgo}),\ }\href {\doibase 10.1103/PhysRevLett.116.061102} {\bibfield
  {journal} {\bibinfo  {journal} {Phys. Rev. Lett.}\ }\textbf {\bibinfo
  {volume} {116}},\ \bibinfo {pages} {061102} (\bibinfo {year}
  {2016})}\BibitemShut {NoStop}%
\bibitem [{\citenamefont {Akiyama}\ \emph {et~al.}(2019)\citenamefont {Akiyama}
  \emph {et~al.}}]{Akiyama:2019cqa}%
  \BibitemOpen
  \bibfield  {author} {\bibinfo {author} {\bibfnamefont {K.}~\bibnamefont
  {Akiyama}} \emph {et~al.} (\bibinfo {collaboration} {Event Horizon
  Telescope}),\ }\href {\doibase 10.3847/2041-8213/ab0ec7} {\bibfield
  {journal} {\bibinfo  {journal} {Astrophys. J.}\ }\textbf {\bibinfo {volume}
  {875}},\ \bibinfo {pages} {L1} (\bibinfo {year} {2019})}\BibitemShut
  {NoStop}%
\bibitem [{\citenamefont {Kerr}(1963)}]{Kerr:1963ud}%
  \BibitemOpen
  \bibfield  {author} {\bibinfo {author} {\bibfnamefont {R.~P.}\ \bibnamefont
  {Kerr}},\ }\href@noop {} {\bibfield  {journal} {\bibinfo  {journal} {Phys.
  Rev. Lett.}\ }\textbf {\bibinfo {volume} {11}},\ \bibinfo {pages} {237}
  (\bibinfo {year} {1963})}\BibitemShut {NoStop}%
\bibitem [{\citenamefont {Gregory}\ and\ \citenamefont
  {Laflamme}(1993)}]{Gregory:1993vy}%
  \BibitemOpen
  \bibfield  {author} {\bibinfo {author} {\bibfnamefont {R.}~\bibnamefont
  {Gregory}}\ and\ \bibinfo {author} {\bibfnamefont {R.}~\bibnamefont
  {Laflamme}},\ }\href {\doibase 10.1103/PhysRevLett.70.2837} {\bibfield
  {journal} {\bibinfo  {journal} {Phys. Rev. Lett.}\ }\textbf {\bibinfo
  {volume} {70}},\ \bibinfo {pages} {2837} (\bibinfo {year}
  {1993})}\BibitemShut {NoStop}%
\bibitem [{\citenamefont {Lehner}\ and\ \citenamefont
  {Pretorius}(2010)}]{Lehner:2010pn}%
  \BibitemOpen
  \bibfield  {author} {\bibinfo {author} {\bibfnamefont {L.}~\bibnamefont
  {Lehner}}\ and\ \bibinfo {author} {\bibfnamefont {F.}~\bibnamefont
  {Pretorius}},\ }\href {\doibase 10.1103/PhysRevLett.105.101102} {\bibfield
  {journal} {\bibinfo  {journal} {Phys. Rev. Lett.}\ }\textbf {\bibinfo
  {volume} {105}},\ \bibinfo {pages} {101102} (\bibinfo {year}
  {2010})}\BibitemShut {NoStop}%
\bibitem [{\citenamefont {Emparan}\ and\ \citenamefont
  {Reall}(2002)}]{Emparan:2001wn}%
  \BibitemOpen
  \bibfield  {author} {\bibinfo {author} {\bibfnamefont {R.}~\bibnamefont
  {Emparan}}\ and\ \bibinfo {author} {\bibfnamefont {H.~S.}\ \bibnamefont
  {Reall}},\ }\href {\doibase 10.1103/PhysRevLett.88.101101} {\bibfield
  {journal} {\bibinfo  {journal} {Phys. Rev. Lett.}\ }\textbf {\bibinfo
  {volume} {88}},\ \bibinfo {pages} {101101} (\bibinfo {year}
  {2002})}\BibitemShut {NoStop}%
\bibitem [{\citenamefont {Pomeransky}\ and\ \citenamefont
  {Sen'kov}(2006)}]{Pomeransky:2006bd}%
  \BibitemOpen
  \bibfield  {author} {\bibinfo {author} {\bibfnamefont {A.~A.}\ \bibnamefont
  {Pomeransky}}\ and\ \bibinfo {author} {\bibfnamefont {R.~A.}\ \bibnamefont
  {Sen'kov}},\ }\href@noop {} {\enquote {\bibinfo {title} {{Black ring with two
  angular momenta}},}\ } (\bibinfo {year} {2006}),\ \Eprint
  {http://arxiv.org/abs/hep-th/0612005} {arXiv:hep-th/0612005 [hep-th]}
  \BibitemShut {NoStop}%
\bibitem [{\citenamefont {Santos}\ and\ \citenamefont
  {Way}(2015)}]{Santos:2015iua}%
  \BibitemOpen
  \bibfield  {author} {\bibinfo {author} {\bibfnamefont {J.~E.}\ \bibnamefont
  {Santos}}\ and\ \bibinfo {author} {\bibfnamefont {B.}~\bibnamefont {Way}},\
  }\href {\doibase 10.1103/PhysRevLett.114.221101} {\bibfield  {journal}
  {\bibinfo  {journal} {Phys. Rev. Lett.}\ }\textbf {\bibinfo {volume} {114}},\
  \bibinfo {pages} {221101} (\bibinfo {year} {2015})}\BibitemShut {NoStop}%
\bibitem [{\citenamefont {Figueras}\ \emph {et~al.}(2016)\citenamefont
  {Figueras}, \citenamefont {Kunesch},\ and\ \citenamefont
  {Tunyasuvunakool}}]{Figueras:2015hkb}%
  \BibitemOpen
  \bibfield  {author} {\bibinfo {author} {\bibfnamefont {P.}~\bibnamefont
  {Figueras}}, \bibinfo {author} {\bibfnamefont {M.}~\bibnamefont {Kunesch}}, \
  and\ \bibinfo {author} {\bibfnamefont {S.}~\bibnamefont {Tunyasuvunakool}},\
  }\href {\doibase 10.1103/PhysRevLett.116.071102} {\bibfield  {journal}
  {\bibinfo  {journal} {Phys. Rev. Lett.}\ }\textbf {\bibinfo {volume} {116}},\
  \bibinfo {pages} {071102} (\bibinfo {year} {2016})}\BibitemShut {NoStop}%
\bibitem [{\citenamefont {Emparan}\ and\ \citenamefont
  {Myers}(2003)}]{Emparan:2003sy}%
  \BibitemOpen
  \bibfield  {author} {\bibinfo {author} {\bibfnamefont {R.}~\bibnamefont
  {Emparan}}\ and\ \bibinfo {author} {\bibfnamefont {R.~C.}\ \bibnamefont
  {Myers}},\ }\href {\doibase 10.1088/1126-6708/2003/09/025} {\bibfield
  {journal} {\bibinfo  {journal} {JHEP}\ }\textbf {\bibinfo {volume} {09}},\
  \bibinfo {pages} {025} (\bibinfo {year} {2003})}\BibitemShut {NoStop}%
\bibitem [{\citenamefont {Emparan}\ \emph {et~al.}(2009)\citenamefont
  {Emparan}, \citenamefont {Harmark}, \citenamefont {Niarchos},\ and\
  \citenamefont {Obers}}]{Emparan:2009cs}%
  \BibitemOpen
  \bibfield  {author} {\bibinfo {author} {\bibfnamefont {R.}~\bibnamefont
  {Emparan}}, \bibinfo {author} {\bibfnamefont {T.}~\bibnamefont {Harmark}},
  \bibinfo {author} {\bibfnamefont {V.}~\bibnamefont {Niarchos}}, \ and\
  \bibinfo {author} {\bibfnamefont {N.~A.}\ \bibnamefont {Obers}},\ }\href
  {\doibase 10.1103/PhysRevLett.102.191301} {\bibfield  {journal} {\bibinfo
  {journal} {Phys. Rev. Lett.}\ }\textbf {\bibinfo {volume} {102}},\ \bibinfo
  {pages} {191301} (\bibinfo {year} {2009})}\BibitemShut {NoStop}%
\bibitem [{\citenamefont {Dias}\ \emph {et~al.}(2009)\citenamefont {Dias},
  \citenamefont {Figueras}, \citenamefont {Monteiro}, \citenamefont {Santos},\
  and\ \citenamefont {Emparan}}]{Dias:2009iu}%
  \BibitemOpen
  \bibfield  {author} {\bibinfo {author} {\bibfnamefont {O.~J.~C.}\
  \bibnamefont {Dias}}, \bibinfo {author} {\bibfnamefont {P.}~\bibnamefont
  {Figueras}}, \bibinfo {author} {\bibfnamefont {R.}~\bibnamefont {Monteiro}},
  \bibinfo {author} {\bibfnamefont {J.~E.}\ \bibnamefont {Santos}}, \ and\
  \bibinfo {author} {\bibfnamefont {R.}~\bibnamefont {Emparan}},\ }\href
  {\doibase 10.1103/PhysRevD.80.111701} {\bibfield  {journal} {\bibinfo
  {journal} {Phys. Rev.}\ }\textbf {\bibinfo {volume} {D80}},\ \bibinfo {pages}
  {111701} (\bibinfo {year} {2009})}\BibitemShut {NoStop}%
\bibitem [{\citenamefont {Dias}\ \emph
  {et~al.}(2010{\natexlab{a}})\citenamefont {Dias}, \citenamefont {Figueras},
  \citenamefont {Monteiro}, \citenamefont {Reall},\ and\ \citenamefont
  {Santos}}]{Dias:2010eu}%
  \BibitemOpen
  \bibfield  {author} {\bibinfo {author} {\bibfnamefont {O.~J.~C.}\
  \bibnamefont {Dias}}, \bibinfo {author} {\bibfnamefont {P.}~\bibnamefont
  {Figueras}}, \bibinfo {author} {\bibfnamefont {R.}~\bibnamefont {Monteiro}},
  \bibinfo {author} {\bibfnamefont {H.~S.}\ \bibnamefont {Reall}}, \ and\
  \bibinfo {author} {\bibfnamefont {J.~E.}\ \bibnamefont {Santos}},\ }\href
  {\doibase 10.1007/JHEP05(2010)076} {\bibfield  {journal} {\bibinfo  {journal}
  {JHEP}\ }\textbf {\bibinfo {volume} {05}},\ \bibinfo {pages} {076} (\bibinfo
  {year} {2010}{\natexlab{a}})}\BibitemShut {NoStop}%
\bibitem [{\citenamefont {Figueras}\ \emph {et~al.}(2017)\citenamefont
  {Figueras}, \citenamefont {Kunesch}, \citenamefont {Lehner},\ and\
  \citenamefont {Tunyasuvunakool}}]{Figueras:2017zwa}%
  \BibitemOpen
  \bibfield  {author} {\bibinfo {author} {\bibfnamefont {P.}~\bibnamefont
  {Figueras}}, \bibinfo {author} {\bibfnamefont {M.}~\bibnamefont {Kunesch}},
  \bibinfo {author} {\bibfnamefont {L.}~\bibnamefont {Lehner}}, \ and\ \bibinfo
  {author} {\bibfnamefont {S.}~\bibnamefont {Tunyasuvunakool}},\ }\href
  {\doibase 10.1103/PhysRevLett.118.151103} {\bibfield  {journal} {\bibinfo
  {journal} {Phys. Rev. Lett.}\ }\textbf {\bibinfo {volume} {118}},\ \bibinfo
  {pages} {151103} (\bibinfo {year} {2017})}\BibitemShut {NoStop}%
\bibitem [{\citenamefont {Shibata}\ and\ \citenamefont
  {Yoshino}(2010{\natexlab{a}})}]{Shibata:2009ad}%
  \BibitemOpen
  \bibfield  {author} {\bibinfo {author} {\bibfnamefont {M.}~\bibnamefont
  {Shibata}}\ and\ \bibinfo {author} {\bibfnamefont {H.}~\bibnamefont
  {Yoshino}},\ }\href {\doibase 10.1103/PhysRevD.81.021501} {\bibfield
  {journal} {\bibinfo  {journal} {Phys. Rev.}\ }\textbf {\bibinfo {volume}
  {D81}},\ \bibinfo {pages} {021501} (\bibinfo {year}
  {2010}{\natexlab{a}})}\BibitemShut {NoStop}%
\bibitem [{\citenamefont {Shibata}\ and\ \citenamefont
  {Yoshino}(2010{\natexlab{b}})}]{Shibata:2010wz}%
  \BibitemOpen
  \bibfield  {author} {\bibinfo {author} {\bibfnamefont {M.}~\bibnamefont
  {Shibata}}\ and\ \bibinfo {author} {\bibfnamefont {H.}~\bibnamefont
  {Yoshino}},\ }\href {\doibase 10.1103/PhysRevD.81.104035} {\bibfield
  {journal} {\bibinfo  {journal} {Phys. Rev.}\ }\textbf {\bibinfo {volume}
  {D81}},\ \bibinfo {pages} {104035} (\bibinfo {year}
  {2010}{\natexlab{b}})}\BibitemShut {NoStop}%
\bibitem [{\citenamefont {Dias}\ \emph {et~al.}(2014)\citenamefont {Dias},
  \citenamefont {Hartnett},\ and\ \citenamefont {Santos}}]{Dias:2014eua}%
  \BibitemOpen
  \bibfield  {author} {\bibinfo {author} {\bibfnamefont {O.~J.~C.}\
  \bibnamefont {Dias}}, \bibinfo {author} {\bibfnamefont {G.~S.}\ \bibnamefont
  {Hartnett}}, \ and\ \bibinfo {author} {\bibfnamefont {J.~E.}\ \bibnamefont
  {Santos}},\ }\href {\doibase 10.1088/0264-9381/31/24/245011} {\bibfield
  {journal} {\bibinfo  {journal} {Class. Quant. Grav.}\ }\textbf {\bibinfo
  {volume} {31}},\ \bibinfo {pages} {245011} (\bibinfo {year}
  {2014})}\BibitemShut {NoStop}%
\bibitem [{\citenamefont {Pretorius}(2005)}]{Pretorius:2004jg}%
  \BibitemOpen
  \bibfield  {author} {\bibinfo {author} {\bibfnamefont {F.}~\bibnamefont
  {Pretorius}},\ }\href {\doibase 10.1088/0264-9381/22/2/014} {\bibfield
  {journal} {\bibinfo  {journal} {Class. Quant. Grav.}\ }\textbf {\bibinfo
  {volume} {22}},\ \bibinfo {pages} {425} (\bibinfo {year} {2005})}\BibitemShut
  {NoStop}%
\bibitem [{\citenamefont {Cook}\ \emph {et~al.}(2016)\citenamefont {Cook},
  \citenamefont {Figueras}, \citenamefont {Kunesch}, \citenamefont {Sperhake},\
  and\ \citenamefont {Tunyasuvunakool}}]{Cook:2016soy}%
  \BibitemOpen
  \bibfield  {author} {\bibinfo {author} {\bibfnamefont {W.~G.}\ \bibnamefont
  {Cook}}, \bibinfo {author} {\bibfnamefont {P.}~\bibnamefont {Figueras}},
  \bibinfo {author} {\bibfnamefont {M.}~\bibnamefont {Kunesch}}, \bibinfo
  {author} {\bibfnamefont {U.}~\bibnamefont {Sperhake}}, \ and\ \bibinfo
  {author} {\bibfnamefont {S.}~\bibnamefont {Tunyasuvunakool}},\ }\bibfield
  {booktitle} {\emph {\bibinfo {booktitle} {{Proceedings, 3rd Amazonian
  Symposium on Physics: Belem, Brazil, September 28-October 2, 2015}}},\ }\href
  {\doibase 10.1142/S0218271816410133} {\bibfield  {journal} {\bibinfo
  {journal} {Int. J. Mod. Phys.}\ }\textbf {\bibinfo {volume} {D25}},\ \bibinfo
  {pages} {1641013} (\bibinfo {year} {2016})}\BibitemShut {NoStop}%
\bibitem [{\citenamefont {Godazgar}\ and\ \citenamefont
  {Reall}(2012)}]{Godazgar:2012zq}%
  \BibitemOpen
  \bibfield  {author} {\bibinfo {author} {\bibfnamefont {M.}~\bibnamefont
  {Godazgar}}\ and\ \bibinfo {author} {\bibfnamefont {H.~S.}\ \bibnamefont
  {Reall}},\ }\href {\doibase 10.1103/PhysRevD.85.084021} {\bibfield  {journal}
  {\bibinfo  {journal} {Phys. Rev.}\ }\textbf {\bibinfo {volume} {D85}},\
  \bibinfo {pages} {084021} (\bibinfo {year} {2012})}\BibitemShut {NoStop}%
\bibitem [{\citenamefont {Alic}\ \emph {et~al.}(2012)\citenamefont {Alic},
  \citenamefont {Bona-Casas}, \citenamefont {Bona}, \citenamefont {Rezzolla},\
  and\ \citenamefont {Palenzuela}}]{Alic:2011gg}%
  \BibitemOpen
  \bibfield  {author} {\bibinfo {author} {\bibfnamefont {D.}~\bibnamefont
  {Alic}}, \bibinfo {author} {\bibfnamefont {C.}~\bibnamefont {Bona-Casas}},
  \bibinfo {author} {\bibfnamefont {C.}~\bibnamefont {Bona}}, \bibinfo {author}
  {\bibfnamefont {L.}~\bibnamefont {Rezzolla}}, \ and\ \bibinfo {author}
  {\bibfnamefont {C.}~\bibnamefont {Palenzuela}},\ }\href {\doibase
  10.1103/PhysRevD.85.064040} {\bibfield  {journal} {\bibinfo  {journal} {Phys.
  Rev.}\ }\textbf {\bibinfo {volume} {D85}},\ \bibinfo {pages} {064040}
  (\bibinfo {year} {2012})}\BibitemShut {NoStop}%
\bibitem [{\citenamefont {Weyhausen}\ \emph {et~al.}(2012)\citenamefont
  {Weyhausen}, \citenamefont {Bernuzzi},\ and\ \citenamefont
  {Hilditch}}]{Weyhausen:2011cg}%
  \BibitemOpen
  \bibfield  {author} {\bibinfo {author} {\bibfnamefont {A.}~\bibnamefont
  {Weyhausen}}, \bibinfo {author} {\bibfnamefont {S.}~\bibnamefont {Bernuzzi}},
  \ and\ \bibinfo {author} {\bibfnamefont {D.}~\bibnamefont {Hilditch}},\
  }\href {\doibase 10.1103/PhysRevD.85.024038} {\bibfield  {journal} {\bibinfo
  {journal} {Phys. Rev.}\ }\textbf {\bibinfo {volume} {D85}},\ \bibinfo {pages}
  {024038} (\bibinfo {year} {2012})}\BibitemShut {NoStop}%
\bibitem [{\citenamefont {Alic}\ \emph {et~al.}(2013)\citenamefont {Alic},
  \citenamefont {Kastaun},\ and\ \citenamefont {Rezzolla}}]{Alic:2013xsa}%
  \BibitemOpen
  \bibfield  {author} {\bibinfo {author} {\bibfnamefont {D.}~\bibnamefont
  {Alic}}, \bibinfo {author} {\bibfnamefont {W.}~\bibnamefont {Kastaun}}, \
  and\ \bibinfo {author} {\bibfnamefont {L.}~\bibnamefont {Rezzolla}},\ }\href
  {\doibase 10.1103/PhysRevD.88.064049} {\bibfield  {journal} {\bibinfo
  {journal} {Phys. Rev.}\ }\textbf {\bibinfo {volume} {D88}},\ \bibinfo {pages}
  {064049} (\bibinfo {year} {2013})}\BibitemShut {NoStop}%
\bibitem [{\citenamefont {Kreiss}\ \emph {et~al.}(1973)\citenamefont {Kreiss},
  \citenamefont {Oliger},\ and\ \citenamefont {Committee}}]{kreiss1973methods}%
  \BibitemOpen
  \bibfield  {author} {\bibinfo {author} {\bibfnamefont {H.}~\bibnamefont
  {Kreiss}}, \bibinfo {author} {\bibfnamefont {J.}~\bibnamefont {Oliger}}, \
  and\ \bibinfo {author} {\bibfnamefont {G.~A. R. P. J.~O.}\ \bibnamefont
  {Committee}},\ }\href {https://books.google.co.uk/books?id=OxMZAQAAIAAJ}
  {\emph {\bibinfo {title} {Methods for the approximate solution of time
  dependent problems}}},\ GARP publications series\ (\bibinfo  {publisher}
  {International Council of Scientific Unions, World Meteorological
  Organization},\ \bibinfo {year} {1973})\BibitemShut {NoStop}%
\bibitem [{\citenamefont {Clough}\ \emph {et~al.}(2015)\citenamefont {Clough},
  \citenamefont {Figueras}, \citenamefont {Finkel}, \citenamefont {Kunesch},
  \citenamefont {Lim},\ and\ \citenamefont {Tunyasuvunakool}}]{Clough:2015sqa}%
  \BibitemOpen
  \bibfield  {author} {\bibinfo {author} {\bibfnamefont {K.}~\bibnamefont
  {Clough}}, \bibinfo {author} {\bibfnamefont {P.}~\bibnamefont {Figueras}},
  \bibinfo {author} {\bibfnamefont {H.}~\bibnamefont {Finkel}}, \bibinfo
  {author} {\bibfnamefont {M.}~\bibnamefont {Kunesch}}, \bibinfo {author}
  {\bibfnamefont {E.~A.}\ \bibnamefont {Lim}}, \ and\ \bibinfo {author}
  {\bibfnamefont {S.}~\bibnamefont {Tunyasuvunakool}},\ }\href {\doibase
  10.1088/0264-9381/32/24/245011} {\bibfield  {journal} {\bibinfo  {journal}
  {Class. Quant. Grav.}\ }\textbf {\bibinfo {volume} {32}},\ \bibinfo {pages}
  {245011} (\bibinfo {year} {2015})},\ \bibinfo {note} {[Class. Quant.
  Grav.32,24(2015)]}\BibitemShut {NoStop}%
\bibitem [{\citenamefont {Adams}\ \emph {et~al.}(2015)\citenamefont {Adams}
  \emph {et~al.}}]{Adams:2015kgr}%
  \BibitemOpen
  \bibfield  {author} {\bibinfo {author} {\bibfnamefont {M.}~\bibnamefont
  {Adams}} \emph {et~al.},\ }\href@noop {} {\  (\bibinfo {year}
  {2015})}\BibitemShut {NoStop}%
\bibitem [{\citenamefont {Pook-Kolb}\ \emph {et~al.}(2019)\citenamefont
  {Pook-Kolb}, \citenamefont {Birnholtz}, \citenamefont {Krishnan},\ and\
  \citenamefont {Schnetter}}]{Pook-Kolb:2018igu}%
  \BibitemOpen
  \bibfield  {author} {\bibinfo {author} {\bibfnamefont {D.}~\bibnamefont
  {Pook-Kolb}}, \bibinfo {author} {\bibfnamefont {O.}~\bibnamefont
  {Birnholtz}}, \bibinfo {author} {\bibfnamefont {B.}~\bibnamefont {Krishnan}},
  \ and\ \bibinfo {author} {\bibfnamefont {E.}~\bibnamefont {Schnetter}},\
  }\href {\doibase 10.1103/PhysRevD.99.064005} {\bibfield  {journal} {\bibinfo
  {journal} {Phys. Rev.}\ }\textbf {\bibinfo {volume} {D99}},\ \bibinfo {pages}
  {064005} (\bibinfo {year} {2019})}\BibitemShut {NoStop}%
\bibitem [{\citenamefont {Cook}\ and\ \citenamefont
  {Sperhake}(2017)}]{Cook:2016qnt}%
  \BibitemOpen
  \bibfield  {author} {\bibinfo {author} {\bibfnamefont {W.~G.}\ \bibnamefont
  {Cook}}\ and\ \bibinfo {author} {\bibfnamefont {U.}~\bibnamefont
  {Sperhake}},\ }\href {\doibase 10.1088/1361-6382/aa5294} {\bibfield
  {journal} {\bibinfo  {journal} {Class. Quant. Grav.}\ }\textbf {\bibinfo
  {volume} {34}},\ \bibinfo {pages} {035010} (\bibinfo {year}
  {2017})}\BibitemShut {NoStop}%
\bibitem [{\citenamefont {Lindblom}\ \emph {et~al.}(2017)\citenamefont
  {Lindblom}, \citenamefont {Taylor},\ and\ \citenamefont
  {Zhang}}]{Lindblom:2017maa}%
  \BibitemOpen
  \bibfield  {author} {\bibinfo {author} {\bibfnamefont {L.}~\bibnamefont
  {Lindblom}}, \bibinfo {author} {\bibfnamefont {N.~W.}\ \bibnamefont
  {Taylor}}, \ and\ \bibinfo {author} {\bibfnamefont {F.}~\bibnamefont
  {Zhang}},\ }\href {\doibase 10.1007/s10714-017-2303-y} {\bibfield  {journal}
  {\bibinfo  {journal} {Gen. Rel. Grav.}\ }\textbf {\bibinfo {volume} {49}},\
  \bibinfo {pages} {139} (\bibinfo {year} {2017})}\BibitemShut {NoStop}%
\bibitem [{\citenamefont {Emparan}\ and\ \citenamefont
  {Reall}(2008)}]{Emparan:2008eg}%
  \BibitemOpen
  \bibfield  {author} {\bibinfo {author} {\bibfnamefont {R.}~\bibnamefont
  {Emparan}}\ and\ \bibinfo {author} {\bibfnamefont {H.~S.}\ \bibnamefont
  {Reall}},\ }\href {\doibase 10.12942/lrr-2008-6} {\bibfield  {journal}
  {\bibinfo  {journal} {Living Rev. Rel.}\ }\textbf {\bibinfo {volume} {11}},\
  \bibinfo {pages} {6} (\bibinfo {year} {2008})}\BibitemShut {NoStop}%
\bibitem [{\citenamefont {Yang}\ \emph {et~al.}(2015)\citenamefont {Yang},
  \citenamefont {Zimmerman},\ and\ \citenamefont {Lehner}}]{Yang:2014tla}%
  \BibitemOpen
  \bibfield  {author} {\bibinfo {author} {\bibfnamefont {H.}~\bibnamefont
  {Yang}}, \bibinfo {author} {\bibfnamefont {A.}~\bibnamefont {Zimmerman}}, \
  and\ \bibinfo {author} {\bibfnamefont {L.}~\bibnamefont {Lehner}},\ }\href
  {\doibase 10.1103/PhysRevLett.114.081101} {\bibfield  {journal} {\bibinfo
  {journal} {Phys. Rev. Lett.}\ }\textbf {\bibinfo {volume} {114}},\ \bibinfo
  {pages} {081101} (\bibinfo {year} {2015})}\BibitemShut {NoStop}%
\bibitem [{\citenamefont {Dias}\ \emph
  {et~al.}(2010{\natexlab{b}})\citenamefont {Dias}, \citenamefont {Figueras},
  \citenamefont {Monteiro},\ and\ \citenamefont {Santos}}]{Dias:2010maa}%
  \BibitemOpen
  \bibfield  {author} {\bibinfo {author} {\bibfnamefont {O.~J.~C.}\
  \bibnamefont {Dias}}, \bibinfo {author} {\bibfnamefont {P.}~\bibnamefont
  {Figueras}}, \bibinfo {author} {\bibfnamefont {R.}~\bibnamefont {Monteiro}},
  \ and\ \bibinfo {author} {\bibfnamefont {J.~E.}\ \bibnamefont {Santos}},\
  }\href {\doibase 10.1103/PhysRevD.82.104025} {\bibfield  {journal} {\bibinfo
  {journal} {Phys. Rev.}\ }\textbf {\bibinfo {volume} {D82}},\ \bibinfo {pages}
  {104025} (\bibinfo {year} {2010}{\natexlab{b}})}\BibitemShut {NoStop}%
\bibitem [{\citenamefont {Andrade}\ \emph {et~al.}()\citenamefont {Andrade},
  \citenamefont {Emparan}, \citenamefont {Licht},\ and\ \citenamefont
  {Luna}}]{RobertoPrivate}%
  \BibitemOpen
  \bibfield  {author} {\bibinfo {author} {\bibfnamefont {T.}~\bibnamefont
  {Andrade}}, \bibinfo {author} {\bibfnamefont {R.}~\bibnamefont {Emparan}},
  \bibinfo {author} {\bibfnamefont {D.}~\bibnamefont {Licht}}, \ and\ \bibinfo
  {author} {\bibfnamefont {R.}~\bibnamefont {Luna}},\ }\href@noop {} {\bibinfo
  {journal} {\textit{private communication}}\ }\BibitemShut {NoStop}%
\bibitem [{\citenamefont {Witek}\ \emph {et~al.}(2010)\citenamefont {Witek},
  \citenamefont {Zilhao}, \citenamefont {Gualtieri}, \citenamefont {Cardoso},
  \citenamefont {Herdeiro}, \citenamefont {Nerozzi},\ and\ \citenamefont
  {Sperhake}}]{Witek:2010xi}%
  \BibitemOpen
\bibfield  {journal} {  }\bibfield  {author} {\bibinfo {author} {\bibfnamefont
  {H.}~\bibnamefont {Witek}}, \bibinfo {author} {\bibfnamefont
  {M.}~\bibnamefont {Zilhao}}, \bibinfo {author} {\bibfnamefont
  {L.}~\bibnamefont {Gualtieri}}, \bibinfo {author} {\bibfnamefont
  {V.}~\bibnamefont {Cardoso}}, \bibinfo {author} {\bibfnamefont
  {C.}~\bibnamefont {Herdeiro}}, \bibinfo {author} {\bibfnamefont
  {A.}~\bibnamefont {Nerozzi}}, \ and\ \bibinfo {author} {\bibfnamefont
  {U.}~\bibnamefont {Sperhake}},\ }\href {\doibase 10.1103/PhysRevD.82.104014}
  {\bibfield  {journal} {\bibinfo  {journal} {Phys. Rev.}\ }\textbf {\bibinfo
  {volume} {D82}},\ \bibinfo {pages} {104014} (\bibinfo {year}
  {2010})}\BibitemShut {NoStop}%
\bibitem [{\citenamefont {Cook}\ \emph {et~al.}(2017)\citenamefont {Cook},
  \citenamefont {Sperhake}, \citenamefont {Berti},\ and\ \citenamefont
  {Cardoso}}]{Cook:2017fec}%
  \BibitemOpen
  \bibfield  {author} {\bibinfo {author} {\bibfnamefont {W.~G.}\ \bibnamefont
  {Cook}}, \bibinfo {author} {\bibfnamefont {U.}~\bibnamefont {Sperhake}},
  \bibinfo {author} {\bibfnamefont {E.}~\bibnamefont {Berti}}, \ and\ \bibinfo
  {author} {\bibfnamefont {V.}~\bibnamefont {Cardoso}},\ }\href {\doibase
  10.1103/PhysRevD.96.124006} {\bibfield  {journal} {\bibinfo  {journal} {Phys.
  Rev.}\ }\textbf {\bibinfo {volume} {D96}},\ \bibinfo {pages} {124006}
  (\bibinfo {year} {2017})}\BibitemShut {NoStop}%
\bibitem [{\citenamefont {Cook}\ \emph {et~al.}(2018)\citenamefont {Cook},
  \citenamefont {Wang},\ and\ \citenamefont {Sperhake}}]{Cook:2018fxg}%
  \BibitemOpen
  \bibfield  {author} {\bibinfo {author} {\bibfnamefont {W.~G.}\ \bibnamefont
  {Cook}}, \bibinfo {author} {\bibfnamefont {D.}~\bibnamefont {Wang}}, \ and\
  \bibinfo {author} {\bibfnamefont {U.}~\bibnamefont {Sperhake}},\ }\href
  {\doibase 10.1088/1361-6382/aae995} {\bibfield  {journal} {\bibinfo
  {journal} {Class. Quant. Grav.}\ }\textbf {\bibinfo {volume} {35}},\ \bibinfo
  {pages} {235008} (\bibinfo {year} {2018})}\BibitemShut {NoStop}%
\bibitem [{\citenamefont {Andrade}\ \emph {et~al.}(2019)\citenamefont
  {Andrade}, \citenamefont {Emparan}, \citenamefont {Licht},\ and\
  \citenamefont {Luna}}]{Andrade:2018yqu}%
  \BibitemOpen
  \bibfield  {author} {\bibinfo {author} {\bibfnamefont {T.}~\bibnamefont
  {Andrade}}, \bibinfo {author} {\bibfnamefont {R.}~\bibnamefont {Emparan}},
  \bibinfo {author} {\bibfnamefont {D.}~\bibnamefont {Licht}}, \ and\ \bibinfo
  {author} {\bibfnamefont {R.}~\bibnamefont {Luna}},\ }\href {\doibase
  10.1007/JHEP04(2019)121} {\bibfield  {journal} {\bibinfo  {journal} {JHEP}\
  }\textbf {\bibinfo {volume} {04}},\ \bibinfo {pages} {121} (\bibinfo {year}
  {2019})}\BibitemShut {NoStop}%
\bibitem [{\citenamefont {Bondi}\ \emph {et~al.}(1962)\citenamefont {Bondi},
  \citenamefont {van~der Burg},\ and\ \citenamefont {Metzner}}]{Bondi:1962px}%
  \BibitemOpen
  \bibfield  {author} {\bibinfo {author} {\bibfnamefont {H.}~\bibnamefont
  {Bondi}}, \bibinfo {author} {\bibfnamefont {M.~G.~J.}\ \bibnamefont {van~der
  Burg}}, \ and\ \bibinfo {author} {\bibfnamefont {A.~W.~K.}\ \bibnamefont
  {Metzner}},\ }\href {\doibase 10.1098/rspa.1962.0161} {\bibfield  {journal}
  {\bibinfo  {journal} {Proc. Roy. Soc. Lond.}\ }\textbf {\bibinfo {volume}
  {A269}},\ \bibinfo {pages} {21} (\bibinfo {year} {1962})}\BibitemShut
  {NoStop}%
\bibitem [{\citenamefont {Sachs}(1962)}]{Sachs:1962wk}%
  \BibitemOpen
  \bibfield  {author} {\bibinfo {author} {\bibfnamefont {R.~K.}\ \bibnamefont
  {Sachs}},\ }\href {\doibase 10.1098/rspa.1962.0206} {\bibfield  {journal}
  {\bibinfo  {journal} {Proc. Roy. Soc. Lond.}\ }\textbf {\bibinfo {volume}
  {A270}},\ \bibinfo {pages} {103} (\bibinfo {year} {1962})}\BibitemShut
  {NoStop}%
\bibitem [{\citenamefont {Tanabe}\ \emph {et~al.}(2011)\citenamefont {Tanabe},
  \citenamefont {Kinoshita},\ and\ \citenamefont {Shiromizu}}]{Tanabe:2011es}%
  \BibitemOpen
  \bibfield  {author} {\bibinfo {author} {\bibfnamefont {K.}~\bibnamefont
  {Tanabe}}, \bibinfo {author} {\bibfnamefont {S.}~\bibnamefont {Kinoshita}}, \
  and\ \bibinfo {author} {\bibfnamefont {T.}~\bibnamefont {Shiromizu}},\ }\href
  {\doibase 10.1103/PhysRevD.84.044055} {\bibfield  {journal} {\bibinfo
  {journal} {Phys. Rev.}\ }\textbf {\bibinfo {volume} {D84}},\ \bibinfo {pages}
  {044055} (\bibinfo {year} {2011})}\BibitemShut {NoStop}%
\bibitem [{\citenamefont {Durkee}\ \emph {et~al.}(2010)\citenamefont {Durkee},
  \citenamefont {Pravda}, \citenamefont {Pravdova},\ and\ \citenamefont
  {Reall}}]{Durkee:2010xq}%
  \BibitemOpen
  \bibfield  {author} {\bibinfo {author} {\bibfnamefont {M.}~\bibnamefont
  {Durkee}}, \bibinfo {author} {\bibfnamefont {V.}~\bibnamefont {Pravda}},
  \bibinfo {author} {\bibfnamefont {A.}~\bibnamefont {Pravdova}}, \ and\
  \bibinfo {author} {\bibfnamefont {H.~S.}\ \bibnamefont {Reall}},\ }\href
  {\doibase 10.1088/0264-9381/27/21/215010} {\bibfield  {journal} {\bibinfo
  {journal} {Class. Quant. Grav.}\ }\textbf {\bibinfo {volume} {27}},\ \bibinfo
  {pages} {215010} (\bibinfo {year} {2010})}\BibitemShut {NoStop}%
\bibitem [{\citenamefont {Chodos}\ and\ \citenamefont
  {Myers}(1984)}]{Chodos:1983zi}%
  \BibitemOpen
  \bibfield  {author} {\bibinfo {author} {\bibfnamefont {A.}~\bibnamefont
  {Chodos}}\ and\ \bibinfo {author} {\bibfnamefont {E.}~\bibnamefont {Myers}},\
  }\href {\doibase 10.1016/0003-4916(84)90039-3} {\bibfield  {journal}
  {\bibinfo  {journal} {Annals Phys.}\ }\textbf {\bibinfo {volume} {156}},\
  \bibinfo {pages} {412} (\bibinfo {year} {1984})}\BibitemShut {NoStop}%
\bibitem [{\citenamefont {Rubin}\ and\ \citenamefont
  {Ordonez}(1985)}]{Rubin:1984tc}%
  \BibitemOpen
  \bibfield  {author} {\bibinfo {author} {\bibfnamefont {M.~A.}\ \bibnamefont
  {Rubin}}\ and\ \bibinfo {author} {\bibfnamefont {C.~R.}\ \bibnamefont
  {Ordonez}},\ }\href {\doibase 10.1063/1.526749} {\bibfield  {journal}
  {\bibinfo  {journal} {J. Math. Phys.}\ }\textbf {\bibinfo {volume} {26}},\
  \bibinfo {pages} {65} (\bibinfo {year} {1985})}\BibitemShut {NoStop}%
\bibitem [{\citenamefont {Higuchi}(1987)}]{Higuchi:1986wu}%
  \BibitemOpen
  \bibfield  {author} {\bibinfo {author} {\bibfnamefont {A.}~\bibnamefont
  {Higuchi}},\ }\href {\doibase 10.1063/1.527513} {\bibfield  {journal}
  {\bibinfo  {journal} {J. Math. Phys.}\ }\textbf {\bibinfo {volume} {28}},\
  \bibinfo {pages} {1553} (\bibinfo {year} {1987})},\ \bibinfo {note}
  {[Erratum: J. Math. Phys.43,6385(2002)]}\BibitemShut {NoStop}%
\bibitem [{\citenamefont {Gualtieri}\ \emph {et~al.}(2008)\citenamefont
  {Gualtieri}, \citenamefont {Berti}, \citenamefont {Cardoso},\ and\
  \citenamefont {Sperhake}}]{Gualtieri:2008ux}%
  \BibitemOpen
  \bibfield  {author} {\bibinfo {author} {\bibfnamefont {L.}~\bibnamefont
  {Gualtieri}}, \bibinfo {author} {\bibfnamefont {E.}~\bibnamefont {Berti}},
  \bibinfo {author} {\bibfnamefont {V.}~\bibnamefont {Cardoso}}, \ and\
  \bibinfo {author} {\bibfnamefont {U.}~\bibnamefont {Sperhake}},\ }\href
  {\doibase 10.1103/PhysRevD.78.044024} {\bibfield  {journal} {\bibinfo
  {journal} {Phys. Rev.}\ }\textbf {\bibinfo {volume} {D78}},\ \bibinfo {pages}
  {044024} (\bibinfo {year} {2008})}\BibitemShut {NoStop}%
\bibitem [{\citenamefont {Berti}\ \emph {et~al.}(2006)\citenamefont {Berti},
  \citenamefont {Cardoso},\ and\ \citenamefont {Casals}}]{Berti:2005gp}%
  \BibitemOpen
  \bibfield  {author} {\bibinfo {author} {\bibfnamefont {E.}~\bibnamefont
  {Berti}}, \bibinfo {author} {\bibfnamefont {V.}~\bibnamefont {Cardoso}}, \
  and\ \bibinfo {author} {\bibfnamefont {M.}~\bibnamefont {Casals}},\ }\href
  {\doibase 10.1103/PhysRevD.73.109902, 10.1103/PhysRevD.73.024013} {\bibfield
  {journal} {\bibinfo  {journal} {Phys. Rev.}\ }\textbf {\bibinfo {volume}
  {D73}},\ \bibinfo {pages} {024013} (\bibinfo {year} {2006})},\ \bibinfo
  {note} {[Erratum: Phys. Rev.D73,109902(2006)]}\BibitemShut {NoStop}%
\end{thebibliography}%
\bibliographystyle{apsrev4-1-noeprint.bst}

\appendix

\section{Asymptotic flatness and gravitational waves}
\label{sec:AFhigherD}

We assume the spacetime to be asymptotically flat in the sense of Ref.~\cite{Godazgar:2012zq}. 
Thus, in the wave zone, the line element is expressed in the Bondi-Sachs form~\cite{Bondi:1962px,Sachs:1962wk}
\beq
ds^2 = - \mathcal{A} e^{\mathcal{B}} du^2 - 2e^{\mathcal{B}} du\, dr + r^2 h_{\alpha \beta}(d\phi^\alpha + {\mathcal C}^\alpha)(d\phi^\beta + {\mathcal C}^\beta),
\eeq
with $u$ a retarded time and, $\mathcal{A}$, $\mathcal{B}$, ${\mathcal C}^\alpha$ and $h_{\alpha \beta}$ functions on the Bondi coordinate $(u,r,\phi^\alpha)$. Moreover, $\det h_{\alpha \beta} = \det \omega_{\alpha \beta}$, where $\omega_{\alpha \beta}$ is the unit metric on the unit $n$-sphere $S^n$. Here, we are interested in the asymptotic expansion~\cite{Tanabe:2011es}
\beq
h_{\alpha \beta}(u,r,\phi^\gamma) = \omega_{\alpha \beta}(\phi^\gamma) + \sum_{s\ge 1} \dfrac{h_{\alpha \beta}^{(s)}(u,\phi^\gamma)}{r^{D/2+s-2}}. 
\eeq 
with $h_{\alpha \beta}^{(1)}$ the Bondi news function.

We consider the higher-dimensional generalization of the Geroch-Held-Penrose formalism~\cite{Durkee:2010xq} and introduce a tetrad basis $(\ell, k, m_{A})$ satisfying $\ell^\mu k_\mu = 1$, $m_{(A)}^\mu m_{(B)}{}_\mu = \delta_{(A)(B)}$, and with vanishing further contractions. 
Asymptotically, one particular choice for the tetrad is
\bea
&\ell = -\partial_r, \quad k = \partial_u - \dfrac{1}{2}\partial_r \nn \\
\label{eq:Tetrad}
&m_{(1)} = r^{-1}\partial_\theta \quad m_{(2)} =(r\sin\theta)^{-1}\partial_\varphi \\
&m_{(a)} = \left(r\cos \theta \prod\limits_{b=4}^{a-1}\sin(\phi^b)\right)^{-1}\partial_{\phi^a} \quad a = 4 \cdots D-1 \nn
\eea  

Gravitational waves are extracted from the projection of the Weyl tensor $C_{\mu\nu\rho\sigma}$ onto the tetrad basis \eqref{eq:Tetrad}. 
Specifically, the news function $h^{(1)}_{\alpha\beta}$ is given by the leading contribution of Ref.~\cite{Godazgar:2012zq},
\bea
\label{eq:OmegaWeyl}
\Omega'_{AB} &=& C_{\mu\nu\rho\sigma}k^\mu m_{(A)}{}^\nu k^\rho m_{(B)}{}^\sigma \\ 
&=&- \dfrac{ m_{(A)}{}^\mu m_{(B)}{}^\nu \ddot{h}^{(1)}_{\mu \nu}}{2\,r^{D/2-1}} + \mathcal{O}(r^{-D/2}).
\eea
From the symmetries of $C_{\mu\nu\rho\sigma}$, one verifies that $\Omega'_{AB}$ is symmetric and traceless.

\section{Spherical harmonics on an $n$-sphere}
\label{sec:Snharmonics}

In this Appendix we collect some well-known properties of spherical harmonics on an $S^n$. 
In Appendixes!\ref{sec:AppS3} and!\ref{sec:AppS4} we specialize to the cases of interest for us, namely, $S^3$ and $S^4$.  

We start by setting up our notation and conventions.\footnote{Uppercase latin indices $A,B,...$ run from $0$ to $D-1$. Lowercase latin letters $i,j,...$ are spatial indices running from $1$ to $D-1$, and greek letters denote angular indices from $2$ to $D-1$. The sphere has dimension $K=D-2$, and the computation domain has $d=3$ spatial dimensions.} 
Following Ref.~\cite{Cook:2016qnt}, we consider spherical coordinates $Y^A=(t,r,\theta_a)$ which relate to a Cartesian coordinate system $X^a=(t,w_i)$ via
\begin{equation}
\begin{aligned}
&w_1 = r \cos(\theta_n)\,,  \\
&w_2 = r \sin(\theta_n)\cos(\theta_{n-1}) \,, \\
&w_3 = r \sin(\theta_n)\sin(\theta_{n-1})\cos(\theta_{n-2})\,,  \\
&\hspace{1cm}\vdots  \\
&w_{D-2} = r \sin(\theta_n) \cdots \sin(\theta_2)\cos(\theta_1) \,, \\
&w_{D-1} = r \sin(\theta_n) \cdots \sin(\theta_2)\sin(\theta_1)\,.
\end{aligned}
\end{equation}
Here, $\theta_1\in[0,2\pi]$ and $\theta_a \in [0,\pi]$ $\forall a=2\,\ldots,n$.
In the three-dimensional computational domain, we parametrize the numerical Cartesian coordinates $(x,y,z)$ by new spherical coordinates $(r, \theta, \varphi)$ via
\begin{equation}
\begin{aligned}
x & = r\cos(\theta_n)\,,\\
y & = r \sin(\theta_n)\cos(\theta_{n-1})\,, \\ 
z & = r \sin(\theta_n)\sin(\theta_{n-1})\,,
\end{aligned}
\label{eq:cartesian}
\end{equation}
with $\theta_n,\theta_{n-1}\in[0,\pi]$.

Now we are in position to collect some properties of scalar harmonics and higher rank tensor harmonics on an arbitrary $n$-sphere $S^n$, with $n>2$. 
Ultimately we are interested in tensor harmonics on the $S^3$ and the $S^4$. 
While the former are well known, see e.g., Ref.~\cite{Lindblom:2017maa} and references therein, our results for the tensor harmonics on the $S^4$ are, to the best of our knowledge, new.

We write down the metric $ds^2_n$ on the unit round  $S^n$ as
\begin{equation}
\begin{aligned}
&ds_1^2 = d\theta_1^2\,,\\
&ds^2_i = d\theta_i^2 + \sin^2\theta_i \, ds^2_{i-1}\,,\quad i=2,\ldots,n\,,
\end{aligned}
\end{equation}
with $0\leq \theta_1\leq 2\pi$ and $0\leq \theta_i\leq \pi$, $\forall i=2,\ldots,n$. 
We refer to this particular form of the metric on the $S^n$ as the ``standard" form and the corresponding basis of unit vectors as the standard basis. 

Recall that spherical harmonics on the unit $S^n$ are defined as eigenfunctions of the Laplace operator on the $S^n$, \footnote{Here, we work on the unit $S^n$. 
One can easily reinstate the radius in our formulas.}
\begin{equation}
\nabla^a\nabla_a \mathbb{S}^{\ell_n,\ldots,\ell_1} = -\ell_n(\ell_n + n -1)\,\mathbb{S}^{\ell_n,\ldots,\ell_1}\,,
\label{eq:scalar}
\end{equation}  
where the integers $\ell_i$ satisfy $\ell_n\geq\ell_{n-1}\geq\ldots\geq \ell_2\geq |\ell_1|$. 
The scalar harmonics are normalized so that 
\begin{equation}
\delta_{\ell_n \ell_n'\ldots \ell_1\ell_1'} = \int_{S^n} \mathbb{S}^{\ell_n,\ldots,\ell_1}\mathbb{S}^{\ell_n',\ldots,\ell_1'}\,,
\end{equation}
where the volume element of $S^n$ is omitted to avoid clutter. 
From the scalar harmonics, on can construct scalar-derived vector harmonics on the $S^n$. 
These are given by
\begin{equation}
 \mathbb{S}^{\ell_n,\ldots,\ell_1}_a = \frac{1}{\sqrt{\ell_n(\ell_n + n -1)}}\,\nabla_a \mathbb{S}^{\ell_n,\ldots,\ell_1}\,,
 \label{eq:scalarV}
\end{equation}
and they satisfy
\begin{align}
&\nabla^b\nabla_b \mathbb{S}^{\ell_n,\ldots,\ell_1}_a = \left[-\ell_n(\ell_n + n -1) + n - 1\right] \mathbb{S}^{\ell_n,\ldots,\ell_1}_a\,,\\
&\nabla^a \mathbb{S}^{\ell_n,\ldots,\ell_1}_a = - \sqrt{\ell_n(\ell_n + n -1)}\,\mathbb{S}^{\ell_n,\ldots,\ell_1}\,,\\
& \delta_{\ell_n \ell_n'\ldots \ell_1\ell_1'} = \int_{S^n}\,g^{ab}\, \mathbb{S}_a^{\ell_n,\ldots,\ell_1}\mathbb{S}_b^{\ell_n',\ldots,\ell_1'}\,,
\end{align}
where $g$ is the metric on the round unit $S^n$. 
On the other hand, divergence-free vector harmonics $\mathbb{V}_a^{\ell_n,\ldots,\ell_1}$ on the $S^n$ satisfy
\begin{align}
&\nabla^b\nabla_b \mathbb{V}^{\ell_n,\ldots,\ell_1}_a = \left[-\ell_n(\ell_n + n -1) + 1\right] \mathbb{V}^{\ell_n,\ldots,\ell_1}_a\,,\\
&\nabla^a \mathbb{V}^{\ell_n,\ldots,\ell_1}_a = 0\,,\\
& \delta_{\ell_n \ell_n'\ldots \ell_1\ell_1'} = \int_{S^n}\,g^{ab}\, \mathbb{V}_a^{\ell_n,\ldots,\ell_1}\mathbb{V}_b^{\ell_n',\ldots,\ell_1'}\,.
\end{align}
One can show that there are $n-1$ linearly independent, orthogonal, and divergence-free vector harmonics on the $S^n$ which, together with the scalar derived vector harmonics \eqref{eq:scalarV}, form a complete basis of vectors on the $S^n$, see Refs.~\cite{Chodos:1983zi,Rubin:1984tc,Higuchi:1986wu}.

From the scalar harmonics \eqref{eq:scalar}, one can obtain two types of symmetric tensor harmonics:
\begin{align}
&\mathbb{S}_{(1)ab}^{\ell_n,\ldots,\ell_1} = \frac{1}{\sqrt{n}}\,g_{ab}\,\mathbb{S}^{\ell_n,\ldots,\ell_1}\,,\\
& \mathbb{S}_{(2)ab}^{\ell_n,\ldots,\ell_1} = \frac{\sqrt{n}}{\sqrt{(n-1)(\ell_n-1)(\ell_n + n -1)}}\nn \\
&\left(\nabla_a \mathbb{S}^{\ell_n,\ldots,\ell_1}_b + \frac{\sqrt{\ell_n(\ell_n+n-1)}}{n}\,g_{ab} \,\mathbb{S}^{\ell_n,\ldots,\ell_1}\right)
\end{align}
These satisfy:
\begin{align}
&\nabla^c\nabla_c \,\mathbb{S}_{(1)ab}^{\ell_n,\ldots,\ell_1} = -\ell_n(\ell_n + n - 1)\, \mathbb{S}_{(1)ab}^{\ell_n,\ldots,\ell_1}\,\\
&\nabla^a \,\mathbb{S}_{(1)ab}^{\ell_n,\ldots,\ell_1} = -\frac{\sqrt{\ell_n(\ell_n + n -1)}}{\sqrt{n}}\, \mathbb{S}_{b}^{\ell_n,\ldots,\ell_1}\,,\\
&g^{ab} \,\mathbb{S}_{(1)ab}^{\ell_n,\ldots,\ell_1} = \sqrt{n}\, \mathbb{S}_{(1)}^{\ell_n,\ldots,\ell_1}\,,\\
& \delta_{\ell_n \ell_n'\ldots \ell_1\ell_1'} = \int_{S^n}\,g^{ac}g^{bd}\, \mathbb{S}_{(1)ab}^{\ell_n,\ldots,\ell_1}\,\mathbb{S}_{(1)cd}^{\ell_n',\ldots,\ell_1'}\,.
\end{align}
On the other hand, we have
\begin{align}
&\nabla^c\nabla_c \,\mathbb{S}_{(2)ab}^{\ell_n,\ldots,\ell_1} = \left[-\ell_n(\ell_n + n - 1) +2\,n\right]\, \mathbb{S}_{(2)ab}^{\ell_n,\ldots,\ell_1}\,\\
&\nabla^a \,\mathbb{S}_{(2)ab}^{\ell_n,\ldots,\ell_1} =- \frac{\sqrt{(n-1)(\ell_n-1)(\ell_n+n)}}{\sqrt{n}}\, \mathbb{S}_{b}^{\ell_n,\ldots,\ell_1}\,,\\
&g^{ab} \,\mathbb{S}_{(2)ab}^{\ell_n,\ldots,\ell_1} = 0\,,\\
& \delta_{\ell_n \ell_n'\ldots \ell_1\ell_1'} = \int_{S^n}\,g^{ac}g^{bd}\, \mathbb{S}_{(2)ab}^{\ell_n,\ldots,\ell_1}\,\mathbb{S}_{(2)cd}^{\ell_n',\ldots,\ell_1'}\,.
\end{align}
Recall that the matrix of Weyl scalars, $\Omega'_{AB}$, encoding the gravitational radiation is symmetric and traceless. 
Therefore, the family $\mathbb{S}_{(1)ab}$ of scalar-derived tensor harmonics cannot contribute to the multipolar expansion of the Weyl scalars. 
Hence, from now on we will only consider the second family, $\mathbb{S}_{(2)ab}$, of scalar-derived tensor harmonics and we shall drop the subscript $_{(2)}$. 

From each family of vector harmonics, one can similarly construct the corresponding family of vector-derived tensor harmonics:
\begin{equation}
\mathbb{V}_{ab}^{\ell_n,\ldots,\ell_1} = \frac{1}{\sqrt{2(\ell_n-1)(\ell_n+n)}}\,\left( \nabla_ a \mathbb{V}^{\ell_n,\ldots,\ell_1}_b + \nabla_b \mathbb{V}^{\ell_n,\ldots,\ell_1}_a\right)\,.
\end{equation}
These satisfy,
\begin{align}
&\nabla^c\nabla_c \mathbb{V}_{ab}^{\ell_n,\ldots,\ell_1} = \left[-\ell_n(\ell_n + n -1) + n + 2\right] \mathbb{V}_{ab}^{\ell_n,\ldots,\ell_1}\,,\\
&\nabla^a \mathbb{V}_{ab}^{\ell_n,\ldots,\ell_1}  = -\frac{\sqrt{(\ell_n-1)(\ell_n + n )}}{\sqrt{2}}\,\mathbb{V}_{b}^{\ell_n,\ldots,\ell_1}\,,\\
&g^{ab} \,\mathbb{V}_{ab}^{\ell_n,\ldots,\ell_1} = 0\,,\\
& \delta_{\ell_n \ell_n'\ldots \ell_1\ell_1'} = \int_{S^n}\,g^{ac}g^{bd}\, \mathbb{V}_{ab}^{\ell_n,\ldots,\ell_1}\,\mathbb{V}_{cd}^{\ell_n',\ldots,\ell_1'}\,.
\end{align}

Finally, transverse traceless symmetric tensor harmonics are defined by
\begin{align}
&\nabla^c\nabla_c \mathbb{T}_{ab}^{\ell_n,\ldots,\ell_1} = \left[-\ell_n(\ell_n + n -1) + 2\right] \mathbb{T}_{ab}^{\ell_n,\ldots,\ell_1}\,,\\
&\nabla^a \mathbb{T}_{ab}^{\ell_n,\ldots,\ell_1}  = 0\,,\\
&g^{ab} \,\mathbb{T}_{ab}^{\ell_n,\ldots,\ell_1} = 0\,,\\
& \delta_{\ell_n \ell_n'\ldots \ell_1\ell_1'} = \int_{S^n}\,g^{ac}g^{bd}\, \mathbb{T}_{ab}^{\ell_n,\ldots,\ell_1}\,\mathbb{T}_{cd}^{\ell_n',\ldots,\ell_1'}\,.
\end{align}
Needless to say, the different families of tensor harmonics are orthogonal to each other and together form a complete basis of tensor harmonics on the $S^n$.

The class of dynamical spacetimes that we consider possesses a transverse round $S^{n-2}$. 
Therefore, to extract the gravitational waveforms form the components of the Weyl scalars we can restrict ourselves to tensor harmonics on the $S^n$ that preserve an internal SO$(n-1)$ symmetry. 
This amounts to assuming that $\ell_{n-2} = \ldots = \ell_1 = 0$ for the quantum numbers labeling the harmonics. 
We shall restrict to this class of harmonics from now on. 

\section{Tensor harmonics on the $S^3$}
\label{sec:AppS3}
In this Appendix we collect some results on tensor harmonics on the $S^3$.

\subsection{Standard basis}

We write the round metric on the unit $S^3$ as:
\begin{equation}
ds^2 = d\theta_3^2 + \sin^2\theta_3 \left(d\theta_2^2 + \sin^2\theta_2\, d\theta_1^2\right)\,,
\label{eq:metricS3std}
\end{equation} 
with $\theta_3,\theta_2\in[0,\pi]$ and $\theta_1\in[0,2\pi]$. We choose the following basis of unit vectors on the $S^3$:
\begin{equation}
m_{(1)} = \frac{\partial}{\partial\theta_3}\,,~  m_{(2)} = \frac{1}{\sin\theta_3}\frac{\partial}{\partial\theta_2}\,,~ m_{(3)} = \frac{1}{\sin\theta_3\sin\theta_2}\frac{\partial}{\partial\theta_1}
\label{eq:S3standardbasis}
\end{equation}

In this basis, the projected components of the harmonics with $\ell_1=0$ are:\footnote{Since tensor harmonics are symmetric $\mathbb{Y}^{\ell\ldots}_{21} = \mathbb{Y}^{\ell\ldots}_{12}$ so we do not need to list all components.}

\noindent
\paragraph{Scalar-derived tensor harmonics}
\begin{itemize}
\item $(\ell_3,\ell_2) = (2,0)$:
\begin{equation}
\begin{aligned}
&\mathbb{S}^{(2,0)}_{11} = -\frac{2}{\pi} \sqrt{\frac{2}{15}} \sin^2(\theta_3 )\,,\\
&\mathbb{S}^{(2,0)}_{22} = \frac{1}{\pi}\sqrt{\frac{2}{15}} \sin^2(\theta_3 )\,,\\
&\mathbb{S}^{(2,0)}_{33} = \frac{1}{\pi}\sqrt{\frac{2}{15}} \sin^2(\theta_3)\,.
\end{aligned}
\end{equation}
\item $(\ell_3,\ell_2) = (2,1)$:
\begin{equation}
\begin{aligned}
& \mathbb{S}^{(2,1)}_{11} = -\frac{1}{\sqrt{5} \pi} \sin (2\theta_3 ) \cos (\theta_2 )\,,\\
& \mathbb{S}^{(2,1)}_{12} = \frac{3}{2 \sqrt{5} \pi } \sin(\theta_3)\sin (\theta_2 )\,,\\
& \mathbb{S}^{(2,1)}_{22} = \frac{1}{2\sqrt{5} \pi} \sin (2\theta_3 ) \cos (\theta_2 )\,,\\
& \mathbb{S}^{(2,1)}_{33} = \frac{1}{2\sqrt{5} \pi} \sin (2\theta_3 ) \cos (\theta_2 )\,.
\end{aligned}
\end{equation}
\item $(\ell_3,\ell_2) = (2,2)$:
\begin{equation}
\begin{aligned}
& \mathbb{S}^{(2,2)}_{11} = -\frac{1}{{4 \sqrt{15} \pi }}(\cos (2 \theta_3 )+2)(3 \cos (2 \theta_2 )+1) \,,\\
& \mathbb{S}^{(2,2)}_{12} = \frac{3}{4 \pi } \sqrt{\frac{3}{5}} \cos (\theta_3 )  \sin (2\theta_2 ) \,,\\
& \mathbb{S}^{(2,2)}_{22} = \frac{1}{8 \sqrt{15} \pi } [3 \cos (2 \theta_2 ) (\cos (2 \theta_3 )+5)+\cos (2 \theta_3 )-7]\,,\\
& \mathbb{S}^{(2,2)}_{33} =\frac{1}{2 \sqrt{15} \pi } \Big[\left(4-3 \cos ^2(\theta_2 )\right) \sin ^2(\theta_3 )+3  \cos ^2(\theta_3 )\Big]\,.
\end{aligned}
\end{equation}
\end{itemize}

\noindent
\paragraph{Vector-derived tensor harmonics}
There are two families of vector-derived tensor harmonics on the $S^3$, but given the class of spacetimes that we consider in this article, only one of them has a nonzero overlap with the Weyl scalars. 
The relevant vector harmonics for us are given by the following
\begin{itemize}
\item $(\ell_3,\ell_2)=(2,1)$
\begin{equation}
\begin{aligned}
& \mathbb{V}^{(2,1)}_{11} = \frac{4}{\sqrt{15} \pi }  \sin (\theta_3 )\cos (\theta_2)\,,\\
& \mathbb{V}^{(2,1)}_{12} =-\frac{1}{2\pi }\sqrt{\frac{3}{5}} \sin(2\theta_3 ) \sin(\theta_2 )\,,\\
& \mathbb{V}^{(2,1)}_{22} = -\frac{2}{\sqrt{15} \pi } \sin(\theta_3 )\cos (\theta_2 )\,,\\
& \mathbb{V}^{(2,1)}_{33} = -\frac{2}{\sqrt{15} \pi } \sin(\theta_3 )\cos (\theta_2 )\,.
\end{aligned}
\end{equation}
\item $(\ell_3,\ell_2)=(2,2)$
\begin{equation}
\begin{aligned}
& \mathbb{V}^{(2,2)}_{11} =  \frac{1}{\sqrt{15} \pi }[3 \cos (2 \theta_2 )+1] \cos (\theta_3 )\,,\\
& \mathbb{V}^{(2,2)}_{12} = -\frac{1}{4 \pi }\sqrt{\frac{3}{5}} \sin (2 \theta_2 ) [3 + \cos(2\theta_3)]\,,\\
& \mathbb{V}^{(2,2)}_{22} = -\frac{1}{\sqrt{15} \pi }[3 \cos (2 \theta_2 )-1]  \cos (\theta_3)\,,\\
& \mathbb{V}^{(2,2)}_{33} = -\frac{2}{\sqrt{15} \pi } \cos (\theta_3 )\,.
\end{aligned}
\end{equation}
\end{itemize}

\noindent
\paragraph{Transverse traceless tensor harmonics}
There are two families of transverse traceless tensor harmonics on the $S^3$, but, as with the vectors, only one of them has a nonzero overlap with the Weyl tensor. 
This is given by the following
\begin{itemize}
\item $(\ell_3,\ell_2)=(2,2)$
\begin{equation}
\begin{aligned}
& \mathbb{T}^{(2,2)}_{11} = -\frac{1}{2 \sqrt{6} \pi }[3 \cos (2 \theta_2 )+1]\,,\\
& \mathbb{T}^{(2,2)}_{12} = \frac{1}{2\pi }\sqrt{\frac{3}{2}} \sin (2\theta_2 ) \cos (\theta_3 ) \,,\\
& \mathbb{T}^{(2,2)}_{22} = \frac{1}{\sqrt{6} \pi }\Big[\cos ^2(\theta_2 )-\frac{1}{2} \sin ^2(\theta_2 ) (3 \cos (2 \theta_3 )+1)\Big]\,,\\
& \mathbb{T}^{(2,2)}_{33} = \frac{1}{4 \sqrt{6} \pi }\Big[6 \sin ^2(\theta_2 ) \cos (2 \theta_3) + 3 \cos(2\theta_2) + 1\Big]\,.
\end{aligned}
\end{equation}
\end{itemize}

\subsection{Adapted basis}
\label{sec:harmonicsS3P}

Given the symmetries of the spacetimes that we consider, it seems more natural to write the metric on the $S^3$ as
\begin{equation}
ds^2 = d\chi^2 + \sin^2\chi\,d\phi^2 + \cos^2\chi\, d\psi^2\,,
\label{eq:metricS3polar}
\end{equation}
with $\chi \in[0,\pi/2]$ and $\phi,\psi\in [0,2\pi]$. Indeed, in this form, the rotation axis in the spacetime coincides with the $\phi$ direction in \eqref{eq:metricS3polar}.  
Written in these coordinates, scalar harmonics on the $S^3$ are given by
\begin{equation}
\begin{aligned}
&Y^{\ell_3,m,\bar m} = C_{\ell_3}^{m_+,m_-} e^{\i(m\phi + \bar m \psi)} \\
&\times (1-\cos2\chi)^\frac{|m|}{2}  (1+\cos2\chi)^\frac{|\bar m|}{2} \\
&\times P_{\frac{\ell_3}{2} - m_+}^{(m,\bar m)}(\cos2\chi)\,,
\end{aligned}
\end{equation}
where $m=m_+ + m_-$, $\bar m = m_+ - m_-$,  $P_n^{(a,b)}(x)$ is the Jacobi polynomial and 
\begin{equation}
C_{\ell_3}^{m_+,m_-} = \frac{1}{2^{m_+}\pi}\sqrt{\frac{\ell_3 + 1}{2}} \sqrt{\frac{(\ell_3/2 + m_+)!(\ell_3 - m_+)!}{(\ell_3/2 + m_-)!(\ell_3 - m_-)!}}
\end{equation}
with $|m_\pm| \leq \frac{\ell_3}{2}$ and $\frac{\ell_3}{2}-m_\pm \in \mathbb N$.

Given the symmetries of spacetimes that we are considering, we are interested in harmonics with $\bar m=0$. We consider the obvious basis of vectors on the $S^3$ given the form of the metric in \eqref{eq:metricS3polar}, 
\begin{equation}
m_{(1)} = \frac{\partial}{\partial\chi}\,,~m_{(2)} = \frac{1}{\sin\chi}\frac{\partial}{\partial\phi}\,,~m_{(3)} = \frac{1}{\cos\chi}\frac{\partial}{\partial\psi}\,,
\label{eq:S3polarbasis}
\end{equation}
In this basis, the scalar-derived tensor harmonics on the $S^3$ with $\ell_3=2$ are given by the following
\begin{itemize}
\item $(\ell_3,m)=(2,0)$
\begin{equation}
\begin{aligned}
&\mathbb{S}^{(2,0)}_{11} = -\frac{1}{\sqrt{10} \pi }\cos (2 \chi )\,,\\
&\mathbb{S}^{(2,0)}_{22} = \frac{1}{2 \sqrt{10} \pi }[\cos (2 \chi )-3]\,,\\
&\mathbb{S}^{(2,0)}_{33} = \frac{1}{2 \sqrt{10} \pi }[\cos (2 \chi )+3]\,.
\end{aligned}
\end{equation}
\item $(\ell_3,m)=(2,2)$
\begin{equation}
\begin{aligned}
&\mathbb{S}^{(2,2)}_{11} = \frac{1}{2 \sqrt{10} \pi }e^{2 \i \phi } [\cos (2 \chi )+2]\,,\\
&\mathbb{S}^{(2,2)}_{12} = \frac{3 \i}{2 \sqrt{10} \pi } e^{2 \i \phi } \cos (\chi )\,,\\
&\mathbb{S}^{(2,2)}_{22} = -\frac{1}{4 \sqrt{10} \pi }e^{2 \i \phi } [\cos (2 \chi )+5]\,,\\
&\mathbb{S}^{(2,2)}_{33} = \frac{1}{2 \sqrt{10} \pi }e^{2 \i \phi } \sin ^2(\chi )\,.
\end{aligned}
\end{equation}
\end{itemize}
The $(\ell_3,m)=(2,-2)$ harmonics are given by the complex conjugate of the $(\ell_3,m)=(2,2)$ harmonics.

As discussed, for example, in Ref.~\cite{Gualtieri:2008ux}, for a fixed $\ell_n$ one can relate the tensor harmonics on given basis to those on a different basis by a simple linear transformation.\footnote{From representation theory, it follows that harmonics with different $\ell_n$ do not mix. Likewise, scalar-derived tensor harmonics in one basis will only mix with the scalar-derived tensor harmonics in the other basis.} 
For the case at hand, we can relate the scalar derived tensor harmonics in the basis \eqref{eq:S3polarbasis} to the harmonics in the other basis \eqref{eq:S3standardbasis} as follows
\begin{equation}
\mathbb{S}_{ab}^{(\ell_3,m)}(\chi,\phi) = R_a^{\phantom a c}R_b^{\phantom b d}\sum_{\ell_2=0}^{\ell_3} A^m_{\phantom m \ell_2} \mathbb{S}_{cd}^{(\ell_3,\ell_2)}(\theta_3,\theta_2)\,,
\end{equation}
where in this expression $(\chi,\phi)$ should be understood as functions of $(\theta_3,\theta_2)$ and $R_a^{\phantom a b}$ is a rotation matrix given by
\begin{equation}
R = \left(
\begin{array}{ccc}
-\frac{\sin (\theta _2) \cos (\theta_3)}{\sqrt{1-\sin ^2(\theta_2) \sin ^2(\theta_3)}} & -\frac{\cos (\theta_2)}{\sqrt{1-\sin ^2(\theta_2) \sin ^2(\theta_3)}} & 0  \\
\frac{\cos (\theta_2)}{\sqrt{1-\sin ^2(\theta_2) \sin ^2(\theta_3)}} & -\frac{\sin (\theta_2) \cos (\theta_3)}{\sqrt{1-\sin ^2(\theta_2) \sin ^2(\theta_3)}} & 0  \\
0 & 0 & 1 
\end{array}
\right)\,.
\end{equation}

For the $\ell_3 = 2$ scalar-derived tensor harmonics, we find that the transformation matrix $A^m_{\phantom m \ell_2}$ is given by:
\begin{equation}
A = \left(
\begin{array}{ccc}
\frac{1}{\sqrt{3}} & 0 & \sqrt{\frac{2}{3}}\\
-\frac{1}{\sqrt{3}} & \frac{\i}{\sqrt{2}} & \frac{1}{\sqrt{6}}\\
-\frac{1}{\sqrt{3}} & -\frac{\i}{\sqrt{2}} & \frac{1}{\sqrt{6}}
\end{array}
\right)\,,
\label{eq:matAS3}
\end{equation}
where the rows correspond to the values of $m=0,2,-2$, respectively, and similarly the columns correspond to the values of $\ell_2 =0,1,2$. 
Furthermore, as shown in Ref.~\cite{Gualtieri:2008ux} reviewed in Appendix~\ref{sec:WeylTransf}, the inverse of this matrix \eqref{eq:matAS3}, determines the transformation of the coefficients in the multipolar expansion of the Weyl scalars in different bases. 

\section{Tensor harmonics on the $S^4$}
\label{sec:AppS4}
In this Appendix we collect some properties of tensor harmonics in $S^4$. 
Furthermore, we list the harmonics that we have used to produce the waveforms in the main text. 

\subsection{Standard basis}

We first write the metric on the $S^4$ in the standard form:
\begin{equation}
ds^2 = d\theta_4^2  + \sin^2\theta_4 ds^2_{S^3}\,,
\label{eq:metricS4std}
\end{equation}
where $ds^2_{S^3}$ is the metric on the $S^3$ written as in \eqref{eq:metricS3std} and $\theta_4\in[0,\pi]$. 
Furthermore, we choose the obvious basis of unit vectors on the $S^4$:
\begin{equation}
\begin{aligned}
&m_{(1)} = \frac{\partial}{\partial\theta_4}\,,~m_{(2)} = \frac{1}{\sin\theta_4}\frac{\partial}{\partial\theta_3}\,,~ m_{(3)} = \frac{1}{\sin\theta_4\sin\theta_3}\frac{\partial}{\partial\theta_2}\,, \\ &m_{(4)} = \frac{1}{\sin\theta_4\sin\theta_3\sin\theta_2}\frac{\partial}{\partial\theta_1}\,.
\end{aligned}
\label{eq:basisS4std}
\end{equation}

The advantage of using this form of the metric on the $S^4$ is that we can easily construct tensor harmonics of any rank following the algorithm of Ref.~\cite{Higuchi:1986wu}. 
To do so, we start from scalar harmonics on the $S^4$. 
Since we are interested in spacetimes that possess an SO$(3)$ in six dimensions, we can restrict ourselves to scalar harmonics that are constant on the $S^2$ sitting inside the $S^3$, that in turn sits inside the $S^4$, \eqref{eq:metricS4std}. 
These (normalized) scalar harmonics are given by:
\begin{equation}
\mathbb Y^{(\ell_4,\ell_3)} = \frac{1}{\sqrt{4\pi}}c^{\ell_4,\ell_3}  P_{\ell_4+1}^{-(\ell_3+1)}(\cos\theta_4) P_{\ell_3+1/2}^{-1/2}(\cos\theta_3)\,,
\end{equation}
where $P_\ell^m(x)$ are the associated Legendre polynomials,
\begin{equation}
c^{\ell_4,\ell_3} = (\ell_3+1)\sqrt{\frac{2\ell_4+3}{2}\frac{(\ell_4+\ell_3+2)!}{(\ell_4-\ell_3)!}}\,,
\end{equation}
is the  normalization constant, and  $\ell_4\geq \ell_3\geq 0$. 
In this basis, the projected components of the tensor harmonics with $\ell_2=\ell_1=0$ are as follows\footnote{Note that for this particular class of harmonics, $\mathbb{Y}^{(\ell_4,\ell_3)}_{44} = \mathbb{Y}^{(\ell_4,\ell_3)}_{33}$, where here $\mathbb Y$ denotes any tensor harmonic.}

\noindent
\paragraph{Scalar-derived tensor harmonics}
\begin{itemize}
\item $(\ell_4,\ell_3) = (2,0)$:
\begin{equation}
\begin{aligned}
&\mathbb{S}^{(2,0)}_{11} = \frac{\sqrt{105}}{16 \pi }\sin ^2(\theta_4)\,,\\
&\mathbb{S}^{(2,0)}_{22} =-\frac{1}{16 \pi }\sqrt{\frac{35}{3}}\sin ^2(\theta_4)\,,\\
&\mathbb{S}^{(2,0)}_{33} =-\frac{1}{16 \pi }\sqrt{\frac{35}{3}}\sin ^2(\theta_4)\,.
\end{aligned}
\end{equation}
\item $(\ell_4,\ell_3) = (2,1)$:
\begin{equation}
\begin{aligned}
& \mathbb{S}^{(2,1)}_{11} = -\frac{1}{8 \pi }\sqrt{\frac{21}{2}} \cos (\theta_3) \sin (2 \theta_4)\,,\\
& \mathbb{S}^{(2,1)}_{12} = \frac{1}{2 \pi } \sqrt{\frac{7}{6}} \sin (\theta_3) \sin (\theta_4)\,,\\
& \mathbb{S}^{(2,1)}_{22} = \frac{1}{8 \pi }\sqrt{\frac{7}{6}}\cos (\theta_3) \sin (2\theta_4)\,,\\
& \mathbb{S}^{(2,1)}_{33} = \frac{1}{8 \pi }\sqrt{\frac{7}{6}}\cos (\theta_3) \sin (2\theta_4) \,.
\end{aligned}
\end{equation}
\item $(\ell_4,\ell_3) = (2,2)$:
\begin{equation}
\begin{aligned}
& \mathbb{S}^{(2,2)}_{11} = \frac{\sqrt{7}}{96 \pi } [2 \cos (2 \theta_3)+1] [3 \cos (2 \theta_4)+5]\,,\\
& \mathbb{S}^{(2,2)}_{12} =  -\frac{\sqrt{7}}{6 \pi } \sin (2\theta_3) \cos (\theta_4)\,,\\
& \mathbb{S}^{(2,2)}_{22} = -\frac{\sqrt{7}}{96 \pi }[2 \cos (2\theta_3) (\cos (2 \theta_4)+7)+\cos (2 \theta_4)-9]\,,\\
& \mathbb{S}^{(2,2)}_{33} = -\frac{\sqrt{7}}{96 \pi } \left[-4 \cos (2 \theta_3) \sin ^2(\theta_4)+\cos (2 \theta_4)+7\right]\,.
\end{aligned}
\end{equation}
\item $(\ell_4,\ell_3) = (4,0)$:
\begin{equation}
\begin{aligned}
& \mathbb{S}^{(4,0)}_{11} = \frac{\sqrt{1155}}{128 \pi } \sin ^2(\theta_4) [9 \cos (2 \theta_4)+7]\,,\\
& \mathbb{S}^{(4,0)}_{12} = -\frac{1}{128 \pi } \sqrt{\frac{385}{3}} \sin ^2(\theta_4) [9 \cos (2 \theta_4)+7]\,,\\
& \mathbb{S}^{(4,0)}_{22} = -\frac{1}{128 \pi } \sqrt{\frac{385}{3}} \sin ^2(\theta_4) [9 \cos (2 \theta_4)+7]\,,\\
& \mathbb{S}^{(4,0)}_{33} = -\frac{1}{128 \pi } \sqrt{\frac{385}{3}} \sin ^2(\theta_4) [9 \cos (2 \theta_4)+7]\,.
\end{aligned}
\end{equation}
\item $(\ell_4,\ell_3) = (4,1)$:
\begin{equation}
\begin{aligned}
& \mathbb{S}^{(4,1)}_{11} = \frac{\sqrt{165}}{128 \pi } \cos (\theta_3) [2 \sin (2 \theta_4)-9 \sin (4 \theta_4)] \,,\\
& \mathbb{S}^{(4,1)}_{12} = \frac{1}{32 \pi } \sqrt{\frac{55}{3}} \sin (\theta_3) \sin (\theta_4) [9 \cos (2 \theta_4)+7]\,,\\
& \mathbb{S}^{(4,1)}_{22} = \frac{1}{64 \pi } \sqrt{\frac{55}{3}} \cos (\theta_3) \sin (\theta_4) [7 \cos (\theta_4)+9 \cos (3 \theta_4)]\,,\\
& \mathbb{S}^{(4,1)}_{33} = \frac{1}{64 \pi } \sqrt{\frac{55}{3}} \cos (\theta_3) \sin (\theta_4) [7 \cos (\theta_4)+9 \cos (3 \theta_4)] \,.
\end{aligned}
\end{equation}
\item $(\ell_4,\ell_3) = (4,2)$:
\begin{equation}
\begin{aligned}
& \mathbb{S}^{(4,2)}_{11} = \frac{1}{1536 \pi } \sqrt{\frac{55}{2}} [2 \cos (2 \theta_3)+1] \\
&\hspace{1cm}\times[12 \cos (2 \theta_4)+81 \cos (4 \theta_4)+35]\,,\\
& \mathbb{S}^{(4,2)}_{12} = -\frac{1}{192 \pi } \sqrt{\frac{55}{2}} \sin (2 \theta_3) [5 \cos (\theta_4)+27 \cos (3 \theta_4)]\,,\\
& \mathbb{S}^{(4,2)}_{22} = \frac{1}{1536 \pi } \sqrt{\frac{55}{2}}\big\{92 \cos (2 \theta_4)-27 \cos (4 \theta_4)+63 \\
&\hspace{1cm}-2 \cos (2 \theta_3) [52 \cos (2 \theta_4)+27 \cos (4 \theta_4)+49]\big\}\,,\\
& \mathbb{S}^{(4,2)}_{33} = - \frac{\sqrt{\frac{55}{2}}}{1536 \pi }  \{52 \cos (2 \theta_4)+27 \cos (4 \theta_4)+49 \\
&\hspace{1cm} - 8 \cos (2 \theta_3) \sin ^2(\theta_4) [27 \cos (2 \theta_4)+17]\}\,.
\end{aligned}
\end{equation}
\item $(\ell_4,\ell_3) = (4,3)$:
\begin{equation}
\begin{aligned}
& \mathbb{S}^{(4,3)}_{11} = \frac{3 \sqrt{55}}{256 \pi }[\cos (\theta_3)+\cos (3 \theta_3)] [2 \sin (2 \theta_4)+3 \sin (4 \theta_4)]\,,\\
& \mathbb{S}^{(4,3)}_{12} = -\frac{\sqrt{55}}{64 \pi } [\sin (\theta_3)+3 \sin (3 \theta_3)] \sin (\theta_4) [3 \cos (2 \theta_4)+1]\,,\\
& \mathbb{S}^{(4,3)}_{22} = \frac{\sqrt{55}}{64\pi } \cos (\theta_3) \sin (2\theta_4) \big\{8-3 \cos (2 \theta_3) [\cos (2 \theta_4)+3]\big\}\,,\\
& \mathbb{S}^{(4,3)}_{33} = \frac{\sqrt{55}}{32 \pi } \cos (\theta_3) \sin (2\theta_4) \big[3 \cos (2 \theta_3) \sin ^2(\theta_4)-2\big]\,.
\end{aligned}
\end{equation}
\item $(\ell_4,\ell_3) = (4,4)$:
\begin{equation}
\begin{aligned}
& \mathbb{S}^{(4,4)}_{11} = \frac{3}{128 \pi } \sqrt{\frac{11}{2}} \sin ^2(\theta_4) [3 \cos (2 \theta_4)+5] \\
&\hspace{1cm} \times [2 \cos (2 \theta_3)+2 \cos (4 \theta_3)+1]\,,\\
& \mathbb{S}^{(4,4)}_{12} = -\frac{3 \sqrt{\frac{11}{2}} }{16 \pi } \sin ^2(\theta_4) \cos (\theta_4) [\sin (2 \theta_3)+2 \sin (4 \theta_3)] \,,\\
& \mathbb{S}^{(4,4)}_{22} = \frac{1}{64 \pi } \sqrt{\frac{11}{2}} \sin ^2(\theta_4) \big\{-3 \cos (4 \theta_3) [\cos (2 \theta_4)+7] \\
&\hspace{1cm}+[6 \cos (2 \theta_3)+7] \sin ^2(\theta_4)+4 \cos ^2(\theta_4)\big\}\,,\\
& \mathbb{S}^{(4,4)}_{33} = -\frac{1}{128 \pi } \sqrt{\frac{11}{2}} \sin ^2(\theta_4) \big\{6 \cos (2 \theta_3) [\cos (2\theta_4)+3] \\
&\hspace{1cm}-12 \cos (4\theta_3) \sin ^2(\theta_4)+3 \cos (2 \theta_4)+13\big\}\,.
\end{aligned}
\end{equation}
\end{itemize}

\noindent
\paragraph{Vector-derived tensor harmonics}
There are three families of vector-derived tensor harmonics on the $S^4$, but with our symmetry assumptions, only one of them has a nonzero overlap with the Weyl scalars. 
The relevant vector harmonics for us are given by the following
\begin{itemize}
\item $(\ell_4,\ell_3)=(2,1)$
\begin{equation}
\begin{aligned}
& \mathbb{V}^{(2,1)}_{11} = -\frac{1}{8 \pi } \sqrt{\frac{35}{2}} \sin (2\theta_4)\,,\\
& \mathbb{V}^{(2,1)}_{12} = \frac{1}{12 \pi } \sqrt{\frac{35}{2}}  \sin (\theta_3) \sin (2\theta_4)\,,\\
& \mathbb{V}^{(2,1)}_{22} = \frac{1}{24 \pi } \sqrt{\frac{35}{2}} \sin (2\theta_4)\,,\\
& \mathbb{V}^{(2,1)}_{33} = \frac{1}{24 \pi } \sqrt{\frac{35}{2}} \sin (2\theta_4)\,.
\end{aligned}
\end{equation}
\item $(\ell_4,\ell_3)=(2,2)$
\begin{equation}
\begin{aligned}
& \mathbb{V}^{(2,2)}_{11} =  \frac{1}{12 \pi } \sqrt{\frac{35}{2}} [2 \cos (2 \theta_3)+1] \cos (\theta_4)\,,\\
& \mathbb{V}^{(2,2)}_{12} = -\frac{1}{24 \pi } \sqrt{\frac{35}{2}} \sin (2\theta _3) [\cos (2 \theta_4)+3]\,,\\
& \mathbb{V}^{(2,2)}_{22} = \frac{1}{12 \pi } \sqrt{\frac{35}{2}} [1-2 \cos (2 \theta_3)] \cos (\theta_4)\,,\\
& \mathbb{V}^{(2,2)}_{33} = -\frac{1}{12 \pi } \sqrt{\frac{35}{2}} \cos (\theta_4)\,.
\end{aligned}
\end{equation}
\item $(\ell_4,\ell_3)=(4,1)$
\begin{equation}
\begin{aligned}
& \mathbb{V}^{(4,1)}_{11} =  -\frac{\sqrt{77}}{32 \pi }\cos (\theta_3) \sin (\theta_4) [9 \cos (2 \theta_4)+7]\,,\\
& \mathbb{V}^{(4,1)}_{12} = \frac{\sqrt{77}}{192 \pi } \sin (\theta_3) \sin (\theta_4) [37 \cos (\theta_4)+27 \cos (3 \theta_4)] \,,\\
& \mathbb{V}^{(4,1)}_{22} = \frac{\sqrt{77}}{96 \pi }  \cos (\theta_3) \sin (\theta_4) [9 \cos (2 \theta_4)+7] \,,\\
& \mathbb{V}^{(4,1)}_{33} = \frac{\sqrt{77}}{96 \pi } \cos (\theta_3) \sin (\theta_4) [9 \cos (2 \theta_4)+7]\,.
\end{aligned}
\end{equation}
\item $(\ell_4,\ell_3)=(4,2)$
\begin{equation}
\begin{aligned}
& \mathbb{V}^{(4,2)}_{11} = \frac{\sqrt{77}}{384 \pi } [2 \cos (2 \theta_3)+1] [5 \cos (\theta_4)+27 \cos (3 \theta_4)] \,,\\
& \mathbb{V}^{(4,2)}_{12} =  -\frac{\sqrt{77}}{1536 \pi } \sin (2 \theta_3) [100 \cos (2 \theta_4)+81 \cos (4 \theta_4)+75] \,,\\
& \mathbb{V}^{(4,2)}_{22} = \frac{\sqrt{77}}{384 \pi } \big\{[23-28 \cos (2 \theta_3)] \cos (\theta_4)\\
&\hspace{1cm}+9 [1-4 \cos (2 \theta_3)] \cos (3 \theta_4)\big\}\,,\\
& \mathbb{V}^{(4,2)}_{33} = \frac{\sqrt{77}}{192 \pi }  \big\{\cos (\theta_4) [18 \cos (2 \theta_3) \sin ^2(\theta_4)-7] \\
&\hspace{1cm}-9 \cos (3 \theta_4)\big\}\,.
\end{aligned}
\end{equation}
\item $(\ell_4,\ell_3)=(4,3)$
\begin{equation}
\begin{aligned}
& \mathbb{V}^{(4,3)}_{11} = \frac{\sqrt{1155} }{64 \pi }[\cos (\theta_3)+\cos (3 \theta_3)] \sin (\theta_4) [3 \cos (2 \theta_4)+1]\,,\\
& \mathbb{V}^{(4,3)}_{12} = -\frac{1}{64 \pi } \sqrt{\frac{77}{15}} \sin (\theta_3) [3 \cos (2 \theta_3)+2]  \\
&\hspace{1cm}\times \sin (\theta_4) [31 \cos (\theta_4)+9 \cos (3 \theta_4)] \,,\\
& \mathbb{V}^{(4,3)}_{22} = \frac{1}{128 \pi } \sqrt{\frac{77}{15}} \big\{\cos (\theta_3) [13 \sin (\theta_4)+9 \sin (3 \theta_4)] \\
&\hspace{1cm}-3 \cos (3 \theta_3) [\sin (\theta_4)+13 \sin (3 \theta_4)]\big\} \,,\\
& \mathbb{V}^{(4,3)}_{33} = \frac{1}{64 \pi } \sqrt{\frac{77}{15}} \cos (\theta_3) \big\{[9 \cos (2 \theta_3)-4] \sin (\theta_4) \\
&\hspace{1cm}-3 [\cos (2 \theta_3)+4] \sin (3 \theta_4)\big\} \,.
\end{aligned}
\end{equation}
\item $(\ell_4,\ell_3)=(4,4)$
\begin{equation}
\begin{aligned}
& \mathbb{V}^{(4,4)}_{11} = \frac{3}{32 \pi } \sqrt{\frac{231}{5}}  \sin ^2(\theta_4) \cos (\theta_4) \\
&\hspace{1cm} \times [2 \cos (2 \theta_3)+2 \cos (4 \theta_3)+1]\,,\\
& \mathbb{V}^{(4,4)}_{12} = -\frac{3}{64 \pi } \sqrt{\frac{231}{5}}\sin (\theta_3) \sin ^2(\theta_4) [\cos (2 \theta_4)+3] \\
&\hspace{1cm} \times [3 \cos (\theta_3)+2 \cos (3 \theta_3)]\,,\\
& \mathbb{V}^{(4,4)}_{22} =  \frac{1}{32 \pi } \sqrt{\frac{231}{5}} [1-6 \cos (4 \theta_3)] \sin ^2(\theta_4) \cos (\theta_4)\,,\\
& \mathbb{V}^{(4,4)}_{33} =  -\frac{1}{32 \pi } \sqrt{\frac{231}{5}} [3 \cos (2 \theta_3)+2] \sin ^2(\theta_4) \cos (\theta_4) \,.
\end{aligned}
\end{equation}
\end{itemize}

\noindent
\paragraph{Transverse traceless tensor harmonics}
Given the symmetries of the spacetimes that we are considering, there is only one family of transverse traceless tensor harmonics on the $S^4$ that can contribute to the multipolar expansion of the Weyl scalars. 
The relevant tensor harmonics are given by: 
\begin{itemize}
\item $(\ell_4,\ell_3)=(2,2)$
\begin{equation}
\begin{aligned}
& \mathbb{T}^{(2,2)}_{11} = \frac{1}{12 \pi } \sqrt{\frac{35}{2}} [2 \cos (2 \theta_3)+1]\,,\\
& \mathbb{T}^{(2,2)}_{12} = -\frac{1}{6 \pi } \sqrt{\frac{35}{2}} \sin (2 \theta_3) \cos (\theta_4)\,,\\
& \mathbb{T}^{(2,2)}_{22} =  \frac{1}{12 \pi } \sqrt{\frac{35}{2}} [4 \sin ^2(\theta_3) \cos ^2(\theta_4)-1]\,,\\
& \mathbb{T}^{(2,2)}_{33} =  -\frac{1}{12 \pi } \sqrt{\frac{35}{2}} [\sin ^2(\theta_3) \cos (2 \theta_4)+\cos ^2(\theta_3)] \,.
\end{aligned}
\end{equation}
\item $(\ell_4,\ell_3)=(4,2)$
\begin{equation}
\begin{aligned}
& \mathbb{T}^{(4,2)}_{11} = \frac{1}{96 \pi } \sqrt{\frac{55}{2}} [2 \cos (2 \theta_3)+1] [9 \cos (2 \theta_4)+7]\,,\\
& \mathbb{T}^{(4,2)}_{12} = -\frac{1}{96 \pi } \sqrt{\frac{55}{2}} \sin (2 \theta_3) \cos (\theta_4) [27 \cos (2 \theta_4)+5] \,,\\
& \mathbb{T}^{(4,2)}_{22} =  \frac{1}{384 \pi } \sqrt{\frac{11}{10}} \big\{4 [26-71 \cos (2 \theta_3)] \cos (2 \theta_4) \\
&\hspace{1cm}+378 \sin ^2(\theta_3) \cos (4 \theta_4)-167 \cos (2 \theta_3)+27\big\}) \,,\\
& \mathbb{T}^{(4,2)}_{33} = -\frac{1}{768 \pi } \sqrt{\frac{11}{10}}  \big\{\cos (2 \theta_3) [76 \cos (2 \theta_4)+113] \\
&\hspace{1cm} + 378 \sin ^2(\theta_3) \cos (4 \theta_4)+284 \cos (2\theta_4)+167\big\} \,.
\end{aligned}
\end{equation}
\item $(\ell_4,\ell_3)=(4,3)$
\begin{equation}
\begin{aligned}
& \mathbb{T}^{(4,3)}_{11} = \frac{3}{16 \pi } \sqrt{\frac{55}{2}} [\cos (\theta_3)+\cos (3 \theta_3)] \sin (2 \theta_4) \,,\\
& \mathbb{T}^{(4,3)}_{12} = -\frac{1}{4 \pi } \sqrt{\frac{11}{10}} \sin (\theta_3) \sin (\theta_4) \\
&\hspace{1cm} \times  [3 \cos (2 \theta_3)+2] [3 \cos (2 \theta_4)+2] \,,\\
& \mathbb{T}^{(4,3)}_{22} =  \frac{1}{16 \pi } \sqrt{\frac{11}{10}}  \cos (\theta_3) \\
&\hspace{1cm} \times \big\{ 21 \sin ^2(\theta_3) \sin (4 \theta_4)-[9 \cos (2 \theta_3)+1] \sin (2 \theta_4)\big\} \,,\\
& \mathbb{T}^{(4,3)}_{33} = \frac{1}{32 \pi } \sqrt{\frac{11}{10}} \cos (\theta_3) \\ 
&\hspace{1cm} \times \big\{[1-21 \cos (2 \theta_3)] \sin (2 \theta_4)-21 \sin ^2(\theta_3) \sin (4 \theta_4)\big\} \,.
\end{aligned}
\end{equation}
\item $(\ell_4,\ell_3)=(4,4)$
\begin{equation}
\begin{aligned}
& \mathbb{T}^{(4,4)}_{11} = \frac{3}{16 \pi } \sqrt{\frac{77}{10}} \sin ^2(\theta_4) [2 \cos (2 \theta_3)+2 \cos (4 \theta_3)+1] \,,\\
& \mathbb{T}^{(4,4)}_{12} =  -\frac{3}{8 \pi } \sqrt{\frac{77}{10}} \sin (\theta_3) \sin ^2(\theta_4) \cos (\theta_4) \\ 
&\hspace{1cm} \times [3 \cos (\theta_3)+2 \cos (3 \theta_3)] \,,\\
& \mathbb{T}^{(4,4)}_{22} =  -\frac{1}{16 \pi } \sqrt{\frac{77}{10}} \sin ^2(\theta_4) \\
&\hspace{1cm} \times \big\{-3 \cos ^2(\theta_3) [9 \cos (2 \theta_4)+7]  \\
&\hspace{1cm}  + 48 \cos ^4(\theta_3) \cos ^2(\theta_4) + 3 \cos (2 \theta_4)+2\big\} \,,\\
& \mathbb{T}^{(4,4)}_{33} =  \frac{1}{256 \pi } \sqrt{\frac{77}{10}} \big\{ 4 [3 \cos (2 \theta_3)+6 \cos (4 \theta_3)+1] \cos (2 \theta_4) \\
&\hspace{1cm}+6 \sin ^2(\theta_3) [4 \cos (2 \theta_3)+3] \cos (4 \theta_4) \\
&\hspace{1cm}-15 \cos (2 \theta_3)-18 \cos (4\theta_3)-7\big\} \,.
\end{aligned}
\end{equation}
\end{itemize}

\subsection{Multipolar expansion of the Weyl scalars in the standard basis}
\label{sec:multipolar_std}
The main advantage of using the standard form of the metric on the $S^4$ \eqref{eq:metricS4std} and the associated basis of vectors to obtain the multipolar expansion of the Weyl tensor, is that one can systematically construct the required tensor harmonics of any rank. 
In this way, we can identify the sector of tensor harmonics that captures most of the signal.  

\begin{figure}[t!]
\begin{center}
\includegraphics[width=7.5cm]{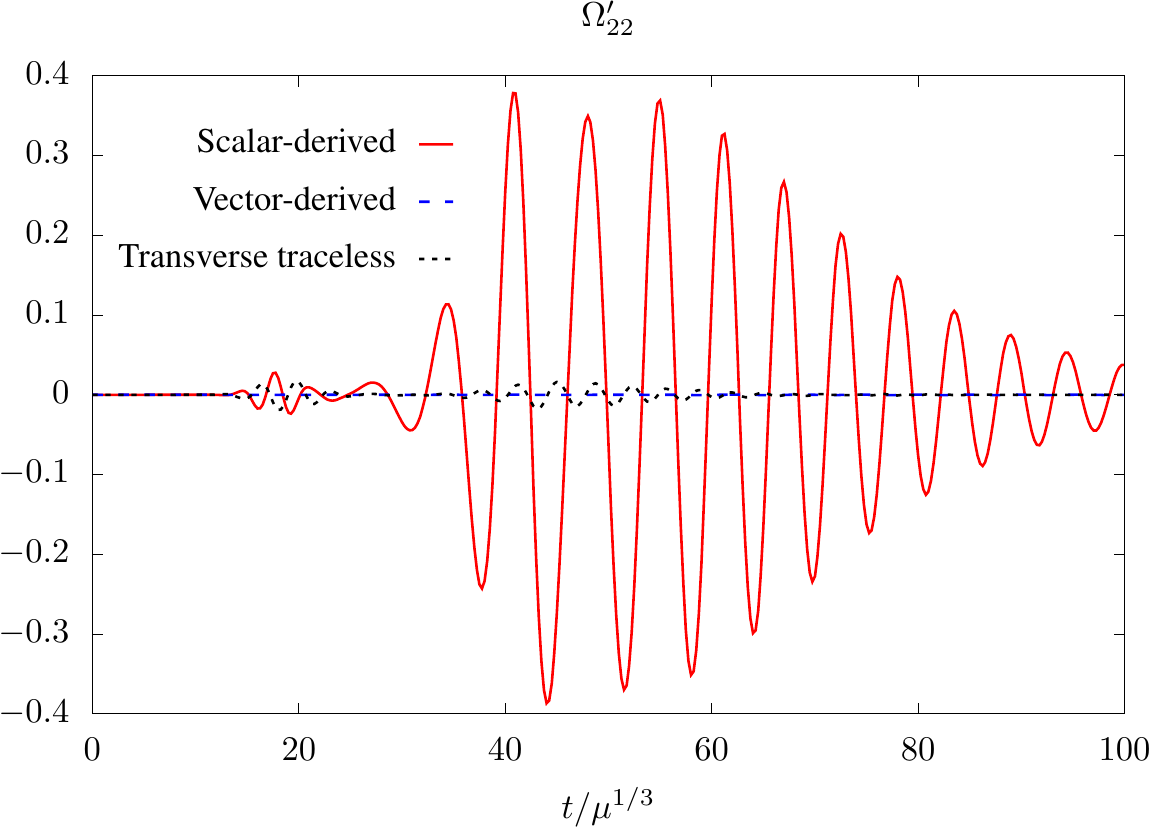}
\end{center}
\caption{Scalar derived, vector derived and transverse traceless tensor projections with $(\ell_4,\ell_3)=(2,2)$ of the Weyl scalars for the same $a/\mu^{1/3} = 1.3$ run as in the main text. This figure illustrates that most of the signal is in the scalar derived sector. The same happens in any of the other simulations that we have performed. }
\label{fig:StableEvol_GW_SVT}
\end{figure}

In Fig.~\ref{fig:StableEvol_GW_SVT} we display the $(\ell_4,\ell_3)=(2,2)$ multipole of the Weyl scalars in the scalar derived, vector derived and transverse traceless tensor harmonics sectors for the same $a/\mu^{1/3} = 1.3$ simulation reported in the main text, Fig. \ref{fig:StableEvol_GW}. 
As this figure illustrates, most the signal is in the scalar derived sector. 
We have checked that this is the case in all our simulations and this is why in the main text we only report on the multipoles from the scalar derived tensor harmonics. 
Perhaps it should not be surprising that, given the symmetries of the spacetimes that we are considering, most the waveforms are captured by the scalar derived tensor harmonics.

\begin{figure}[t!]
\begin{center}
\includegraphics[width=7.5cm]{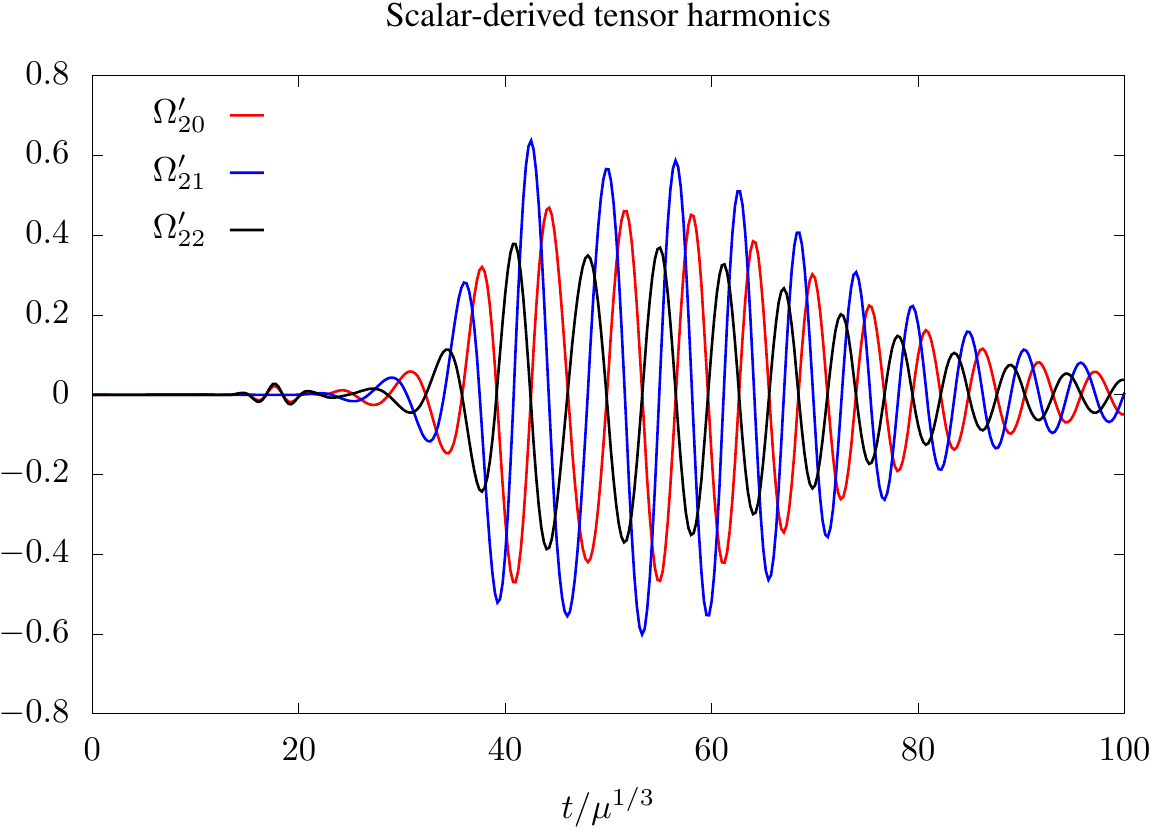}
\end{center}
\caption{Scalar derived tensor multipoles with $(\ell_4,\ell_3)=(2,0)$, $(2,1)$ and $(2,2)$ of the Weyl scalars for the same $a/\mu^{1/3} = 1.3$ run as in the main text. This figure shows the mixing of modes with the same $\ell_4$ and different $\ell_3$ due to the misalignment between the chosen basis of angular vectors on the $S^4$ and the rotation axis of the black hole. }
\label{fig:StableEvol_GW_Mixing}
\end{figure}

The drawback of using the basis \eqref{eq:metricS4std} is that it is not aligned with the axis of rotation of the black hole spacetimes that we have considered. 
This has the consequence that the various harmonics with the same $\ell_4$ but different $\ell_3$ mix and one cannot accurately extract the frequencies and growth/decay rates of the various modes.\footnote{This phenomenon was observed in \cite{Gualtieri:2008ux} in 4$D$, for instance, when extracting gravitational waves from head on collisions of black holes and considering an extraction frame that is not suitably aligned with the collision axis. We would like to thank Ulrich Sperhake for pointing this out to us.} 
This mixing of modes with the same $\ell_3$  and different $\ell_2$'s can be seen in Fig. \ref{fig:StableEvol_GW_Mixing}. 
In this plot we display the $(\ell_4,\ell_3)=(2,0)$, $(2,1)$ and $(2,2)$ multipoles of the Weyl scalars for the same $a/\mu^{1/3} = 1.3$ run as in the main text. 
As this plot suggests, the various modes appear to have similar frequencies and decay rates. 
To disentangle the various modes one would have to rotate the basis or consider an adapted basis. 
We chose the latter option.

\subsection{Adapted basis}
As in the 5$D$ case, it turns out to be convenient to write the metric on the $S^4$ in a form that makes the symmetries and the rotation plane of spacetimes that we consider manifest:
\begin{equation}
ds^2 = d\chi^2 + \sin^2\chi\,d\phi^2 + \cos^2\chi\,(d\theta^2 + \sin^2\theta\,d\psi^2)\,,
\label{eq:metricS4P}
\end{equation}
with $\chi\in [0,\pi/2]$, $\theta\in [0,\pi]$ and $\phi,\psi\in [0,2\pi]$. 
In these coordinates, the angle $\phi$ on the extraction $S^4$ coincides with the angle on the rotation plane of the full spacetime. 
We choose the obvious basis of angular vectors on the $S^4$ written as in \eqref{eq:metricS4P}:
\begin{equation}
\begin{aligned}
&m_{(1)} = \frac{\partial}{\partial\chi}\,,~m_{(2)} = \frac{1}{\sin\chi} \frac{\partial}{\partial\phi}\,,~m_{(3)} = \frac{1}{\cos\chi}\frac{\partial}{\partial\theta}\,, \\
&m_{(4)} = \frac{1}{\cos\chi \sin\theta}\frac{\partial}{\partial\psi}\,.
\end{aligned}
\label{eq:basisS4P}
\end{equation} 

In the previous subsection \S\ref{sec:multipolar_std}, we have shown that the leading waveforms are in the sector of scalar derived tensor harmonics. 
Therefore, to accurately extract the leading modes that govern the dynamics of the black holes that we are interested in, we only need to concentrate on this sector of tensor harmonics.  
Scalar harmonics on the $S^4$ in the coordinates \eqref{eq:metricS4P} can written as \cite{Berti:2005gp}
\begin{equation}
\begin{aligned}
&\mathbb{Y}^{\ell_4,m,\ell_2,\bar m} = N (\sin\chi)^{|m|}(\cos\chi)^{\ell_2} \, e^{\i m \phi}\, \mathbb{Y}^{\bar m}_{\ell_2}(\theta,\psi) \\
&\hspace{1cm}_{2}F_1\big(\ell_2+|m|-k, \textstyle{k+\frac{3}{2}},\ell_2+\frac{3}{2}; \cos^2\chi\big)
\end{aligned}
\label{eq:scalarYS4P}
\end{equation}
where $N$ is a normalization constant, $\mathbb{Y}^{\bar m}_{\ell_2}(\theta,\psi)$ are the standard spherical harmonics on the $S^2$, $_{2}F_1$ is the ordinary hypergeometric function and $k$ is a positive integer related to the eigenvalue $\ell_4$ by
\begin{equation} 
\ell_4 = 2 k - (\ell_2 + |m|)\,.
\end{equation} 
Since we are considering spacetimes with a manifest SO$(3)$ symmetry, we only need to consider harmonics with $\ell_2=0$ (and consequently $\bar m =0)$ and hence we will drop the corresponding labels from now on.

Writing the $S^4$ as in \eqref{eq:metricS4P} and in the basis \eqref{eq:basisS4P}, the projected scalar derived tensor harmonics obtained from \eqref{eq:scalarYS4P} are:
\begin{itemize}
\item $(\ell_4,m) = (2,0)$:
\begin{equation}
\begin{aligned}
&\mathbb{S}^{(2,0)}_{11} = \frac{1}{48 \pi } \sqrt{\frac{35}{2}} [3 \cos (2 \chi )+1]\,,\\
&\mathbb{S}^{(2,0)}_{22} = -\frac{1}{48 \pi } \sqrt{\frac{35}{2}} [\cos (2 \chi )-5]\,,\\
&\mathbb{S}^{(2,0)}_{33} = -\frac{1}{48 \pi } \sqrt{\frac{35}{2}} [\cos (2 \chi )+3]\,.
\end{aligned}
\end{equation}
\item $(\ell_4,m) = (2,2)$:
\begin{equation}
\begin{aligned}
&\mathbb{S}^{(2,2)}_{11} = \frac{1}{32 \pi }  \sqrt{\frac{7}{3}} \, e^{2 \i \phi } [3 \cos (2 \chi )+5]\,,\\
&\mathbb{S}^{(2,2)}_{21} = \frac{\i }{4 \pi } \sqrt{\frac{7}{3}} \, e^{2 \i \phi } \cos (\chi )\,,\\
&\mathbb{S}^{(2,2)}_{22} = -\frac{1}{32 \pi } \sqrt{\frac{7}{3}} \, e^{2 \i \phi } [\cos (2 \chi )+7]\,,\\
&\mathbb{S}^{(2,2)}_{33} = \frac{1}{16 \pi } \sqrt{\frac{7}{3}} \, e^{2 \i \phi } \sin ^2(\chi )\,,
\end{aligned}
\end{equation}
\item $(\ell_4,m) = (4,0)$:
\begin{equation}
\begin{aligned}
&\mathbb{S}^{(4,0)}_{11} = -\frac{\sqrt{77}}{1536 \pi }  [12 \cos (2 \chi )+81 \cos (4 \chi )-29]\,,\\
&\mathbb{S}^{(4,0)}_{22} = \frac{\sqrt{77}}{1536 \pi } [-92 \cos (2 \chi )+27 \cos (4 \chi )+1]\,,\\
&\mathbb{S}^{(4,0)}_{33} = \frac{\sqrt{77}}{1536 \pi } [52 \cos (2 \chi )+27 \cos (4 \chi )-15] \,.
\end{aligned}
\end{equation}
\item $(\ell_4,m) = (4,2)$:
\begin{equation}
\begin{aligned}
&\mathbb{S}^{(4,2)}_{11} = -\frac{1}{256 \pi } \sqrt{\frac{55}{6}} \,e^{2 \i \phi } [12 \cos (2 \chi )+27 \cos (4 \chi )-7] \,,\\
&\mathbb{S}^{(4,2)}_{21} = \frac{\i }{64 \pi } \sqrt{\frac{55}{6}} \,e^{2 \i \phi } [\cos (\chi )-9 \cos (3 \chi )] \,,\\
&\mathbb{S}^{(4,2)}_{22} = \frac{1}{256 \pi } \sqrt{\frac{55}{6}} \,e^{2 \i \phi } [4 \cos (2 \chi )+9 \cos (4 \chi )+19]\,,\\
&\mathbb{S}^{(4,2)}_{33} = -\frac{1}{64 \pi } \sqrt{\frac{55}{6}} e^{2 \i \phi } \sin ^2(\chi ) [9 \cos (2 \chi )+11] \,,
\end{aligned}
\end{equation}
\item $(\ell_4,m) = (4,4)$:
\begin{equation}
\begin{aligned}
&\mathbb{S}^{(4,4)}_{11} =  \frac{3}{128 \pi }\sqrt{\frac{55}{2}}\, e^{4 \i \phi } \sin ^2(\chi ) [3 \cos (2 \chi )+5]\,,\\
&\mathbb{S}^{(4,4)}_{21} = \frac{3 \i }{16 \pi }\sqrt{\frac{55}{2}} \,e^{4 \i \phi } \sin ^2(\chi ) \cos (\chi )\,,\\
&\mathbb{S}^{(4,4)}_{22} = -\frac{3}{128 \pi } \sqrt{\frac{55}{2}} \,e^{4 \i \phi } \sin ^2(\chi ) [\cos (2 \chi )+7] \,,\\
&\mathbb{S}^{(4,4)}_{33} = \frac{3}{64 \pi } \sqrt{\frac{55}{2}} \,e^{4 \i \phi } \sin ^4(\chi )\,.
\end{aligned}
\end{equation}
\end{itemize}
The $(\ell_4,m)=(2,-2)$, $(4,-2)$, $(4,-4)$ harmonics are given by the complex conjugate of the $(2,2)$, $(4,2)$, $(4,4)$ respectively. 

As we have already seen in Appendix~\ref{sec:harmonicsS3P}, for a fixed $\ell_4$ we can relate the harmonics in this basis \eqref{eq:basisS4P} to those in the standard basis \eqref{eq:basisS4std} by a linear transformation,
\begin{equation}
\mathbb{S}_{ab}^{(\ell_4,m)}(\chi,\phi) = R_a^{\phantom a c}R_b^{\phantom b d}\sum_{\ell_3=0}^{\ell_4} A^m_{\phantom m \ell_3} \mathbb{S}_{cd}^{(\ell_4,\ell_3)}(\theta_4,\theta_3)\,,
\end{equation}
where in the lhs of this expression $(\chi,\phi)$ should be understood as functions of $(\theta_4,\theta_3)$, and $R_a^{\phantom a b}$ is a rotation matrix given by
\begin{equation}
R = \left(
\begin{array}{cccc}
-\frac{\sin (\theta _3) \cos (\theta_4)}{\sqrt{1-\sin ^2(\theta_3) \sin ^2(\theta_4)}} & -\frac{\cos (\theta_3)}{\sqrt{1-\sin ^2(\theta_3) \sin ^2(\theta_4)}} & 0 & 0  \\
\frac{\cos (\theta_3)}{\sqrt{1-\sin ^2(\theta_3) \sin ^2(\theta_4)}} & -\frac{\sin (\theta_3) \cos (\theta_4)}{\sqrt{1-\sin ^2(\theta_3) \sin ^2(\theta_4)}} & 0  & 0 \\
0 & 0 & 1 & 0 \\
0 & 0 & 0 & 1
\end{array}
\right)\,.
\end{equation}

For the $\ell_4=2$ scalar-derived tensor harmonics, the transformation matrix is given by
\begin{equation}
A^m_{\phantom m \ell_3} = 
\left(
\begin{array}{ccc}
 \frac{\sqrt{3}}{2\sqrt{2}} & 0 & \frac{\sqrt{5}}{2\sqrt{2}} \\
 \frac{\sqrt{5}}{4} & \frac{\i}{\sqrt{2}} & -\frac{\sqrt{3}}{4} \\
 \frac{\sqrt{5}}{4} & -\frac{\i}{\sqrt{2}} & -\frac{\sqrt{3}}{4} \\
\end{array}
\right)\,,
\end{equation}
where the rows correspond to $m=0,2,-2$ and the columns correspond to $\ell_3=0,1,2$. Similarly, for the $\ell_4=4$ harmonics, we find
\begin{equation}
A^m_{\phantom m \ell_3} = 
\left(
\begin{array}{ccccc}
 \frac{\sqrt{15}}{8} & 0 & \frac{\sqrt{35}}{8\sqrt{2}} & 0 & \frac{3}{8} \sqrt{\frac{7}{2}}\\
 \frac{\sqrt{7}}{4\sqrt{2}} & \frac{\i}{2 \sqrt{2}} & \frac{\sqrt{3}}{8} & \frac{\i}{2} \sqrt{\frac{3}{2}} & -\frac{\sqrt{15}}{8} \\
 \frac{\sqrt{7}}{4\sqrt{2}} & -\frac{\i}{2 \sqrt{2}} & \frac{\sqrt{3}}{8} & -\frac{\i}{2} \sqrt{\frac{3}{2}} & -\frac{\sqrt{15}}{8} \\
 \frac{\sqrt{21}}{8\sqrt{2}} & \frac{\i}{2} \sqrt{\frac{3}{2}} & -\frac{9}{16} & -\frac{\i}{2 \sqrt{2}} & \frac{\sqrt{5}}{16} \\
 \frac{\sqrt{21}}{8\sqrt{2}} & -\frac{\i}{2} \sqrt{\frac{3}{2}} & -\frac{9}{16} & \frac{i}{2 \sqrt{2}} & \frac{\sqrt{5}}{16} \\
\end{array}
\right)\,,
\end{equation}
where the rows correspond to $m=0,2,-2,4,-4$ and the columns correspond to $\ell_3=0,1,2,3,4$ respectively.

\section{Transformation of the Weyl multipoles under changes of basis}
\label{sec:WeylTransf}
In this Appendix, we review the transformations of the Weyl multipoles under rotations of the basis vectors and relate them to the transformation properties of the tensor harmonics. 
This is a straightforward generalization of the results in Ref.~\cite{Gualtieri:2008ux} to the higher dimensions. 

The expansion of the Weyl scalars in multipoles is given by
\begin{equation}
\Omega_{AB}'(\theta) = \sum_{\ell} \psi_{\ell} \mathbb{Y}^{\ell}_{AB}(\theta)
\end{equation}
where $\ell$ is a collective label  that specifies each of the tensor harmonics on the $S^n$ at infinity, $\theta$ denotes the collection of angles on such a sphere and $\psi_l$ are the corresponding multipoles. 
Under a rotation, $\Omega_{AB}'(\theta)$ transforms as
\begin{equation}
\Omega_{AB}'(\theta') = R_{A}^{\phantom A C} R_{B}^{\phantom A D}\, \Omega_{CD}'(\theta(\theta'))\,,
\end{equation}
where the $R_{A}^{\phantom A B}$ are rotation matrices. Given the transformation of the harmonics under rotations, 
\begin{equation}
\mathbb{Y}^{\ell'}_{AB}(\theta') = R_{A}^{\phantom A C} R_{B}^{\phantom A D}\sum_\ell  A^{\ell'}_{\phantom{\ell'}\ell} \mathbb{Y}^{\ell}_{CD}(\theta)\,,
\end{equation}
it follows that
\begin{equation}
\psi_{\ell'} = \sum_{\ell}\psi_{\ell}\,(A^{-1})^{\ell}_{\phantom \ell \ell'}
\end{equation}
This transformation rule allows us to compute the Weyl multipoles in the \eqref{eq:basisS4P} from the multipoles computed in the standard basis \eqref{eq:basisS4std}.

\section{Apparent horizon and contours of $\chi$}
\label{sec:AppHrznChi}

\begin{figure}[h!]
\begin{center}
\includegraphics[width=7.5cm]{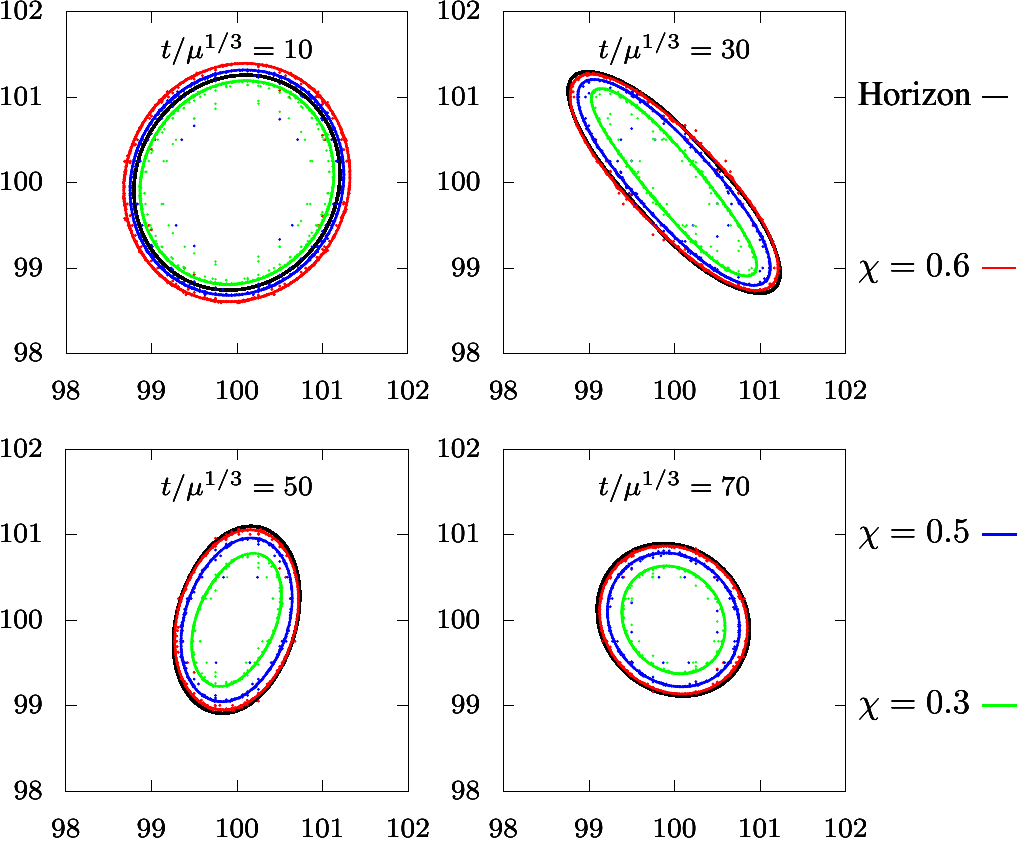}
\end{center}
\caption{Several $\chi$ contours for a 6D MP BH with $a/\mu^{1/3} = 1.3$ and an initial $m=2$ deformation. The AH can be tracked throughout the evolution, and is described well by the $\chi \sim 0.4$ contour at early times, and by the $\chi \sim 0.6$ contour at late times.
}
\label{fig:hrznchi_a1p3}
\end{figure}

In cases in which the AH is a star-shaped surface, we find that contours of the conformal factor $\chi$ closely track the AH obtained by our apparent horizon finder.
Figure~\ref{fig:hrznchi_a1p3} demonstrates this for a 6D MP black hole of dimensionless spin $a/\mu^{1/3} = 1.3$ an initial $m=2$ deformation in $\chi$ described by \eqref{eq:chi_deformation1} and \eqref{eq:chi_deformation2}.
For this simulation, the AH can be tracked by our apparent horizon finder for the entire evolution. 
The AH at early times is roughly followed by the $\chi \sim 0.4$ contour, and as the gauge evolution proceeds, the AH at late times is most closely followed by the $\chi \sim 0.6$ contour.
In particular, at time $t/\mu^{1/3} \sim 30$ when the AH is most elongated, the AH remains described well by the $\chi \sim 0.6$ contour.

\begin{figure}[h!]
\begin{center}
\includegraphics[width=7.5cm]{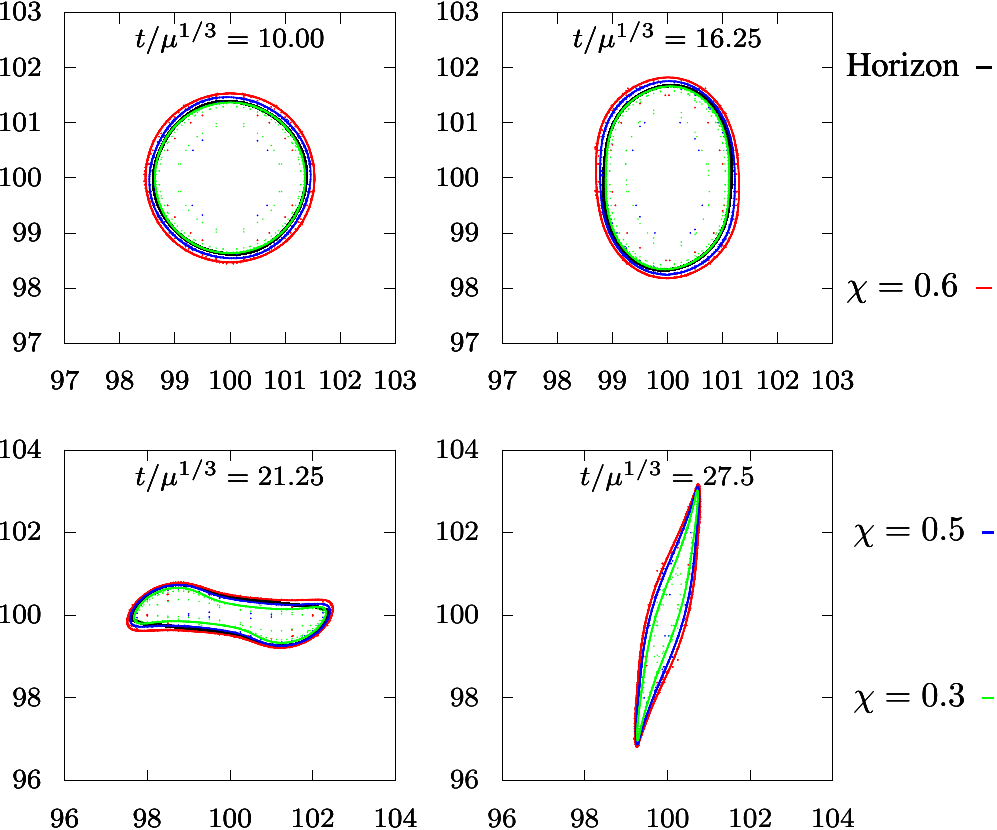}
\end{center}
\caption{Several $\chi$ contours for a 6D MP BH with $a/\mu^{1/3} = 1.5$ and an initial $m=2$ deformation. The AH is well-described by the $\chi \sim 0.4$ contour at early times (see first two panels), and by $\chi \sim 0.5$ at the last time slice when the apparent horizon finder can still track the AH location (see third panel). 
}
\label{fig:hrznchi_a1p5_m2}
\end{figure}

\begin{figure}[h!]
\begin{center}
\includegraphics[width=7.5cm]{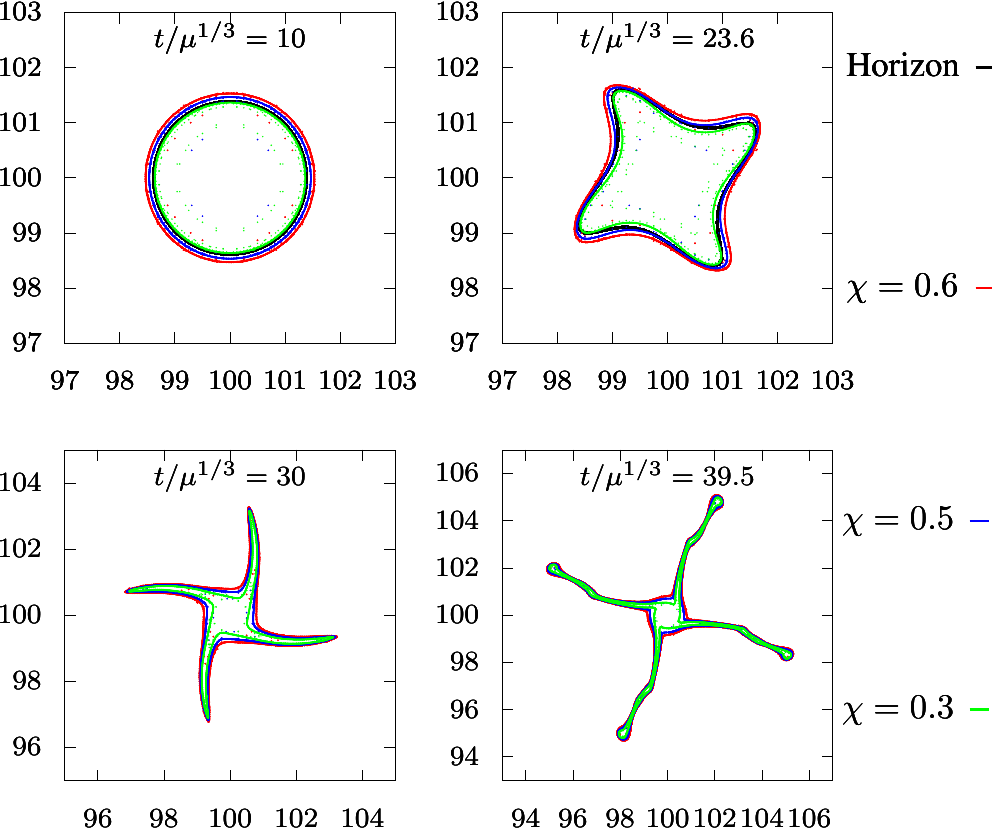}
\end{center}
\caption{Several $\chi$ contours for a 6D MP BH with $a/\mu^{1/3} = 1.5$ and an initial $m=4$ deformation. The AH is well-described by the $\chi \sim 0.4$ contour at early times (see the first panel), and by $\chi \sim 0.5$ at the last time slice when the apparent horizon finder can still track the AH location (see the second panel). }
\label{fig:hrznchi_a1p5_m4}
\end{figure}

Following $\chi$ contours is particularly useful when the AH ceases to be star shaped, and can no longer be tracked by our apparent horizon finder (though, see Refs.~\cite{Figueras:2017zwa, Pook-Kolb:2018igu} for different apparent horizon finder implementations that get around the star-shaped requirement). 
Figures~\ref{fig:hrznchi_a1p5_m2} and~\ref{fig:hrznchi_a1p5_m4} show the $\chi$ contours for two such cases of 6D MP black holes of dimensionless spin $a/\mu^{1/(D-3)} = 1.5$ with initial $m=2$ and $m=4$ deformations in $\chi$, respectively, along with their AH shapes for as long as they can be tracked.

Figure~\ref{fig:hrznchi_a1p5_m2} corresponds to an initial $m=2$ deformation, and Fig.~\ref{fig:hrznchi_a1p5_m4} corresponds to an initial $m=4$ deformation.
In both cases, the AH at early times is roughly followed by the $\chi \sim 0.4$ contour, and the gauge appears to evolve in such a way that the AH at late times is described well by the $\chi \sim 0.5$ contour. 
Thus, we continue to use the $\chi \sim 0.5$ contour as a proxy for the AH shape even at late times when the AH can no longer be found.
With this proxy, the MP BH with an initial $m=2$ deformation develops a thin bar shape (see the last panel of Fig.~\ref{fig:hrznchi_a1p5_m2}), and the MP BH with an initial $m=4$ deformation develops four thin arms (see the last panel of Fig.~\ref{fig:hrznchi_a1p5_m4}). 

\section{$m=3$ perturbation}
\label{sec:Pert_m3}
\begin{figure}[h!]
\begin{center}
\includegraphics[scale=0.25]{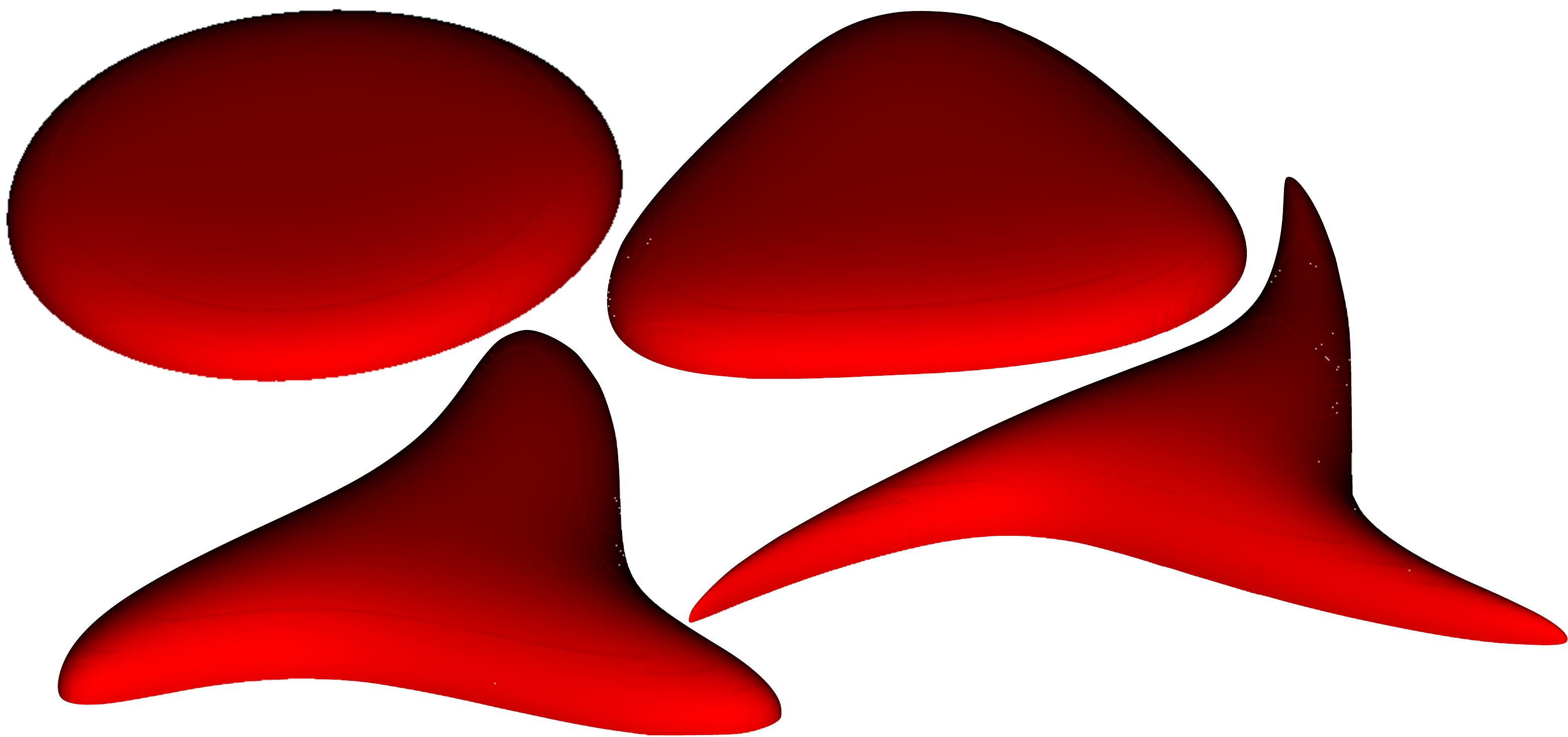}
\end{center}
\caption{Snapshots of the $\chi=0.5$ contour during the evolution of an unstable MP BH with initial $a/\mu^{1/3} = 1.3$ and a $m=3$ perturbation. During the evolution, the black hole develops a triangle shape, and its tips eventually grow into long arms. The latter eventually become GL unstable, as discussed in Sec.~\ref{sec:mp6d_large_spin}}
\label{fig:Evol_Chi_a130_m3}
\end{figure}

We display here for completeness the evolution of the $6D$ MP BH with initial spin $a/\mu^{1/3}=1.3$ perturbed with a $m=3$ deformation, see eq.~\eqref{eq:chi_deformation2}. 
Fig.~\ref{fig:Evol_Chi_a130_m3} shows the evolution of the $\chi=0.5$ contour, which has qualitatively the same features as the $m=4$ perturbation discussed sec.~\ref{sec:mp6d_large_spin}. 
The initial deformation of the black hole horizon into a triangle shape is characteristic of the $m=3$ mode. 
The corners of the triangle grow into three arms, which become elongated and develop sharp features at the edges. 
As in the $m=4$ case, the appearance of these sharp features precedes the formation of long and thin arms, which eventually become GL unstable.  

\section{Convergence Test}
\label{sec:AppConv}

\begin{figure}[h!]
\begin{center}
\includegraphics[width=7.cm]{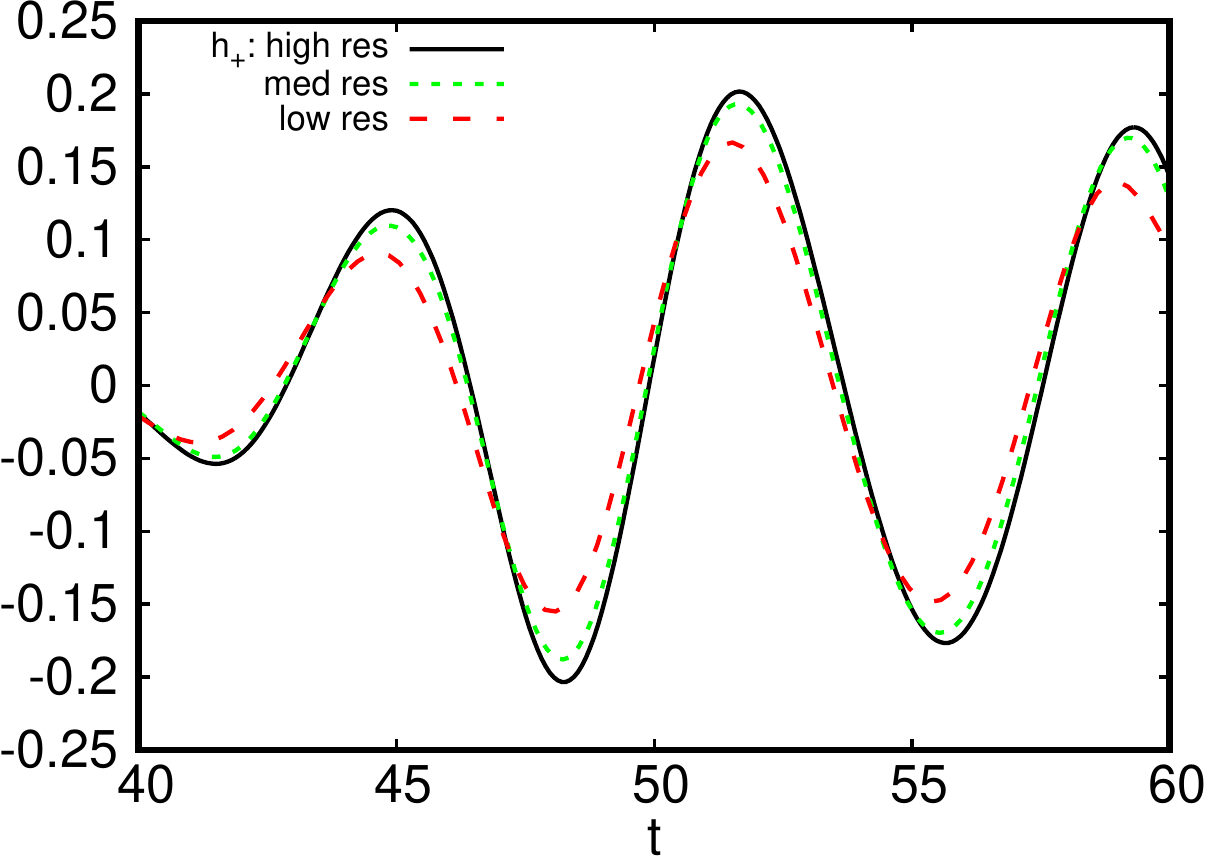}
\end{center}
\caption{Convergence test of the gravitational wave data from a 6D MP BH 
Here, we have zoomed in on the time interval in which the wave is most prominent. The coarsest grid of the low-, medium-, and high-resolution simulations covers each linear dimension by $80$, $120$, and $160$ points, respectively. 
}
\label{fig:hplus}
\end{figure}

Here we present a numerical test taken from a simulation of a 6D MP BH perturbed at $t/\mu^{1/3} = 10$ by an $m=2$ deformation in $\chi$ described by Eqs.~\eqref{eq:chi_deformation1} and \eqref{eq:chi_deformation2}. 
Figure~\ref{fig:hplus} shows the resulting gravitational waves via the $h_+$ component of the metric perturbation given by \eqref{eq:metric_perturbation}, extracted on the $z$ axis at $z/\mu^{1/3} = 29$.
This is done at three different resolutions, for which each linear dimension is covered by $80$, $120$, and $160$ points, respectively. 
The pointwise difference in the waveform between subsequent resolutions yields a convergence factor of approximately $3$, which indicates a rate convergence that is between second and third order.

\end{document}